\documentclass[useAMS,usenatbib]{mnras}
\usepackage[british]{babel}             
\usepackage{newtxtext}               
\usepackage[slantedGreek]{newtxmath}    
\let\la=\lesssim 
\usepackage[T1]{fontenc}   
\usepackage{ae,aecompl}
\usepackage{graphicx}
\usepackage{caption}
\usepackage{amsmath}
\usepackage{hyperref}

\graphicspath{./{figures/}}
\DeclareGraphicsExtensions{.pdf}

\newcommand{\Hii}{H{\sc ii}\ } 
\newcommand{\Te}{T$_\text{e}$}

\newcommand{\msun}{M$_{\odot}$\ }
\newcommand{\mstar}{M$_{\star}$\ }
\newcommand{\oiii}{[\ion{O}{iii}]$\lambda5007$}
\newcommand{\oii}{[\ion{O}{ii}]$\lambda3727,29$}
\newcommand{\sii}{[\ion{S}{ii}]$\lambda6717,31$}
\newcommand{\nii}{[\ion{N}{ii}]$\lambda6584$}

\def\arcsec{\hbox{$^{\prime\prime}$}}

\title[New Te-based, strong-line MZR and FMR]{The Mass-Metallicity and the Fundamental Metallicity Relation revisited on a fully Te-based abundance scale for galaxies}

\author[M. Curti et al.]{Mirko Curti$^{1,2}$,
Filippo Mannucci,$^{3}$
Giovanni Cresci,$^{3}$
Roberto Maiolino$^{1,2}$ \\
\\ 
$^{1}$Cavendish Laboratory, University of Cambridge, 19 J. J. Thomson Ave., Cambridge CB3 0HE, UK\\
$^{2}$Kavli Institute for Cosmology, University of Cambridge, Madingley Road, Cambridge CB3 0HA, UK\\
$^{3}$INAF - Osservatorio Astrofisico di Arcetri, Largo E. Fermi 5, I-50125, Firenze, Italy\\
}

\begin{document}

	\date{Accepted 2019 October 14. Received 2019 October 14; in original form 2019 January 30}
	\pagerange{\pageref{firstpage}--\pageref{fig:lastfig}} \pubyear{2019}
	\maketitle
	\label{firstpage}

%%%%%%%%%%%%%%%%%%%%% ABSTRACT %%%%%%%%%%%%%%%%%%%
\begin{abstract}
The relationships between stellar mass, gas-phase metallicity and star formation rate (i.e. the Mass-Metallicity, MZR, and the Fundamental Metallcity Relation, FMR) in the local Universe are revisited by fully anchoring the metallicity determination for SDSS galaxies on the T$_{\text{e}}$ abundance scale defined exploiting the strong-line metallicity calibrations presented in \cite{Curti:2017aa}.
Self-consistent metallicity measurements allow a more unbiased assessment of the scaling relations involving M, Z and SFR, which provide powerful constraints for the chemical evolution models. We parametrise the MZR with a new functional form which allows us to better characterise the turnover mass. 
The slope and saturation metallicity are in good agreement with previous determinations of the MZR based on the Te method, while showing significantly lower normalisation compared to those based on photoionisation models.
The Z-SFR dependence at fixed stellar mass is also investigated, being particularly evident for highly star forming galaxies, where the scatter in metallicity is reduced up to a factor of $\sim30\%$.
A new parametrisation of the FMR is given by explicitly introducing the SFR-dependence of the turnover mass into the MZR. 
The residual scatter in metallicity for the global galaxy population around the new FMR is $0.054$ dex.
The new FMR presented in this work represents a useful local benchmark to compare 
theoretical predictions and observational studies (of both local and high-redshift galaxies) whose metallicity measurements are tied to the abundance scale defined by the \Te\ method, hence allowing to properly assess its evolution with cosmic time.
\end{abstract}


%%%%%%%%%%%%%%%%%%% INTRODUCTION %%%%%%%%%%%%%%%%%%%
\section{Introduction}
\label{sect:intro}

Galaxies continuously undergo chemical evolution, as heavy elements produced in stars are dispersed into the interstellar medium (ISM) and gas flows regulate the level of metal content by diluting its gas-phase or directly expelling the enriched gas out of the galactic potential well.
The metal content of the gas-phase of the ISM (i.e. the \textit{gas metallicity}) is therefore one of the key physical quantities that has to be considered in galaxy evolution studies, as it is strongly sensitive to all the processes that drive and regulate the baryon cycle within galaxies (see \citealt{Maiolino:2019aa} for a review on the topic).
On global scales, this incessant interplay is naturally reflected by the presence of a number of scaling relations, which encode important informations about the evolutionary stage of galaxies.

Among them, the relation between the stellar mass content (M$_{\star}$) of galaxies and their ISM metallicity (Z), known as the Mass-Metallicity Relation (MZR, \citealt{Lequeux:1979aa}), is probably one of the most famous and thoroughly investigated.
The advent of the Sloan Digital Sky Survey (SDSS, \citealt{York:2000aa}) led to a dramatic improvement in the statistics of both stellar mass and metallicity measurements in the local Universe, allowing to assess the existence of such a relation with much more significance \citep[e.g.][]{Tremonti:2004aa}.
A look to the MZR indubitably reveals how more massive systems appear to be more chemically enriched.
This relation holds for a broad range of stellar masses (from $\sim10^{7}$\msun to $\sim10^{12}$\msun ), but its shape change with varying \mstar: the MZR is steep at low masses, then its slope change in correspondence of a characteristic value of \mstar (the \textit{turnover mass}), asymptotically flattening towards a saturation metallicity.
The interpretation of the MZR involves both secular and dynamical processes: on one side, it may simply imply that massive galaxies represent a more advanced stage of chemically evolution (``chemical downsizing", \citealt{Somerville:2015aa}) whereas, on the other, that they are more capable to retain metals (thanks to their deeper gravitational potential) compared to low mass systems, whose enriched gas can be effectively expelled by winds and outflows \citep[e.g.][]{Chisholm:2015aa}. 
The MZR has been observed to hold also at redshifts out to 3 and beyond, nonetheless showing clear signs of evolution \citep{Erb:2006ad,Maiolino:2008aa,Mannucci:2009aa,Zahid:2011aa,Zahid:2014ab,Yabe:2015aa,Ly:2016ab,Sanders:2018aa}. In fact, high-z galaxies are observed to be less enriched than local ones for a given stellar mass, an effect that is more predominantly observed at low masses.
This evolution can be ascribed to several factors, including an increase in the relative gas content of high-z galaxies (i.e. earlier evolutionary stage or higher dilution by inflows, e.g. \citealt{Erb:2006ac,Lagos:2016aa}), a higher efficiency of gas outflows \citep[e.g.][]{Chisholm:2018aa} and a reduction in the stellar yields driven by a mass dependence of the IMF \citep[e.g.][]{Lian:2018ac}.
Unveiling the origin, the intrinsic properties of the MZR (i.e. its slope, scatter, turnover mass, normalization) and its redshift evolution is therefore crucial to break the degeneracy on the relative contribution that different physical processes play in driving galaxy evolution.

The scatter in the MZR has been observed to correlate with different galaxy properties. \cite{Ellison:2008aa} first showed an anti-correlation between metallicity and specific star formation rate for galaxies at a fixed stellar mass. \cite{Mannucci:2010aa} clearly observed a secondary dependence of the MZR on the SFR in a large sample of SDSS galaxies (then extended toward lower masses by \citealt{Mannucci:2011aa}), with highly star forming galaxies showing lower metallicities at fixed stellar mass, and first proposed that local galaxies are distributed on a tight surface in 3D space defined by mass, metallicity and SFR. 
%(i.e. the so-called ``Fundamental Metallicity Relation'', FMR).
%whose galaxies differently populate at any epoch driven by their secular evolution and reflecting the effects of gas flows.
Perhaps even more interestingly, no significant evolution in this scaling relation is seen up to redshift $\sim3$;
therefore, it is often referred to as the `Fundamental Metallicity Relation'' (FMR).
Although hotly debated in the literature, the result is generally confirmed when the measurements of all the quantities involved (especially metallicity) are performed self-consistently and the associated observational uncertainties are properly taken into account (see the discussion in \citealt{Cresci:2018aa} and references therein).
This suggests that the evolution of galaxies is regulated by smooth secular processes and that an equilibrium condition is set between the involved physical mechanisms over cosmic time.

Many theoretical frameworks managed to reproduce the observed relation by means of the interplay between the infall of pristine gas, the so triggered star formation activity and the amount of enriched material expelled through outflows 
\citep[e.g.][]{Dave:2011aa,Dayal:2013aa,Lilly:2013aa,De-Rossi:2015aa,Dave:2017aa,Torrey:2018aa}.
In this context, the redshift evolution of the MZR would naturally arise by sampling different regions of the non-evolving FMR, given the higher average star formation rate of high-z galaxies.
In the context of gas-equilibrium models, the metallicity-SFR dependence is likely the by-product of a more fundamental relationship between the gas metallicity and the gas fraction: this has been indeed observed in small samples of local galaxies \citep{Bothwell:2013aa, Bothwell:2016aa, Bothwell:2016ab}, as well as predicted by cosmological simulations \citep[e.g.][]{Lagos:2016aa,De-Rossi:2017ab}.

This mutual relationship between \mstar, metallicity and SFR has been observed and confirmed by several authors \citep{Yates:2012aa,Andrews:2013aa,Nakajima:2014aa,Salim:2014aa,Grasshorn-Gebhardt:2016aa,Hunt:2016aa, Hirschauer:2018aa}. 
However, different prescriptions on the methodology of data analysis, including both the way in which SFR and metallicity are measured and the effects related to different sample selection criteria, have led to the assessment of M-Z-SFR relations characterised by very different shapes and properties. 
Furthermore, some recent studies conducted with integral field spectroscopy suggested that the observed global scaling relations may be driven by more local processes involving the gas metallicity, the stellar mass surface density and the surface density of SFR  \citep{Barrera-Ballesteros:2016aa, Gao:2018aa, Sanchez-Almeida:2018ab, Sanchez-Menguiano:2019aa}. 

The impact of different SFR and metallicity measurements, as well as of the potential biases introduced by sample selection in terms of signal-to-noise cuts, on the overall shape of the FMR has been nicely discussed, for instance, by \cite{Telford:2016aa} and more recently by \cite{Cresci:2018aa}.
In particular, one of the largest contributions to the differences reported in the literature about the properties of the FMR can be related to the choice of the metallicity diagnostics and calibrations, especially when comparing abundances determined from the \Te\ method and different methods (i.e. strong-line diagnostics based on predictions from grids of photoionisation models), which are well known to be affected by systematic discrepancies (e.g. \citealt{Kewley:2008aa,Lopez-Sanchez:2012aa}).
These may significantly change the strength of the observed dependencies, in a way that correlates with all the parameters involved \citep{Yates:2012aa, Andrews:2013aa,Telford:2016aa}.
%For example, the systematic metallicity overestimate provided by method calibrated on photoionisation model predictions compared to \Te estimates can modify the 
For this reason, when trying to interpret the observations of galaxies in the framework of the FMR predictions, and especially when comparing samples selected at different redshifts, within different environments etc..., it is of primary importance to adopt a consistent set of metallicity diagnostics, calibrated on the same abundance scale as those used to derive the benchmark scaling relations.
%observational results for 
% of galaxies at different redshifts, environment in light of the FMR framework, it is of primary importance to report all the measurements to the same abundance scale used in the definition of the benchmark relationships.
%want to evaluate whether galaxies at different redshifts, environments, etc.  lie on the FMR we need to be sure we use consistent abundances when we make the comparison. 

Although affected by a certain number of weaknesses, mainly attributed to the presence of temperature and chemical inhomogeneities in the nebulae \citep{Peimbert:1967qv,Stasinska:2002aa,Stasinska:2005aa} and/or to the contribution of diffuse gas to line emission when dealing with global galaxy spectra \citep{Sanders:2017aa,Zhang:2017aa}, the metallicity measurements based on the \Te\ method have proven to be in better agreement with independent measurements of stellar metallicities performed in young (i.e. $< 10$ Myr) stars \citep[see e.g.][]{Bresolin:2016aa,Davies:2017ab}, thus constituting a more reliable absolute scale for chemical abundances compared to that defined by many photoionisation models.
Adopting the proper abundance scale is indeed crucial to reliably compare observational results with predictions of chemical evolution models and simulations, as many of the observed features in the metallicity scaling relations may change when considering different abundance scales.
For example, the very different overall normalisation of the MZR, as provided by various metallicity calibrations, largely affects the correct assessment of the asymptotic metallicity, a quantity that has strong implications in the determination of the effective yields of the stellar populations and the capability of galaxies to retain the produced metals.

In this paper we aim at revisiting the M-Z-SFR relations in the local Universe by adopting a set of strong-line diagnostics,  presented in \cite{Curti:2017aa}, which are self-consistently calibrated on the abundance scale defined by the \Te\  method over the full range of stellar mass and SFR spanned by SDSS galaxies.
%The use of \Te-based metallicities over the full range of stellar mass and SFR spanned by the entire galaxy sample 
This allows us to reduce the possible systematics introduced in the determination of the M$_{\star}$-Z and SFR-Z dependences by non self-consistent metallicity calibrations.
The scaling relations derived in this work will therefore constitute useful benchmarks for forthcoming local studies exploiting metallicity measurements based on the same method, as well as for high redshift studies aimed at investigating their cosmic evolution.

For the purposes of this work, we leverage on the use of strong-line diagnostics to maintain a large statistical significance in all the regions of the parameter space and hence derive more representative properties for the whole galaxy population.   
Considering only subsamples of individual galaxies with auroral line detections would strongly limit the effective range probed 
in \mstar, SFR and metallicity, while stacked spectra (in bins of \mstar and SFR) would not preserve the statistical information needed to assess the effective role of secondary dependences (e.g. preventing a proper evaluation of the reduction of the scatter in metallicity 
in different stellar mass bins when accounting for the secondary dependence on SFR). 
Moreover, adopting a different scheme as e.g. in \cite{Curti:2017aa} (i.e. stacking in bins of [\ion{O}{iii}]/H$\upbeta$ vs [\ion{O}{ii}]/H$\upbeta$) would combine galaxies of different stellar mass and SFR, making it impossible to use composite spectra to assess dependencies in the latter parameters.
%In addition, binning in M-SFR bins assuming a priori the existence of the FMR (otherwise, one could not claim that stacked spectra are representative of the average metallicity of the galaxies falling in such bins), hence biasing the entire analysis.
%Finally, metallicity inferred from stacked spectra are weighted averages on the intensity of the auroral lines, therefore biasing each metallicity determination from M-SFR stacked spectra towards galaxy with the brightest auroral lines, espcially at low masses where the number of galaxies per bin strongly decreases.

The paper is organised as follows.
The sample selection and the methodology used to derive galaxy properties (in particular the gas-phase metallicity) are described in Section~\ref{sec:selection}. % and ~\ref{sec:metallicity} respectively.
In Section~\ref{sec:mzr} we present the analysis of the mass-metallicity relation, while in Section~\ref{sec:fmr} we analyse its dependence on SFR following different approaches and defining a new analytical parametrisation for the FMR. 
Finally, Section~\ref{sec:disc} summarizes our main results.
In Table~\ref{tab:line_ratios}, we report the notations used throughout the paper to indicate the line ratios adopted in our analysis.
%R$_{2}$ = \oii/H$\beta$, R$_{3}$ = \oiii/H$\beta$,\newline
%N$_{2}$ = \nii/H$\alpha$, S$_{2}$ = \sii/H$\alpha$.\newline
%R$_{23}$ = R$_{2}$+R$_{3}$, O$_{3}$O$_{2}$ = \oiii/\oii, \newline
%RS$_{32}$ = R$_{3}$+S$_{2}$, O$_{3}$S$_{2}$ = R$_{3}$/S$_{2}$, \newline
%O$_{3}$N$_{2}$ = R$_{3}$/N$_{2}$. \newline
In this work we adopt a standard $\Lambda$CDM cosmology, assuming the parameters presented by \cite{Planck-Collaboration:2016aa}.
%\textbf{To be completed...}

\begin{table}
	\centering
	\caption{ Definition of line ratios adopted throughout this paper 
	}
	\begin{tabular}{|@{} l|c @{}|}\hline
		 \text{Notation} & \text{Line Ratio} \\
		\hline
	R$_{2}$  &  \oii/H$\beta$ \\
	R$_{3}$ & \oiii/H$\beta$\\
	N$_{2}$ & \nii/H$\alpha$ \\
	S$_{2}$ & \sii/H$\alpha$\\
	R$_{23}$ & (\oii + [\ion{O}{iii}]$\lambda4959,5007$)/H$\beta$ \\
	O$_{3}$O$_{2}$ &  \oiii/\oii \\
	RS$_{32}$ & \oiii/H$\beta$ +  \sii/H$\alpha$ \\
	O$_{3}$S$_{2}$ & \oiii/H$\beta$ / \sii/H$\alpha$  \\
	O$_{3}$N$_{2}$ & \oiii/H$\beta$ / \nii/H$\alpha$  \\
	\hline
	\end{tabular}
	
	\label{tab:line_ratios}
\end{table}

\section{Sample Selection and measured quantities}
\label{sec:selection}

\subsection{Sample selection}
\label{sec:sample}
Our parent sample is drawn from the seventh data release (DR7) of the Sloan Digital Sky Survey \citep{Abazajian:2009aa}, whose galaxy properties and emission line fluxes are provided by the MPA/JHU catalog\footnote{available at \href{http://www.mpa-garching.mpg.de/SDSS/DR7/}{http://www.mpa-garching.mpg.de/SDSS/DR7/}}.
The criteria followed to define our final sample are described in the following.

We required galaxies to be classified as star forming, according to their position on the {[\ion{N}{ii}]-BPT diagram and following the classification scheme by \cite{Kauffmann:2003aa}.
We applied a redshift selection of $\text{z}>0.027$ to ensure the presence of the [\ion{O}{ii}]$\lambda3727$ emission line  within the wavelength coverage of the SDSS spectrograph. This allows us to exploit this particular emission line in the metallicity determination while keeping, at the same time, the sample redshift-consistent with that analysed by \cite{Curti:2017aa}.
However, such a low redshift cut would imply including galaxies with very small sampling inside the SDSS fiber, as $3\arcsec$ are equivalent to a projected physical distance of only $\sim 1.6$ kpc at z$=0.027$, thus sampling only the most inner regions.
On one side, this would introduce more uncertainties when applying the aperture corrections for the SFR (see below) and, on the other side, it would make the metallicity measured within the fibre less representative of the global galaxy metallicity, being more sensitive to the presence of metallicity gradients.
To mitigate this problem, we decided to include in our analysis only those galaxies with a covering factor of at least $10\%$, as inferred from the fraction of the total light that goes into the fibre. 
 %This criterion removes $12,252$ galaxies.
In addition, we discarded all galaxies whose catalogue flags indicates unreliable SFR estimates, which includes also all those galaxies 
showing unphysical aperture correction factors lower than $1$.
%this removed another $\sim 17,088$ objects.

%Moreover, applying a lower redshift threshold reduce the biases introduced by aperture effects  
%\citep{Kewley:2005aa}; setting the redshift limit to $0.04$ as proposed in that work, however, may bias our sample towards 
%higher stellar masses,
%due to the decrease in the number of low mass galaxies, preventing in particular a proper analysis of the low mass end of the M-Z-SFR relation.
%In any case, a different choice in redshift selection than z$> 0.027$ does not significantly change the results of this work.

In terms of signal-to-noise cuts on emission lines, following \cite{Mannucci:2010aa} we have applied a SNR threshold\footnote{applying the re-scaled uncertainties provided by the MPA/JHU group, which include both the uncertainties on the spectrophotometry and continuum subtraction} of $15$ on the H$\upalpha$ flux only.
We do not apply any cut on oxygen, nitrogen and/or sulfur lines, to minimize possible biases in metallicity determination as caused by removing galaxies with low SNR on the emission lines involved in the metallicity diagnostics.
In particular, low metallicity galaxies have low SNR on nitrogen lines, while high metallicity galaxies have low SNR on oxygen lines. 
This means that, for instance, introducing a SNR cut on the \oiii\ would translate in removing, in the high metallicity regime (thus preferentially at high masses), more metal rich than metal poor galaxies at fixed stellar mass, in a way that correlates with SFR, hence biasing our determination of the MZR and the M-Z-SFR relation. 
As an example, it has been shown that a SNR cut on metal lines may contribute to the apparent intersection between the various MZR curves at fixed SFR (see e.g. \citealt{Yates:2012aa,Kashino:2016ab}), which could be in principle interpreted as an inversion in the nature of the metallicity-SFR dependence, but that is likely just the consequence of combining selection effects with different SFR and metallicity estimates \citep{Cresci:2018aa}.
For a more in-depth discussion on how S/N cuts on different emission lines could introduce biases in the metallicity measurements as a function of different parameters (like \mstar and SFR) in SDSS galaxies, see \cite{Telford:2016aa}.
%$3$ on H$\upbeta$, [\ion{O}{ii}]$\lambda3726,3729$, [\ion{O}{iii}]$\lambda5007$ and 
%[\ion{N}{ii}]$\lambda6584$, and of $15$ to H$\upalpha$.

To reliably compute the dust attenuation correction however, a minimum SNR of $3$ on H$\upbeta$ was also imposed (this removes only additional $70$ galaxies). 
All emission lines were corrected for reddening from the measured Balmer Decrement, assuming the case B recombination (H$\upalpha$/H$\upbeta$=2.87) and adopting the \cite{Cardelli:1989aa} extinction law.
We then also discarded from the analysed sample all galaxies characterized by extreme extinction, i.e. with values of E(B-V) higher than $0.8$.
%as inferred from the Balmer decrement assuming the case B recombination (H$\upalpha$/H$\upbeta$=2.87), and adopting the \cite{Cardelli:1989aa} MW extinction law.

Finally, we cross-matched DR7 objects with photometric flags from DR12 (which differ from those reported in the DR7 
catalogue) and removed galaxies whose photometric flags include $\rm{DEBLEND\_NOPEAK}$ and $\rm{DEBLEND\_AT\_EDGE}$.
We also removed galaxies whose stellar mass correction factors are lower than $1$, i.e. where the stellar mass derived from the total photometry is lower that the stellar mass derived from the photometry within the fibre.
In addition, we have also visually inspected all the objects with log(\mstar) < 8.6 and manually removed the residual poorly deblended systems, which account for another $3\%$.
After applying all these criteria, the total analysed sample is reduced to $153,452$ galaxies.

\subsection{Stellar mass and star formation rate}
\label{sec:mass_sfr}
The stellar masses for our sample are provided by the MPA/JHU catalog and have been estimated from fits to the photometry, following the prescription of \cite{Kauffmann:2003ab} and \cite{Salim:2007aa}. 
Star formation rates used in this work are derived from the extinction corrected H$\upalpha$ luminosity inside the fibre, adopting the calibration proposed by \cite{Kennicutt:2012aa}. 
We then apply the aperture corrections provided by the MPA/JHU catalog, which build on the work of \cite{Salim:2007aa} to improve those originally provided by \cite{Brinchmann:2004aa}, to compute the total SFR for our galaxies.
Both stellar masses and SFRs estimates are re-scaled to a common \cite{Chabrier:2003aa} IMF.

We choose to adopt the total SFR in our framework, despite the uncertainties potentially introduced by the aperture corrections \citep[e.g.][]{Richards:2016aa}, in order to give a global picture of the mutual relationships between \mstar, SFR and metallicity and to facilitate the comparison with IFU-based studies, both in the local and high-redshift Universe.
However, the use of fiber-based SFR measurements, despite sampling only the central $3$\arcsec of galaxies and thus systematically underestimating the total SFR, is still valuable to characterize the global star formation activity and investigate trends with other quantities.
%without introducing any of the uncertainties and possible biases related to the aperture correction procedure \citep[e.g.][]{Richards:2016aa}.
Moreover, if the metallicity dependence on the SFR is driven by a local dilution effect, being related to the local gas reservoir, linking the metallicity and the SFR measured in the same area (i.e. into the fibre) can be considered more physically meaningful. 
For this reason, in this paper we present the results of our analysis for both cases, i.e. adopting a total-and a fibre-based SFR.
The main differences between the two scenarios will be discussed throughout the paper.
Note that \cite{Mannucci:2010aa} adopted fiber SFRs, but applying a much higher redshift cut to the sample (i.e. z$>0.07$), 
thus selecting galaxies with an higher coverage fraction within the $3\arcsec$ fiber.

\subsection{Gas-phase Metallicity}
\label{sec:metallicity}

\begin{figure*}
	\centering
	\includegraphics[width=0.98\textwidth]{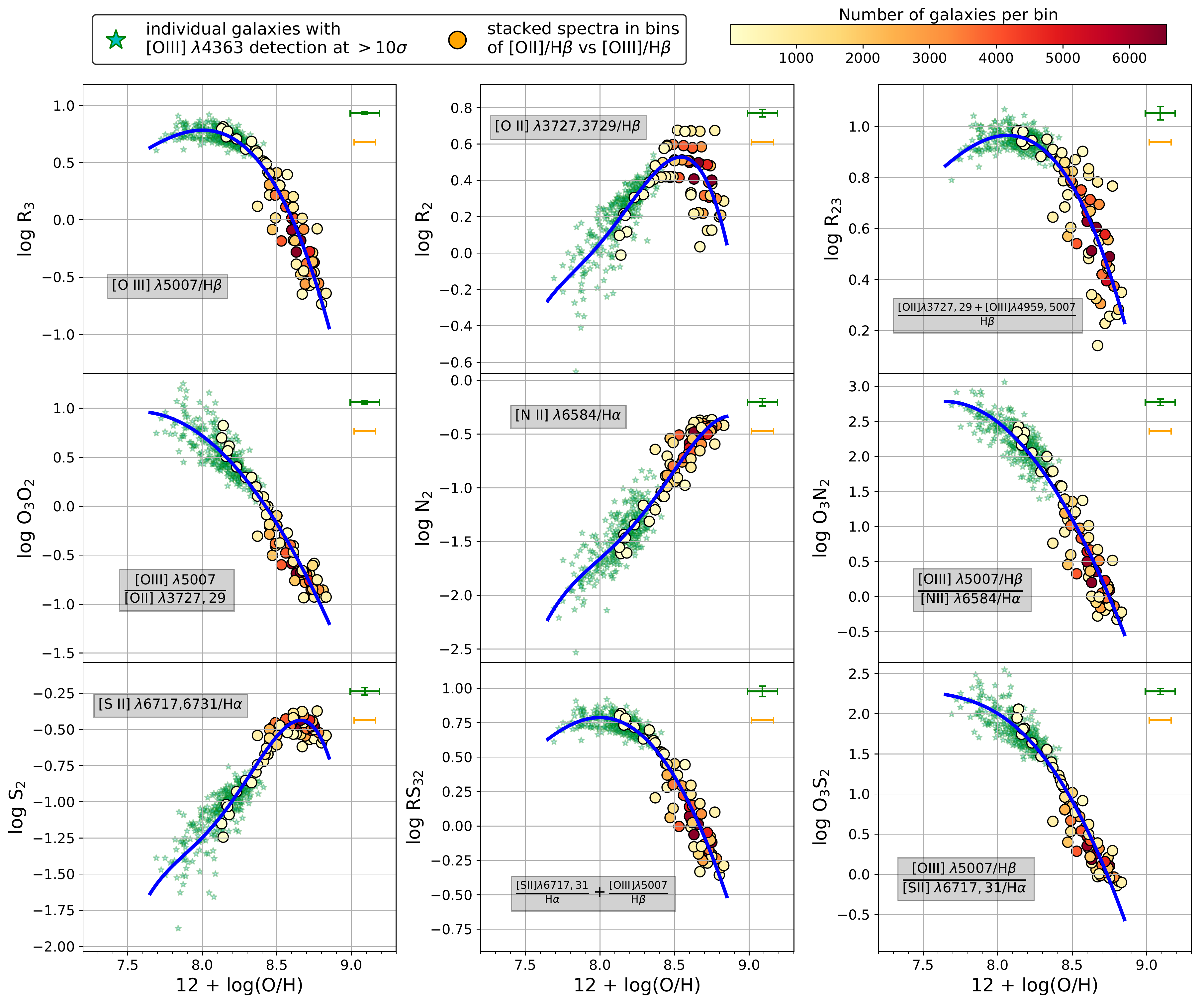}
	\caption{The set of strong-line metallicity diagnostics used in this work. All of them have been calibrated from Te-based measurements of oxygen abundance in a combined sample of individual galaxies (small green stars) and stacked spectra (circular points, color coded by the number of galaxies per stack) in bins of log([\ion{O}{ii}]/H$\beta$) - log([\ion{O}{iii}]/H$\beta$) )(see \citealt{Curti:2017aa} for details on the stacking procedure and analysis). 
		The blue curves represent the polynomial fit that defines the calibration for each diagnostic. }
	\label{fig:met_cal_recap}
\end{figure*}
\begin{table*}
	\centering
	\caption{
		Best-fit coefficients of the polynomial functions ($\text{log(R)} = \sum_{N}c_{n}x^{n}$, with x=Z-8.69) defining the metallicity calibrations presented in Fig.~\ref{fig:met_cal_recap}.
		The RMS column report the root-mean-square of the fit for each calibrator, while $\sigma$ represents an estimate of the dispersion of the calibration along the metallicity axis.}
	\begin{tabular}{|@{} l|c|c|c|c|c|c|c|c|c @{}|}\hline
		
		\text{Diagnostic} & c$_{0}$ & c$_{1}$ & c$_{2}$ & c$_{3}$ & c$_{4}$ & c$_{5}$ & \text{RMS} & $\sigma$ \\% & \text{Range} \\
		\hline
		R$_{2}$ 	&  $0.435 $ &  $-1.362$  & $ -5.655$ &  $-4.851$ &  $-0.478$ &  $0.736$    & $0.11$ & $ 0.10 $ \\ %& $[7.6, 8.4]$ \\
		R$_{3}$ 	&  $-0.277$ &  $-3.549$  & $ -3.593$ &  $-0.981$ & &		 			   & $0.09$ & $ 0.07 $ \\%& $[8.0,8.9]$ \\
		O$_{3}$O$_{2}$  &  $-0.691$ &  $-2.944$  & $ -1.308$ & & &	  			   & $0.15$ & $ 0.14 $ \\%& $[7.6, 8.9]$ \\
		R$_{23}$    &  $0.527 $ &  $-1.569$  & $ -1.652$ &  $-0.421$ & &			   & $0.06$ & $ 0.12 $ \\%& $ [8.1,8.9]$ \\
		N$_{2}$    &  $-0.489$ &  $1.513 $  & $ -2.554$ &  $-5.293$ & $-2.867$ &			   & $0.16$ & $ 0.10 $ \\%& $[7.6, 8.9]$ \\
		O$_{3}$N$_{2}$ 		 &  $0.281 $ &  $-4.765$  & $ -2.268$ & & &			   & $0.21$ & $ 0.09 $ \\%& $[7.6, 8.9]$ \\
		S$_{2}$  	 &  $-0.442$ &  $-0.360$  & $ -6.271$ & $-8.339$  & $-3.559$ &			   & $0.11$& $ 0.06 $ \\%& $[7.6, 8.65]$ \\
		RS$_{32}$   &  $-0.054$ &  $-2.546$  & $-1.970 $ & $0.082 $  &  $0.222$ &			   & $0.07$ & $ 0.08 $ \\%& $[8.0, 8.9]$ \\
		O$_{3}$S$_{2}$ 		 &  $0.191 $ &  $-4.292$  & $ -2.538$ & $0.053 $  &  $0.332$ & 			   & $0.17$ & $ 0.11 $ \\%& $[7.6, 8.9]$ \\
		
		\hline
	\end{tabular}
	
	\label{tab:cal_coeff}
\end{table*}

To measure the metallicity we use a combination of different strong-line diagnostics, assuming the calibrations presented in \cite{Curti:2017aa} which are consistently defined on the T$_{\text{e}}$-based abundance scale over the entire metallicity range spanned by SDSS galaxies.
%Stacked spectra in bins of \oii/H$\upbeta$-\oiii/H$\upbeta$ have been exploited to obtain high SNR detection of temperature sensitive auroral lines and derive the oxygen abundances. 
The full set of metallicity indicators calibrated with this method is presented in Fig.~\ref{fig:met_cal_recap}.
Each calibration has been derived from a set of individual low-metallicity galaxies with auroral lines detection together with stacks of high metallicity galaxies where auroral lines are detected in composite spectra; in this way, the oxygen abundance is self-consistently measured via the \Te\ method for the entire calibration sample.
Compared to the set of diagnostics originally presented in \cite{Curti:2017aa}, here we include three additional calibrated  line ratios involving sulfur lines: these are in particular S$_{2}$ (\sii/H$\alpha$), RS$_{32}$ (\oiii/H$\beta$ + \sii/H$\alpha$) and O3S2 (\oiii/H$\beta$ / \sii/H$\alpha$).
The first indicator is similar to the N$_{2}$ (\nii/H$\alpha$) diagnostic, saturates at high metallicities but can be useful when dealing with low S/N detections of the [\ion{N}{ii}]$\lambda6584$ emission line or with low resolution spectra where this line is blended with H$\alpha$. 
%The latter indicator analogous to O$_{3}$O$_{2}$.
RS$_{32}$ is instead similar to R$_{23}$ (i.e. \oii/H$\beta$ + \oiii/H$\beta$), given the similar ionization potential of the S$^{+}$ and O$^{+}$ ions,  but has the advantages of being unaffected by dust extinction and involves a set of emission lines which are more easily observable even in high-z galaxies, as they fall in the main near-infrared bands for a large range of redshifts.
Finally, the latter indicator is similar to O$_{3}$N$_{2}$ (i.e. \oiii/H$\beta$ / \nii/H$\alpha$), being also similarly unaffected by dust attenuation.

Table~\ref{tab:cal_coeff} summarizes all the coefficients $c_{n}$ of the polynomial functional forms defining the calibrations presented in Fig.~\ref{fig:met_cal_recap}; each calibrator is presented in the form $\text{log(R)} = \sum_{N}c_{n}x^{n}$, where R is the considered diagnostic and $x$ is the oxygen abundance normalised to the solar value (Z$_{\odot}=8.69$, \citealt{Allende-Prieto:2001aa}).
The table also reports the root-mean-square (RMS) of the residuals of the fit, which can be assumed as an estimate of the dispersion of the calibrations along the y-axis (i.e. the dispersion in line ratios at fixed metallicity), together with an estimate of the dispersion of the calibration along the x-axis (i.e. dispersion in metallicity at fixed line ratio), labeled as $\sigma_{\text{O/H}}$. 
This latter quantity is estimated comparing the expected metallicity (applying a calibration to the measured line ratio) and the true metallicity (computed with the \Te\ method) for each point of the calibration sample, and can be assumed as the minimum uncertainty that should be associated with metallicity measurements obtained by using the corresponding diagnostic individually.
All the diagnostics proposed in this work have been calibrated against metallicity in the range 12+log(O/H) $\in$ [$7.6$,$8.9$] and can be safely applied within this range, whereas applications outside this range would rely on extrapolations of the polynomial functions, which may lead to spurious metallicity measurements.
For the purpose of this work we use different combinations of diagnostics, according to the availability and SNR of the involved lines,  following the scheme presented in Table~\ref{tab:diag_used}.
For sake of clarity, in Fig.~\ref{fig:mzr_all_diags} we plot each galaxy on the \mstar-log(O/H) plane color-coded according to the different combination of emission lines involved in its metallicity calculation (upper panel).
We also show, within $0.5$ \rm{dex} wide stellar mass bins, the histograms of the metallicity distribution for each galaxy subsample associated to a different set of emission lines (bottom panels).
We note that for the vast majority of the sample, both globally and at each fixed stellar mass, it is possible to include all the emission lines (i.e. \oii, \oiii, \nii, \sii) in the metallicity calculation, as revealed by the predominance of sky-blue points and bars. 
However, with increasing stellar mass the relative fraction of galaxies whose metallicity has been inferred from a different combination of lines increases, as a primary consequence of the \oiii\ line falling below the detection threshold.

%However, some of these diagnostics, as known, are affected by degeneracies, with multiple metallicities provided for a given line ratio, or suffer from saturation over a given abundance value; for this reason, the last column in the table report the metallicity interval 
%The last column of Table~\ref{tab:cal_coeff}
\begin{table}
	\centering
	\caption{Combination of metallicity diagnostics adopted in this work. Composite diagnostics (in parenthesis) are only used when individual diagnostics approaches their saturation values, e.g. for log(R$_{3}$) > 0.5 and/or log(R$_{2}$) > 0.45.	}
\resizebox{\columnwidth}{!}{
	\begin{tabular}{|@{} l|c|c @{}|}\hline
		\text{Lines detected at $\geq 3 \sigma$} & \text{Diagnostics used} & \# \text{galaxies} \\
		\hline
$[\ion{O}{iii}],[\ion{O}{ii}],[\ion{N}{ii}],[\ion{S}{ii}]$    & R3, R2, N2, S2, (O3N2,O3O2)         & $115005$ \\  
$[\ion{O}{iii}],[\ion{N}{ii}],[\ion{S}{ii}] $					  & R3,  N2, S2, (O3N2,O3S2)              		    & $5292$ \\
$[\ion{O}{ii}],[\ion{N}{ii}],[\ion{S}{ii}] 	$				      & R2, N2, S2               				 	     & $14917$ \\
$[\ion{O}{iii}],[\ion{O}{ii}],[\ion{S}{ii}] $					  & R3, R2, S2, (O3S2)          				 & $108$ \\
$[\ion{O}{iii}],[\ion{O}{ii}],[\ion{N}{ii}] $				      & R3, R2, N2, (O3N2)         				   & $1853$ \\
$[\ion{N}{ii}],[\ion{S}{ii}] $									      & N2, S2         								   & $14547$ \\
$[\ion{O}{ii}],[\ion{O}{iii}] $									      & R2, R3, (O3O2)         			    					   & $8$ \\
%$[\ion{O}{ii}],[\ion{N}{ii}] $									      & R2, N2         			    					  & $?$ \\
$[\ion{O}{iii}],[\ion{N}{ii}] $							              & R3, N2, (O3N2)       			   			 & $502$ \\
		\hline
%\multicolumn{3}{c}{} 		
	\end{tabular}
}
	\label{tab:diag_used}
\end{table}

\begin{figure}
	\centering
	\includegraphics[width=0.975\linewidth]{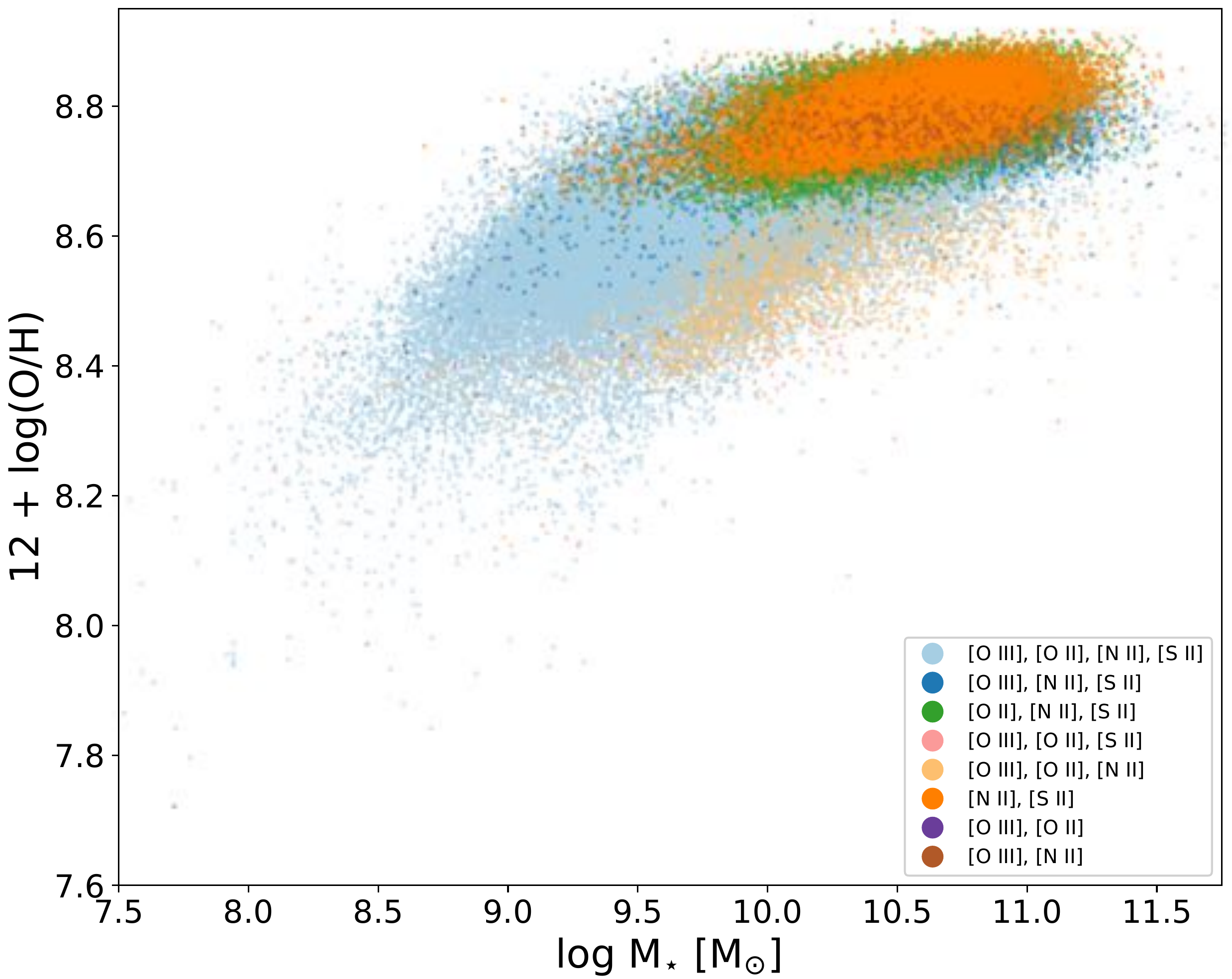}\\
	\vspace{0.3cm}
	\centering
	\includegraphics[width=0.99\linewidth]{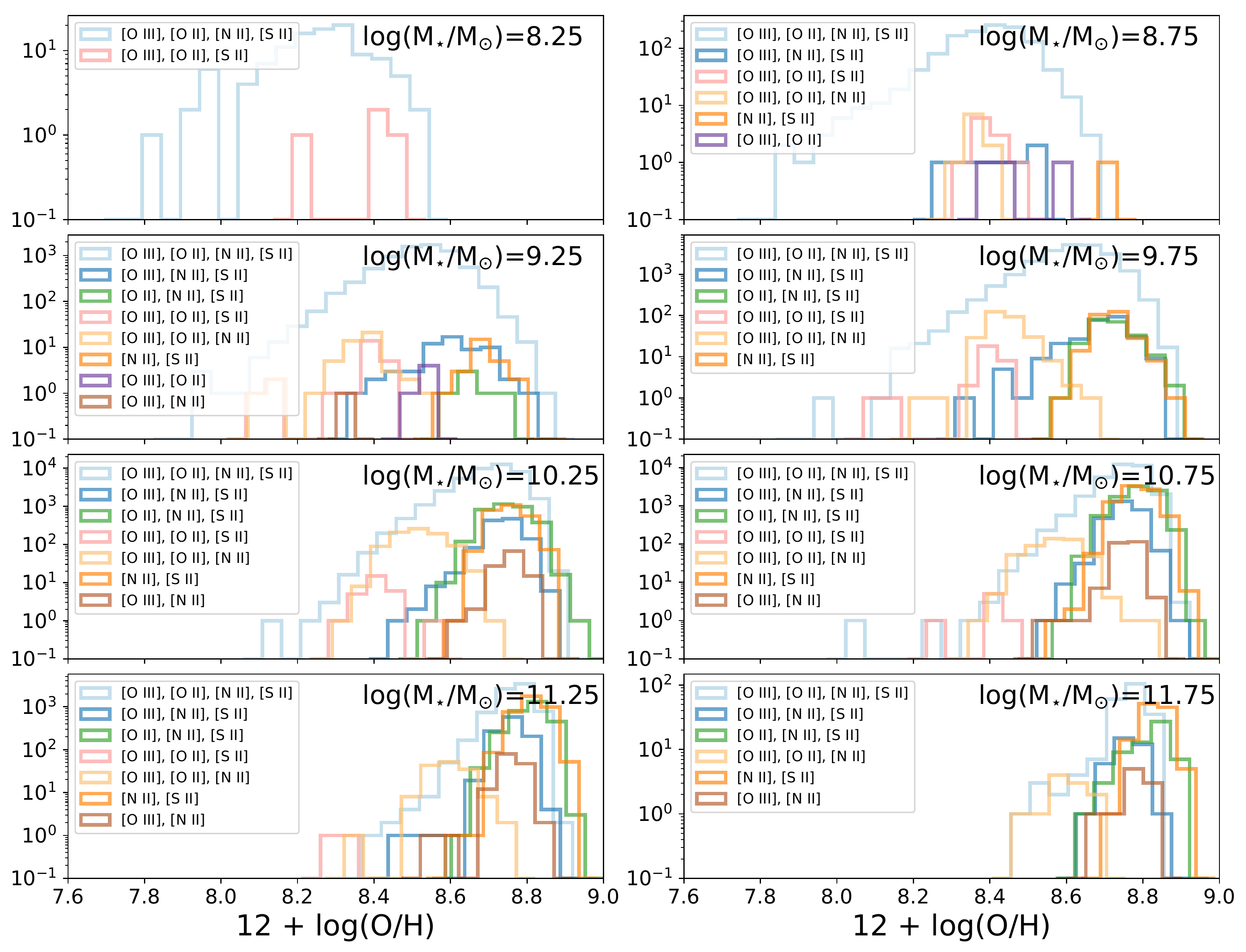}\\
	
	\caption{\textit{Upper Panel:} Distribution of galaxies in our sample on the stellar mass-metallicity plane. 
		Each galaxy on the plot is color-coded according to the different combination of emission lines involved in its metallicity calculation, following the scheme presented in Table~\ref{tab:diag_used}.\newline
	\textit{Bottom Panels:} Histograms (in log-scale) of the metallicity distribution, within \mstar bins of $0.5$ \rm{dex}, for each of the subsamples whose metallicty has been derived adopting a different set of emission lines .
}
	\label{fig:mzr_all_diags}
\end{figure}

%When all the metal lines (\oii, \oiii, \nii and \sii) are detected at more than $3 \sigma$, we exploit simultaneously
%this four, independent diagnostics:  R3, O3O2, S2 and N2. 
Using multiple diagnostics at the same time is crucial to break the degeneracies affecting the calibrations of double-branched indicators and to exploit the information provided by multiple emission lines, whose (both direct and indirect) dependence on O/H can vary in different metallicity ranges, hence setting tighter constraints on the final abundance measurement.
Nevertheless, in our procedure we avoid using double-branched diagnostics when their calibrations approach the region of the ``plateau", hence always choosing the best possible combination of independent and monotonic metallicity indicators.
In practice, this translates in avoiding the use of R$_{3}$ when log(R$_{3}$)$>0.5$ and R$_{2}$  when log(R$_{2}$)$>0.45$, encoding the information from those emission lines only within other diagnostics (like O$_{3}$O$_{2}$ or O$_{3}$N$_{2}$).
However, this criterion affects overall only a small number of galaxies, preferentially at low \mstar.

The metallicity of each galaxy in the sample is computed by searching for the value 
that minimizes the chi-square defined simultaneously by the selected diagnostics as :
\begin{equation}
\chi^{2} = \sum_{i} \frac{(R_{\text{obs}_{i}} - R_{\text{cal},i})^{2}}{\sigma_{\text{obs}}^{2} + \sigma_{R_{\text{cal},i}}^{2}} \ ,
\end{equation}
where R$_{\text{obs}_{i}}$ are the observed line ratios while R$_{\text{cal},i}$ are the values predicted by the calibration for a given metallicity.
Both the uncertainty on the observed line ratio $\sigma_{\text{obs}}$  and the intrinsic dispersion of calibration $\sigma_{R_{\text{cal},i}}$ are taken into account in the procedure. 
A Monte Carlo Markov Chain (MCMC) is run, varying line fluxes according to a Gaussian distribution centred on the measured line ratio and $\sigma$ equal to the measurement error, and the chi-square is minimised at each step to generate a log(O/H) distribution.
%compute the uncertainties on the metallicity estimates. 
All line ratios have been preliminary corrected for reddening (assuming the case B recombination and the \citealt{Cardelli:1989aa} law), 
so extinction is not a free parameter in the procedure.
However, the error on the reddening correction is propagated on the uncertainty of each observed line ratio.
From the generated log(O/H) distribution, the median value is assumed as the inferred metallicity while the $1 \sigma$ interval (computed from the 16th and 84th percentiles) is adopted as an estimate of the associated uncertainties.
We discarded galaxies whose uncertainty on the final metallicity exceeded $0.3$ dex; therefore, the final number of galaxies with robust oxygen abundance determination is $151,862$.

A very important consideration that we want to stress here is that the main results presented throughout the paper
are robust against the choice of different combinations of line ratios, as proven also by the good internal consistency of the proposed calibrations (see Figure 10 and 11 of \citealt{Curti:2017aa}).
Nonetheless, some slight differences in the metallicity determination may arise due to small systematics between the nitrogen-based and oxygen-based diagnostics, with \nii-based diagnostics preferentially underestimating metallicity compared to purely oxygen-based ones, especially in the low metallicity regime (where the calibrations are less constrained and more uncertain due to the smaller statistics).
However, this does not significantly affect or hide in any manner the presence of trends between \mstar, metallicity and SFR, 
although it can change the strength of their mutual dependencies (see also the discussions in Section~\ref{sec:mzr_comparison} and Section~\ref{sec:m_z_sfr_params}).
For this reason, in Appendix B we present and discuss the differences between the MZR and M-Z-SFR relations derived adopting only nitrogen- and only oxygen-based metallicity calibrations respectively.
The simultaneous use of multiple diagnostics is indeed aimed at minimising the impact of these potential systematics.

%\subsection{A calibration for N/O}
%
%...

\section{The Mass-Metallicity Relation}
\label{sec:mzr}

\subsection{A new parametrisation for the MZR}
\label{sec:new_param_mzr}

%In the left panel of Figure~\ref{fig:mzr_1} 
We study here the distribution of our galaxy sample in the \mstar vs log(O/H) plane, i.e. the mass-metallicity relation (MZR).
%To investigate its properties, 
In order to derive the representative properties of this scaling relation, we sort the sample in $0.15$ dex wide stellar mass bins and compute the median and standard deviation of the metallicity distribution in each bin; we limit the analysis only to \mstar bins including at least 25 galaxies, i.e. for $7.95 < $log(M$_{\star}$)$ < 11.85$,  in order to maintain a meaningful statistical representation.
%; the fit to the MZR is then to be considered robust only within this range.
\begin{figure*}
	\centering
	\includegraphics[width=0.65\textwidth]{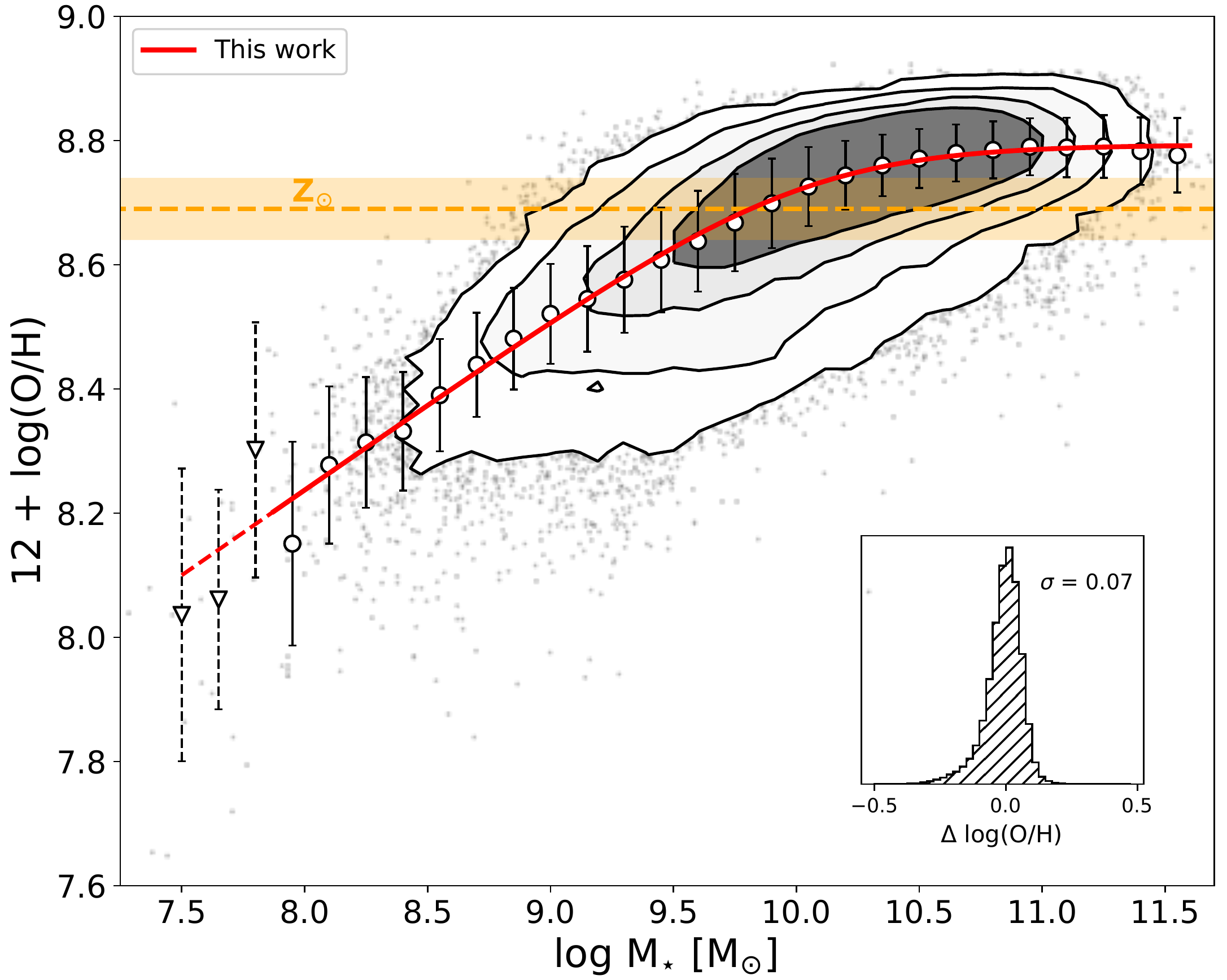} \\
	\includegraphics[width=0.65\textwidth]{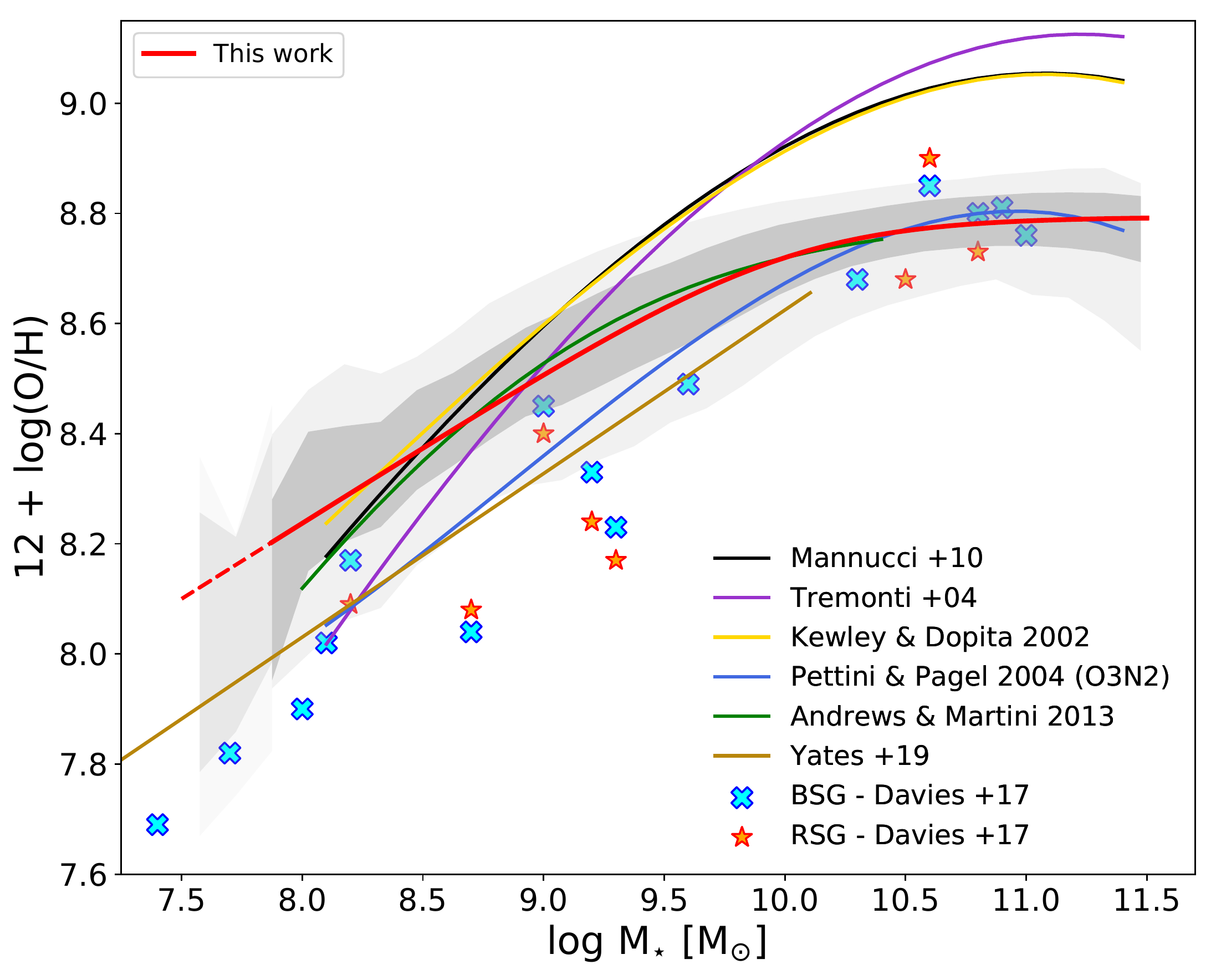} \\
		\caption{\textit{Upper Panel}: Mass-Metallicity relation (MZR) for the sample of SDSS galaxies analysed in this work, as derived by the set of metallicity calibrations presented in Fig.~\ref{fig:met_cal_recap}.
		Grey points represent individual galaxies, while the filled regions encompass the $1$,$2$ $3$ and $4\sigma$ levels of the density contours of the distribution in the log(M$_{\star}$) - log(O/H) plane. 
	   White circles (and associated errorbars) are median metallicities (and dispersions) in narrow $0.15$ dex bins of stellar mass, while the solid red curve represents the best fit to those median points according to the MZR parametrization given in Eq.~\ref{eq:mzr}. 
	   The dashed red part instead is the extrapolation of the MZR fit in the low-mass regime, with the low-mass bins (i.e. with less than $25$ objects) represented by white triangles with dashed error bars.
	   The dashed orange band marks the value assumed for the solar abundance (i.e. Z$_{\odot} = 8.69 \pm 0.05$, \citealt{Allende-Prieto:2001aa}).
	  The small box in the lower right part of the figure show the distribution of the metallicity dispersion of the individual galaxies around the best-fit MZR, whose $1\sigma$ dispersion is equal to $0.07$ {\rm dex}.
    \textit{Bottom Panel}: Comparison between the MZR derived in this work (red curve, with grey areas encompassing the $1\sigma$ and  $2\sigma$ dispersion in each log(M$_{\star}$) bin) and different predictions of the MZR from previous studies in the literature, color coded as reported in the legend. Strong-line MZRs have been re-derived by applying each different calibration method to our sample.
    In particular, the \citealt{Kewley:2002aa, Tremonti:2004aa,Mannucci:2010aa} curves are anchored to abundance scale defined by different grids of photoionization models, while the \citealt{Andrews:2013aa}, \citealt{Pettini:2004aa} and \citealt{Yates:2019aa} 
    curve are based on the T$_{\text{e}}$-based abundance scale. Red stars and blue crosses represent the abundances derived in nearby galaxies from stellar spectroscopy of red an blue supergiants respectively, as collected by \citealt{Davies:2017ab}.}
	\label{fig:mzr_1}
\end{figure*}

The \textit{median} MZR is then parametrized with the following functional form :  
\begin{equation}
\label{eq:mzr}
\text{12 + log(O/H)} = \text{Z}_{0}  - \gamma/\beta * \text{log}\bigg(1 + \bigg(\frac{\text{M}}{\text{M}_{0}} \bigg)^{-\beta} \bigg)\ .
\end{equation}
In this equation, Z$_{0}$ is the metallicity at which the relation saturates, quantifying the asymptotic upper metallicity limit, while M$_{0}$ is the characteristic turnover mass above which the metallicity asymptotically approaches the upper metallicity limit, Z$_{0}$. 
At stellar masses M$_{\star}$ < M$_{0}$, the MZR reduces to a power law of index $\upgamma$. 
Compared to previous works (e.g. \citealt{Moustakas:2011aa, Zahid:2014ab}), the presence of the $\upbeta$ parameter allows us to better control the width of the transition region between the two extremes, providing a better constraint on the turnover mass and overall a better fit to our median points. 
In other words,  $\upbeta$ quantifies how ``fast'' the curve approaches its saturation value: the smaller the value of  $\upbeta$, the broader the knee, and viceversa.
We perform a robust fit of our median MZR relation adopting the the above functional form and using the python-based LMFIT package \citep{lmfit:2014}. 
The data are weighted in the fit according to the metallicity dispersion and the number of objects in each individual \mstar bin
and the errors on the parameters are estimated from the $1\sigma$ confidence levels based on an MCMC simulation.
The best-fit parameters with the associated errors are given in Table ~\ref{tab:mzr_fit}. 
The scatter of individual galaxies around the best-fit relation is $0.07$ {\rm dex}, somewhat lower than in previous determinations 
(which found a scatter around $\sim 0.1$ {\rm dex}, e.g.  \citealt{Tremonti:2004aa}).
%Compared to previous works (e.g. \citealt{Moustakas:2011aa, Zahid:2014ab}), the presence of the \textit{a} parameter allows to better control the width of the turnover knee, providing a different estimate of the turnover mass and a better fit to our median points; in particular, it quantifies how ``fast'' the curve approaches its saturation value: the smaller the value of \textit{a}, the broader the knee, and viceversa.
%Assuming this parametrization, M$_{0}$ represents the mass of knee and $\upgamma$ is the slope of the linear part in the low-mass end of the relation.
%In particular, adopting a slightly modified version of the \cite{Zahid:2014ab} functional form (i.e. substituting the exponential function inside the 10-based logarithm with a power of ten), we obtain a consistent value for the turnover ($10.18$)
%However, we fit to our data even with the functional form proposed by \cite{Zahid:2014ab} (equation $5$ of their paper), whose results are also summarized in Table ~\ref{tab:mzr_fit}.
%\begin{equation}
%\label{eq:mzr}
%\text{12 + log(O/H)} = \text{Z}_{0} + \text{log} \Bigg(1 - \text{exp}\Bigg( \frac{M_{\star}}{M_{0}} \Bigg)^{\gamma} \Bigg)\ .
%\end{equation}

The upper panel of Figure~\ref{fig:mzr_1} shows our new strong-line MZR: small grey points are individual galaxies, while grey filled contours encompass the $68\%$, $84\%$, $95\%$ and $99\%$ of the galaxy distribution in the log(M$_{\star}$) - log(O/H) plane.  
White points are the median metallicities in $0.15$ dex stellar mass bins (with black error bars marking the metallicity dispersion in each bin), while the red curve represents the best-fit \textit{median} MZR according to the functional form of equation ~\ref{eq:mzr}.
The assumed value for solar abundance ($8.69 \pm 0.05$, \citealt{Allende-Prieto:2001aa}) is marked by the horizontal orange stripe.
Our best-fit median MZR asymptotes at 12+log(O/H) $= 8.793 \pm 0.005$ (i.e. $\sim 1.27$ times the solar abundance), presents a turnover at log(\mstar/M$_{\odot}$)$ = 10.02 \pm 0.09$ and is characterized by a low-mass end slope of $\upgamma = 0.28 \pm 0.02$.
%with respect to previous findings in the literature (e.g. \citealt{Andrews:2013aa, Zahid:2014ab}).
%However, we stress here that the low-mass end slope is sensitive to selection effects.
%\textbf{For example, because the SDSS sample selection is apparent magnitude limited, increasing the minimum redshift threshold removes a larger number  of low mass galaxies than high mass galaxies and, at the same time, increases the average SFR of the sample in the low-mass regime.  
%This causes a decrease of the the mean metallicity at fixed stellar mass (due to the effect of the FMR) and, therefore, a steepening in the low-mass end slope of the MZR. Changing the threshold in H$\upalpha$ SNR also produces similar effects.} 
Different functional forms still provide a good representation of the data, although the best-fit values of the parameters can 
be different.
In particular, adopting a slightly modified version of the \cite{Zahid:2014ab} functional form (by substituting the exponential term inside the 10-base logarithm in their equation $5$ with a power of ten), i.e.
\begin{equation}
\label{eq:zahid14}
\text{12 + log(O/H)} = \text{Z}_{0}  +  \text{log} ( 1 - 10^{- \bigl(  \frac{\text{M}}{\text{M}_{0}}\bigl)^{\gamma}  } ) \ ,
\end{equation}
we obtain a slightly higher value for the turnover mass ($10.26 \pm 0.06$), but a steeper low-mass-end slope 
($0.38 \pm 0.09$), simply due to the absence of the $\upbeta$ parameter which causes the width of the knee to be fixed to a much larger value, allowing the purely linear part of the relation to occur at very low masses, largely outside the effective sampled mass range.
%let width of the curvature unconstrained, allowing the purely linear part of the relation to occur at lower masses.
However, in the range of stellar masses probed by our sample, the two representations are almost identical.

%When assuming the \cite{Zahid:2014ab} parametrization of the MZR we find the turnover mass (which, in this case, is representative of the mass associated to the initial variation of the slope) to occur at log M$_{0} = 9.20$, while the best-fit value for the low-mass end slope is $\gamma = 0.36$.

\begin{table}
	\centering
	\caption{Best-fit values for the parameters of the MZR derived with the new set of T$_{\text{e}}$ based calibrations of Fig.~\ref{fig:met_cal_recap} on the SDSS galaxy sample.
	The upper row assumes the new parametrization of equation~\ref{eq:mzr} proposed in this work. 
	The bottom row assumes instead a modified version of the parametrization proposed by \citealt{Zahid:2014aa}, as given by equation~\ref{eq:zahid14}. 
	}
\resizebox{\columnwidth}{!}{%
	\begin{tabular}{|@{} l | c|c|c|c @{}|}
		
%		\multicolumn{5}{c}{MZR - Equation~\ref{eq:mzr} - This work}\\ 
		\hline
		& Z$_{0}$ & log(M$_{0}$/M$_{\odot}$) & $\gamma$ & $\beta$   \\ 
		\hline
%		8.794 0.004 10.104 0.064 0.258 0.013 1.272 0.181
%		8.813 0.004 9.739 0.064 0.331 0.013 0.713 0.181
		
		%\midrule
		\text{Equation (2)} & 8.793 $\pm$ 0.005 & 10.02 $\pm$ 0.09 & 0.28 $\pm$ 0.02 & 1.2 $\pm$ 0.2  \\ 
%		\text{(*)} & 8.813 $\pm$ 0.004 & 9.74 $\pm$ 0.07  & 0.33 $\pm$ 0.03 & 0.7 $\pm$ 0.2   \\
		
		\text{Equation (3)} & 8.792 $\pm$ 0.003 & 10.26 $\pm$ 0.06 & 0.38 $\pm$ 0.08 & -- \\ 	
%		\text{(*)} & 8.796 $\pm$ 0.003 & 10.32 $\pm$ 0.07 & 0.37 $\pm$ 0.09 & -- \\ \hline	
	%	\\ 	
		\hline
	\end{tabular}%
}
	\label{tab:mzr_fit}
\end{table}

\subsection{Comparison with different MZR from the literature}
\label{sec:mzr_comparison}

The bottom panel of Figure~\ref{fig:mzr_1} presents a comparison between our best-fit MZR and previous estimates of the MZR from the literature, based on different methods for metallicity determination. 
The grey shaded areas here mark the $1$ and $2\sigma$ deviations from the median values in each \mstar bin.
All the mass-metallicity relations shown in this plot have been re-derived by applying the different methods and metallicity calibrations adopted in each reference work to the sample considered in this work, in order to minimise the systematics induced by different sample selection criteria. Additional cuts in signal-to-noise (i.e. at SNR$=3$) on different emission lines are implemented when required by the relative calibration method.  
In particular, the metallicity derived with the method described in \cite{Tremonti:2004aa} are already provided in the MPA/JHU catalog.
The \cite{Kewley:2002aa} relation is based on the recursive technique (involving R$_{23}$, O$_{3}$O$_{2}$ and N$_{2}$O$_{2}$) presented in their paper and then revised by \cite{Kobulnicky:2004aa}, while the \cite{Mannucci:2010aa} MZR is based on the R$_{23}$ and N$_{2}$ semi-empirical calibrations presented in \cite{Maiolino:2008aa}.
Similarly to \cite{Tremonti:2004aa}, the calibrations adopted in the latter works provide oxygen abundances based on 
the predictions of different grids of photoionisation models.
More precisely, for \cite{Maiolino:2008aa} calibrations this is true only for $\text{12+log(O/H)}>8.4$, while in the low-metallicity regime they are based on a sample of galaxies with \Te\ measurements.
The \cite{Pettini:2004aa} curve is instead derived by means of their O$_{3}$N$_{2}$ calibration, which is built on \Te\ metallicity measurements in individual \Hii regions.
Finally, the \cite{Andrews:2013aa} and \cite{Yates:2019aa} curves are directly taken from the literature, as they are based on very different samples and/or techniques.
Specifically, \cite{Andrews:2013aa} provide a  \Te-based version of the MZR by measuring the electron temperatures (and hence the metallicities) from SDSS stacked spectra in bins of stellar mass, whereas \cite{Yates:2019aa} adopt a revised version of the \Te\ method on a complied sample of galaxies (both from the literature and from the MANGA survey) with auroral line detections.

In addition, we plot the oxygen abundance measurements derived for nearby galaxies in the local Universe (including the Milky Way) which exploits stellar spectroscopy of young ($\sim 10-50$ Myr) red and blue supergiants (RSG, BSG) to probe the chemical enrichment level \citep{Gazak:2015aa,Davies:2015aa,Bresolin:2016aa,Kudritzki:2016aa}; the data points are taken from the compilation presented in Table 3 of \cite{Davies:2017ab}(see also references therein).
Metallicity measurements from blue supergiants are plotted as blue crosses, while abundances measured from red supergiants are plotted as red stars; a few galaxies in the sample have both measurements, which always agree within $0.1$ dex.
These measurements trace the abundances of the ISM with an independent approach compared to studies targeting \Hii regions, but are sensitive to a similar time-scale of chemical enrichment. 

The abundances probed by means of BSG and RSG spectroscopy are in better agreement with the curves based on \Te\ metallicity measurements rather than with the theoretical derivations the MZR, perfectly matching the normalisation of our new MZR at high masses, but slightly deviating at lower masses.
At log(M$_{\star}$) lower than $9.5$, they are systematically offset towards lower abundances compared to our median values, although being still in agreement within $2\sigma$ considering the large scatter of the galaxy distribution in the \mstar-O/H plane in the low-mass, low-metallicity regime.
It is also worth noting here that the stellar metallicity is mainly traced by the abundance of iron-peak elements, while the metallicity of the gas-phase of the ISM is traced by the oxygen abundance (and more rarely by other $\alpha$-elements); therefore, different $\alpha$/\ion{Fe}\ ratios might contribute to the observed discrepancy.

Our median MZR presents considerable differences in slope and normalization, as expected, from those derived with theoretical strong-line calibrations, especially for what concern the normalisation of the high-mass regime and the saturation metallicity. This is easily explained considering the discrepancy between the abundance scale defined by the \Te\ method and that adopted by the photoionisation models which are at the base of the respective strong-line calibrations.
Overall, the agreement of our best-fit median MZR with the one from \cite{Andrews:2013aa} is instead quite remarkable over the entire range of stellar masses. 
The two relations only slightly deviate between $7.95 < $log(\mstar)$ < 8.5$, with our MZR characterised by a shallower slope.
This lead to a divergence in the prediction of the two relations when relying on their extrapolations at lower masses.
A possible explanation resides in the fact that the oxygen abundance inferred from stacked spectra (as in \cite{Andrews:2013aa}) is a weighted average on the intensity of the auroral lines, which might bias low the metallicity determination in the M$_{\star}$-stacks at low masses, where the number of galaxies per bin strongly decreases.

\begin{figure}
	\centering
	\includegraphics[width=0.98\linewidth]{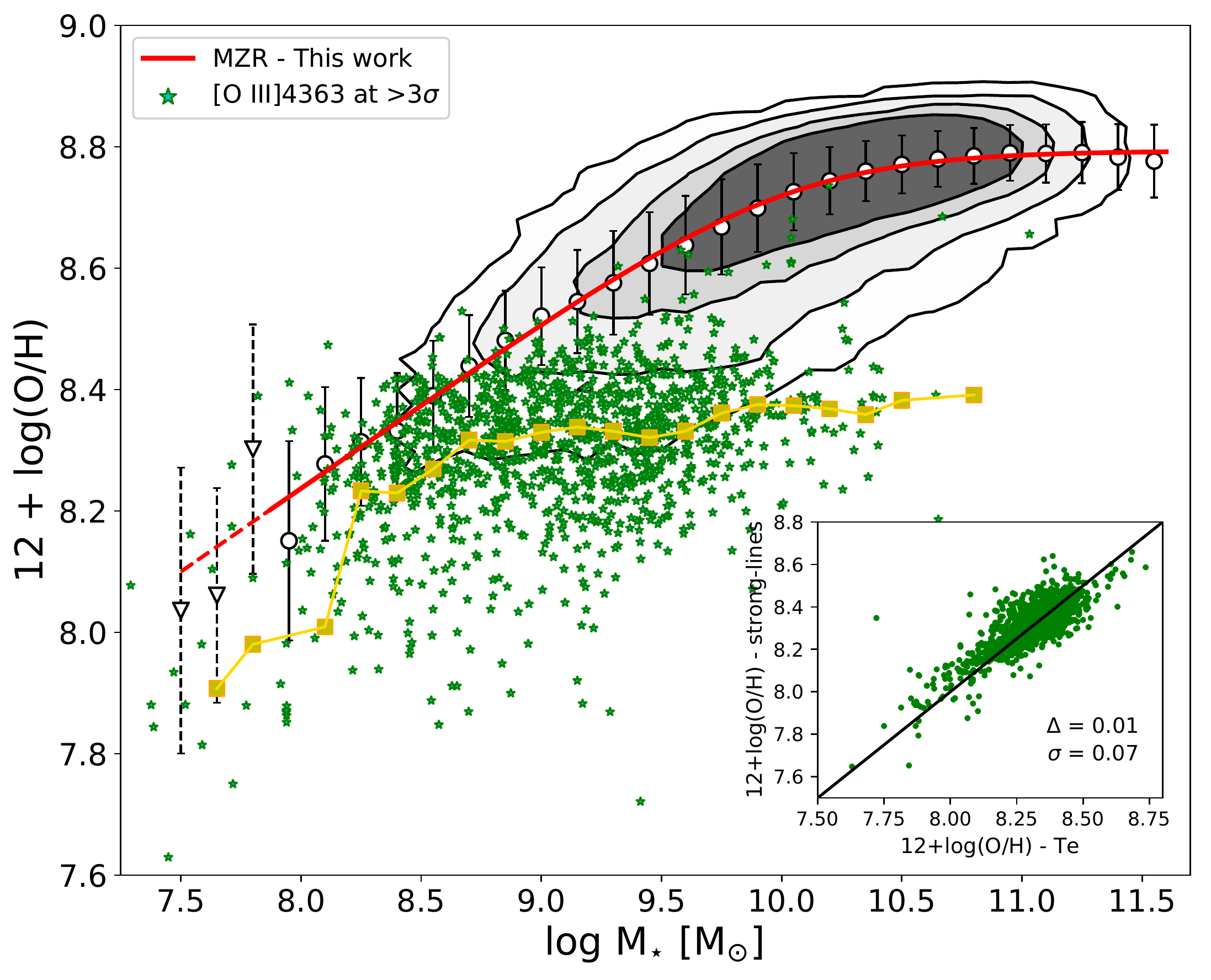}
	\caption{The position of galaxies with [\ion{O}{iii}]$\lambda4363$ detection above $3\sigma$ (with derived \Te-based metallicities) is plotted on the  mass-metallicity plane and compared with the full distribution of SDSS galaxies (based on strong-line metallicities) and our new median-MZR (symbols and colors are as in Fig.~\ref{fig:mzr_1}). 
	The median metallicity in bins of \mstar for the [\ion{O}{iii}]$\lambda4363$-galaxies is marked by the yellow squares.
	At fixed stellar mass, requiring an [\ion{O}{iii}]$\lambda4363$ detection preferentially selects the most metal poor galaxies, potentially biasing the assessment of the MZR. 	
	In the small, inner box, we compare the Te-based and the strong-line-based metallicities for these galaxies. 
	They show good consistency, with an offset and a dispersion from the 1:1 relation of $0.01$ and $0.07$ \rm{dex} respectively. 
%	This is not surprising, as these galaxies are part of the sample exploited in the calibration of the diagnostics presented in \citet{Curti:2017aa} and Fig.~\ref{fig:met_cal_recap}.
}
	\label{fig:mzr_4363}
\end{figure}

More recently, \cite{Yates:2019aa} derived the MZR for a sample of $118$ local, intermediate and low-mass 
(i.e. \mstar$\lesssim10^{10}$\msun) galaxies, exploiting a revised version of the classical \Te\ method; their MZR is characterised by a lower normalisation (on average $\sim 0.2$ {\rm dex} in the overlapping \mstar range) compared to the MZR presented in this work. 
One possible explanation is based on the bias introduced by the requirement of a [\ion{O}{iii}]$\lambda4363$ detection, and can be easily tested within the SDSS sample.
%This may be caused, on one hand, by biasing the sample towards the most metal poor galaxies at fixed stellar mass and SFR, i.e. selecting only those galaxies with auroral lines bright enough to be detected while, on the other hand, by an average overestimate of the true oxygen abundance from the strong-line methods at low masses.
%The abundances probed by means of BSG and RSG spectroscopy are in better agreement with the curves based on \Te\ metallicity measurements rather than with the theoretical derivations the MZR, perfectly matching the normalisation of our new MZR at high masses, but slightly deviating at lower masses.
%At log(M$_{\star}$) lower than $9.5$, they are systematically offset towards lower abundances compared to our median values, although being still in agreement within $2\sigma$ considering the large scatter of the galaxy distribution in the \mstar-O/H plane in the low-mass, low-metallicity regime.
Not surprisingly indeed, the galaxies in the SDSS-DR7 with [\ion{O}{iii}]$\lambda4363$ detection (i.e. those objects for which it is possible to derive \Te\ metallicities) all belong to the large scattered region below the median-MZR at low stellar masses (i.e. below log(M$_{\star}$)$=10$).
This is due to the fact that, at fixed stellar mass, it is easier to detect the  [\ion{O}{iii}]$\lambda4363$ line in low metallicity galaxies.
This is shown in Fig.~\ref{fig:mzr_4363}, where the position of galaxies with [\ion{O}{iii}]$\lambda4363$ detection above $3\sigma$ is plotted on the mass-metallicity plane and compared with the $5\sigma$-level density contours of the full SDSS galaxies distribution and the new median-MZR derived in this Section, clearly demonstrating how biased the SDSS-MZR could be when selecting exclusively these objects. 
The oxygen abundance for these galaxies is calculated following the scheme described in \cite{Curti:2017aa}, 
for consistency with the method implemented in the calibrations of the metallicity diagnostics; in particular, the [\ion{O}{iii}]$\lambda4363$ emission line is exploited to compute the electron temperature of the O$^{++}$ zone (t$_{3}$), while  
the ff-relations are adopted to infer the flux of the [\ion{O}{ii}]$\lambda7320,7330$ auroral line and derive the temperature of the O$^{+}$ zone (t$_{2}$).
We also check that the strong-line metallicity scheme adopted in this work is not introducing any clear systematics in the abundance determination for these galaxies by showing, in the small inner box within Fig.~\ref{fig:mzr_4363}, that the \Te-based and strong-line-based metallicities for such objects are fully consistent, with an average offset of only $0.01$ \rm{dex} and a scatter of $0.07$ \rm{dex}. This is not surprising, as many of these galaxies belong to the sample exploited in the calibration of the diagnostics presented in \citet{Curti:2017aa} and Fig.~\ref{fig:met_cal_recap}.
It is also worth noting here that this is the region where the effects of the FMR are known to be more prominent (see e.g. Section~\ref{sec:m-z-sfr}), hence where the analysis of galaxy samples characterised by different average SFRs can produce MZR with different slopes.
This effect, combined with small statistics and the requirement of strong [\ion{O}{iii}]$\lambda4363$ detections (which preferentially selects among the most metal poor galaxies at fixed \mstar and SFR) might explain the offset observed between the MZR derived in this work and the one presented by \cite{Yates:2019aa}.

The slope of the MZR is also sensitive to other types of selection effects.
For example, because the SDSS sample is apparent magnitude limited, increasing the minimum redshift threshold removes a larger fraction of low mass galaxies than high mass galaxies and, at the same time, increases the average SFR of the sample in the low-mass regime.  
This causes a decrease of the the mean metallicity at fixed stellar mass (due to the effect of the FMR) and, therefore, a steepening in the low-mass end slope of the MZR. Modifying the threshold in H$\upalpha$ SNR also produces similar effects. 
Nonetheless, sample selection effects might be not enough to fully explain the observed discrepancy, as systematics between the different methods (modified-\Te method vs strong-line calibrations) might be present and will be subject to further investigation.

%\textbf{Finally, the abundances probed by means of BSG and RSG spectroscopy are generally better agreement with the curves based on \Te\ metallicity measurements rather than with the theoretical derivations the MZR, perfectly matching the normalisation of our new MZR at high masses, but slightly deviating at lower masses.
%At log(M$_{\star}$) lower than $9.5$, they are systematically offset towards lower abundances compared to our median values, although being still in agreement within $2\sigma$ considering the large scatter of the galaxy distribution in the \mstar-O/H plane in the low-mass, low-metallicity regime.}

As mentioned at the beginning of Sect.~\ref{sec:new_param_mzr}, we have so far considered in our analysis only stellar mass bins with more than $25$ objects, which correspond to log(\mstar)$> 7.95$. Therefore, the best-fit MZR discussed so far has to be considered trustworthy only in the $7.95 < $log(M$_{\star}$)$ < 11.85$ range, whereas extrapolation outside this range is unsafe and can bring to spurious results.
Below this threshold in fact, when entering the low-mass regime, the poor statistics prevents a robust determination of the properties of the MZR. 
An additional big issue is related to the intrinsic uncertainties associated to the stellar mass measurements itself.
As the specific star formation rate increases, the relative contribution from the old stellar population (which makes most of the stellar mass) to the total light in the z-band can be $\ll 1\%$, hence largely affecting the accuracy of the \mstar determination in these galaxies.
This effect is likely to be more prominent at low masses, where the contamination of spurious high mass objects can have a bigger impact on the assessment of the statistical properties of the population. 

However, we try to extend the analysis presented above to the low-mass end by removing the threshold of $25$ objects per bin in this regime, hence computing median metallicities down to log(\mstar)$=7.5$. This means that the lowest mass bins are now populated by less than $10$ objects.
These points are shown as triangles (with dashed error bars) in the upper panel of Figure~\ref{fig:mzr_1}, while the extrapolation of the best-fit MZR is shown by the dashed red line.
The extrapolation produces a slight overestimate of the metallicity compared to the median points at the lowest masses, although being fully consistent within $1\sigma$; including these points in the MZR fit procedure does not strongly affect the overall shape of the relation, producing only a small steepening of the low mass-end slope (up to $\gamma = 0.29$).
%,with the corresponding best-fit MZR indicated by the dashed red line.
%In the overlapping \mstar regime, it can be seen that the two MZR are practically identical. 
%However, we derive a slightly steeper slope when including the low-mass bins in our analysis ($\gamma = 0.33$), compared to the ``original" MZR.  
%This produces a small divergence between the extrapolation of the ``original" MZR and the ``low-mass" MZR, with the latter predicting lower abundances (by $\sim 0.06 $ {\rm dex} at log(M$_{\star}$)$=7.5$) compared to the former, in better agreement with the extrapolation of the \cite{Andrews:2013aa} relation, the \cite{Yates:2019aa} curve and the abundances derived from BSG and RSG in this regime.
%The best-fit parameters of the ``low-mass" MZR are also given in Table~\ref{tab:mzr_fit}.

%One possible issue is certainly related to the intrinsic uncertainties associated to the stellar mass measurements.
%As the specific star formation rate increases, the relative contribution from the old stellar population (which makes most of the stellar mass) to the total light in the z-band can be $\ll 1\%$, hence largely affecting the accuracy of the \mstar determination in these galaxies.
%This effect is likely to be more prominent at low masses, where the contamination of spurious high mass objects can have a bigger impact on the assessment of the average properties of the population. 
%which contribute most of the mass make up << 1  of the light in the z-band

We finally note that, as already mentioned in Sect.~\ref{sec:metallicity}, small systematics are present between nitrogen-based and oxygen-based diagnostics within our strong-line calibrations set, especially at low metallicities (hence preferentially at low masses).
Given the poor sampling of this region of the mass-metallicity plane, this is where such effects can have the larger impact in the determination of the slope of the MZR.
To have an estimate of the amplitude of this effect, we refer the reader to the analysis presented in Appendix B.

\begin{figure*}
	\centering
	\includegraphics[width=0.95\linewidth]{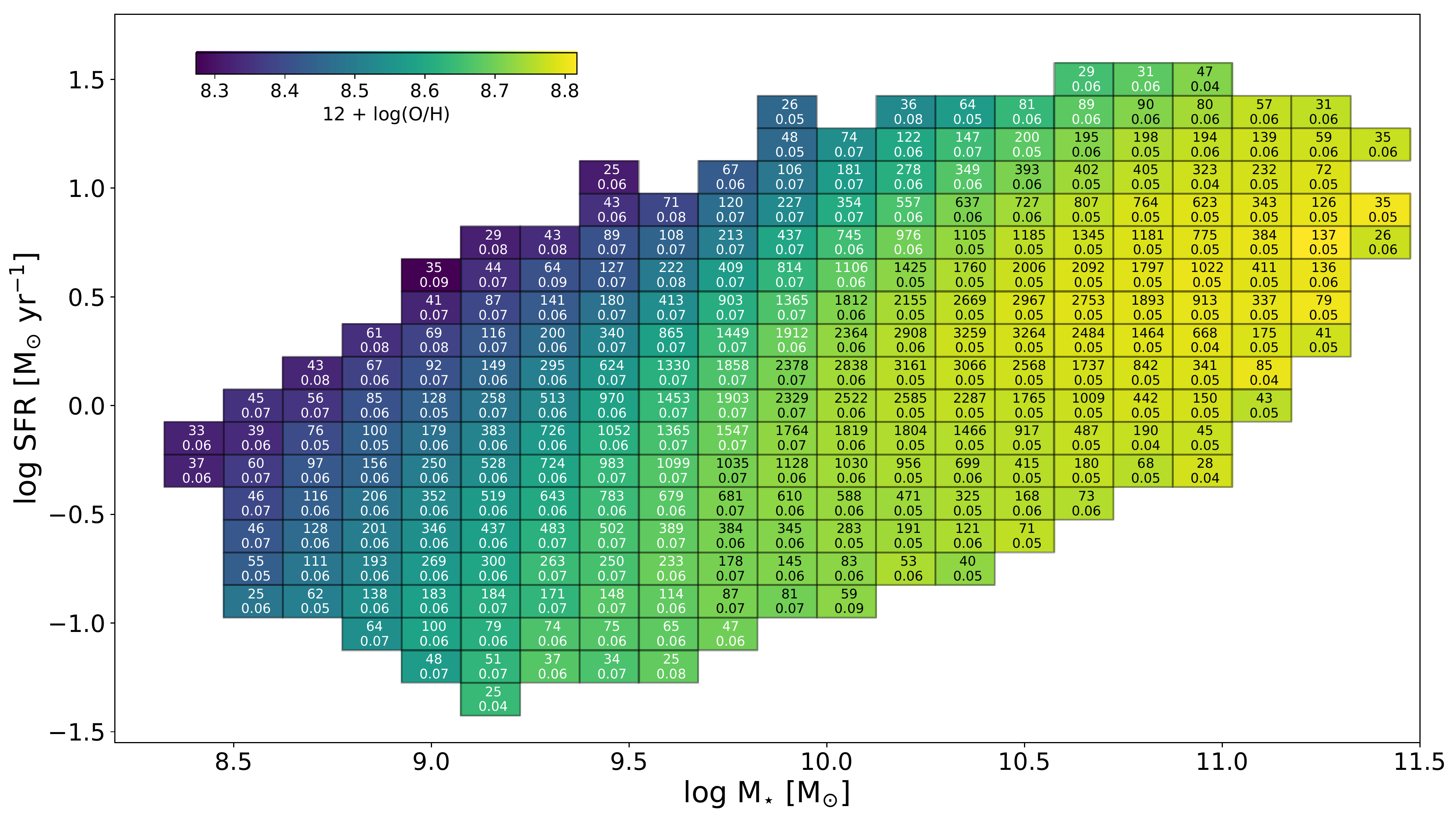}
	\caption{The binning scheme exploited in this work to investigate the M$_{\star}$-Z-SFR relation.
	Each $0.15$ dex wide bin in \mstar and SFR is color-coded by its median metallicity, while the number of galaxies and the internal metallicity dispersion are reported for each of them. Only those bins including more than $25$ galaxies are considered in this analysis.}
	\label{fig:masssfrplane}
\end{figure*}

%\section{The M$_{\star}$-Z-SFR relation}
\section{The Fundamental Metallicity Relation}
\label{sec:fmr}

\subsection{The correlation between M$_{\star}$, Z and SFR }
\label{sec:m-z-sfr}

We now want to consider the mutual relations between stellar mass, metallicity and star formation rate, the M-Z-SFR relation.
First, we explore the dependence of the MZR on the total SFR. 
To do so, we sorted the sample in $0.15$ dex bins of stellar mass and $0.15$ dex in total SFR and computed the median metallicity (and dispersion) in each bin. 
As in the previous section, we limit the analysis only to those bins including at least $25$ galaxies, to sample as much as possible the low-mass, high-SFR regime while keeping a reasonable statistics inside each bin at the same time.

Figure \ref{fig:masssfrplane} shows our binning grid in the \mstar-SFR plane: each bin is color-coded by its median metallicity, while the metallicity dispersion and the number of objects within each bin is reported within.
The upper left panel of Fig.~\ref{fig:fmrhacorrfinalmcmc} shows instead different mass-metallicity relations at fixed SFR (color coded for their total SFR values). % and sSFR. 
The fit to the (global) median MZR (presented in Sect.~\ref{sec:new_param_mzr}) is shown in black, while grey contours trace the $1\sigma$ metallicity dispersion in each mass bin.
A clear segregation in SFR is visible, with highly star forming galaxies characterized by lower metallicities compared to low star forming galaxies of the same stellar mass. % confirming the results previously reported by many studies \citep[][e.g.]{Mannucci:2010aa,}.
Figure \ref{fig:fmrhacorrfinalmcmc} also reveals that the tightness of the observed secondary dependence of the MZR on the SFR strongly decreases in the high mass regime (i.e. above log(M$_{\star}$) $\gtrsim 10$), with all the different MZRs flattening towards the same saturation value Z$_{0}$.
%A mild intersection between the different curves is also visibile, which could suggest an inversion in the nature of the metallicity-SFR dependence, or descend from low-SFR mass-metallicity relations characterized by lower turnover masses.
%However, we know that the strength of this effect could also be strongly dependent both on sample selection effects and on the choice of the metallicity diagnostic.
We can easily visualize this trend by directly tracing the metallicity dependence on SFR at fixed stellar mass (top right panel of Fig.~\ref{fig:fmrhacorrfinalmcmc}). 
Lines of constant stellar mass flatten with increasing \mstar, with the dependence of metallicity on SFR strongly weakening for curves corresponding to log(\mstar)$\gtrsim 10.5$M$_{\odot}$.
The strength of the Z-SFR dependence is also a function of the SFR itself and becomes stronger at high SFRs, as it can be seen from the clear steepening of the different curves in the high-SFR regime at almost all stellar masses. 
%different curves (at fixed \mstar) which clearly steepen at higher values of SFR.

It is also interesting to investigate the mutual relationship between \mstar, metallicity and specific star formation rate ($\text{sSFR}=\text{SFR}/\text{M}_{\star}$, bottom panels of Fig.~\ref{fig:fmrhacorrfinalmcmc}), which describes the relative weight of recent star formation over the global star formation history of a galaxy. 
%In particular, we here define the sSFR as the ratio of the SFR derived from the H$\upalpha$ flux inside the fiber to the total stellar mass of the galaxy; in this way, despite retaining the dependence on galaxy distance, we avoid mixing different type of stellar mass measurements in the same analysis, which would bias our sample 
%considering that the ratio of the fiber-to-total stellar mass of a galaxy depend on its SFR \citep{Telford:2016aa}.
%We use this mix of fiber and total quantities rather than using the stellar mass within the fiber to avoid introducing bias into our sample. The ratio of the fiber to total stellar mass varies systematically with SFR, in the sense that higher SFR galaxies have higher fiber masses at fixed total mass, so mixing the two different types of stellar mass measurements in the same analysis of the Må–Z–SFR relation is problematic. The sSFRs reported for the galaxies in our sample are lower than the true values of sSFR by a factor of ∼2–3, since the total stellar mass includes more of the galaxy than the star formation rate measurement. Again, because we primarily use sSFR as a ranking parameter, our analysis is robust to systematic offsets such as those that affect sSFR.
At low masses, the dependence is present over the entire sSFR range, while weakening for increasing masses at sSFR $< 10^{-9.5}$ yr$^{-1}$, until almost completely disappearing for the highest mass bins.
However, when describing the relation in terms of sSFR, the change in slope  of the Z-sSFR anti-correlation become increasingly evident in the high sSFR regime (i.e. for sSFR $\gtrsim 10^{-9.5}$ yr$^{-1}$, which comprises $\sim 16\%$ of the entire sample), at almost any stellar mass.
These plots confirm the trend originally found by \cite{Mannucci:2010aa} and suggest that a proper description of M-Z-SFR relationships should allow for non-linear trends with (s)SFR, 
as also previously suggested by other studies \citep[e.g.][]{Salim:2014aa} using different metallicity calibrations than the ones used in this work.

A close inspection of Fig.~\ref{fig:fmrhacorrfinalmcmc} reveals a slight intersection of the different MZR curves at high masses and low SFR or, equivalently, that there's an inversion in the trend of O/H vs SFR curves at fixed, high stellar masses. 
This effect has been already reported before, although much more prominently as e.g. in \citealt{Yates:2012aa}, as mainly driven in that case by different SFR measurements and the use of \cite{Tremonti:2004aa} metallicities (see the discussion in \citealt{Cresci:2018aa}).
In our specific case, however, a possible explanation is related to considering total SFRs rather than fibre SFRs in our analysis.
In fact, when applying aperture corrections to fibre-SFRs, metallicity and stellar masses stay unchanged, while we associate the same galaxy to a higher total-SFR bin. 
Since the aperture correction factor correlates with fibre-SFR and, at fixed stellar mass, fibre-SFR correlates with metallicity (as an effect of the FMR), the relative fraction of galaxies which change SFR bin in the high-mass, low-fibre-SFR region is preferentially constituted by the most metal rich galaxies in those bins.
%some of the most metal rich galaxies in low fibre-SFR, high-\mstar bins are shifted towards higher total SFRs, 
%because of their larger 
This has the effect of lowering the median metallicity in high-\mstar, low-total SFR bins whereas increasing it in high-\mstar, high-total SFR bins, compared to the analysis conducted on fibre-based SFR.
Another possible explanation is related to the impact of metallicity gradients at high \mstar, where they are generally steeper \citep{Belfiore:2017aa}, which would make the central fibre-metallicity less representative of the global metallicity of the galaxy and therefore the comparison with the total SFR less fair.
Finally, the uncertainties associated to the derivation of aperture corrections is introducing spurious noise in our relationships.
Indeed, when considering SFRs measured within the fibre (as shown in Fig.\ref{fig:m-sfr-z-no_ap}), the inverted trends fully disappear, and the O/H vs SFR curves at high masses are flat, as expected (i.e. no clear secondary dependence of the MZR on SFR at high masses).
Recently, \cite{Vale-Asari:2019aa} suggested that removing the contamination by diffuse ionised gas (DIG) from the SFR and O/H measurement (in particular when adopting the N2 diagnostic) might reveal the presence of such an inversion at high \mstar even for fibre-based star formation rates.
In our case, the magnitude of such effect is nonetheless small enough to not affect any of the subsequent analysis.
%A visual inspection already reveals that both the turnover mass and the slope of the different MZRs seems not constant, showing a dependence on SFR.% steepening being the high-SFR curves steeper than the low-SFR ones.
%this may have implications in the definition of the projection of minimum scatter of the FMR, which appears to be SFR-dependent. 
\begin{figure*}
	\centering
	\includegraphics[width=0.45\textwidth]{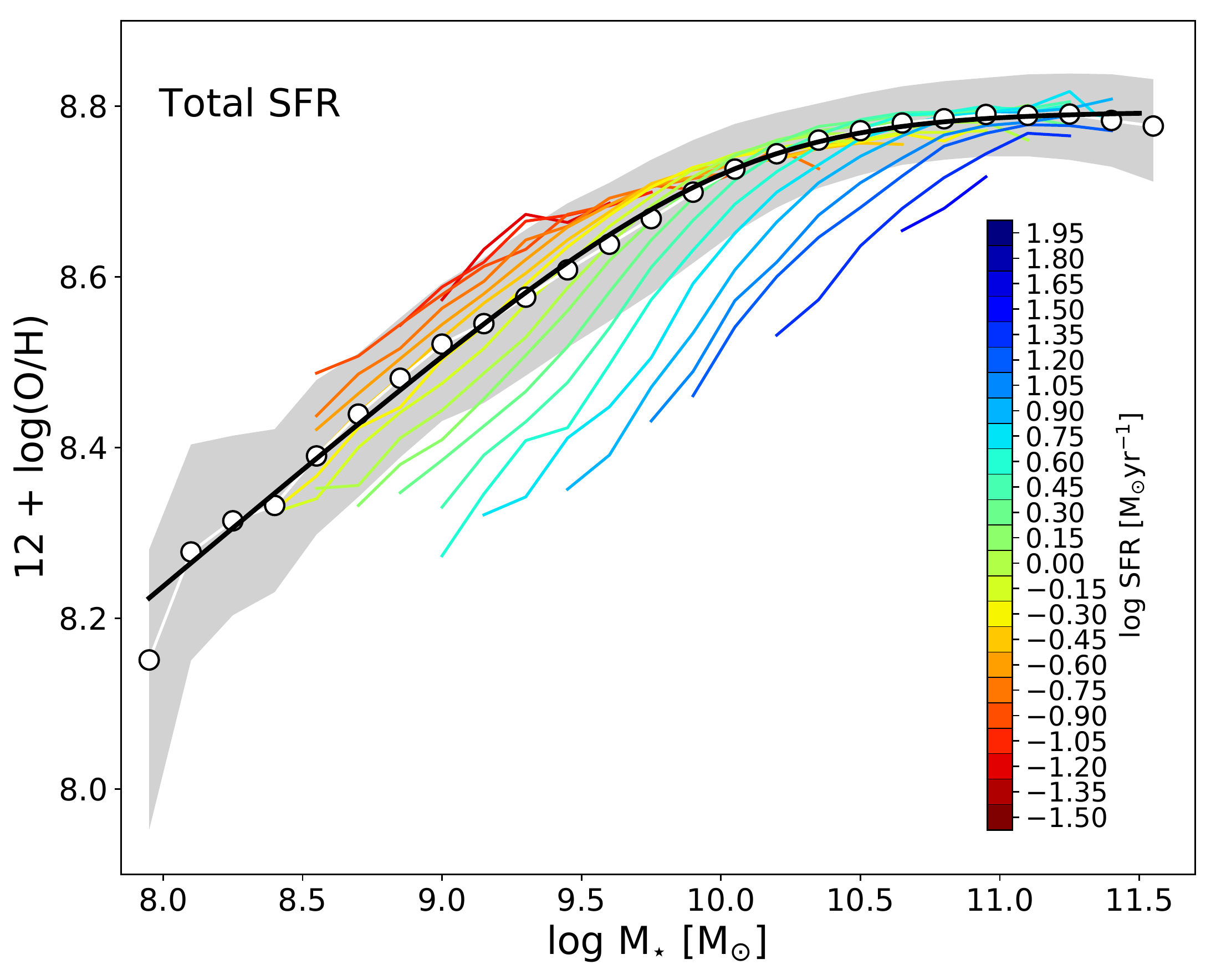}
	\includegraphics[width=0.45\textwidth]{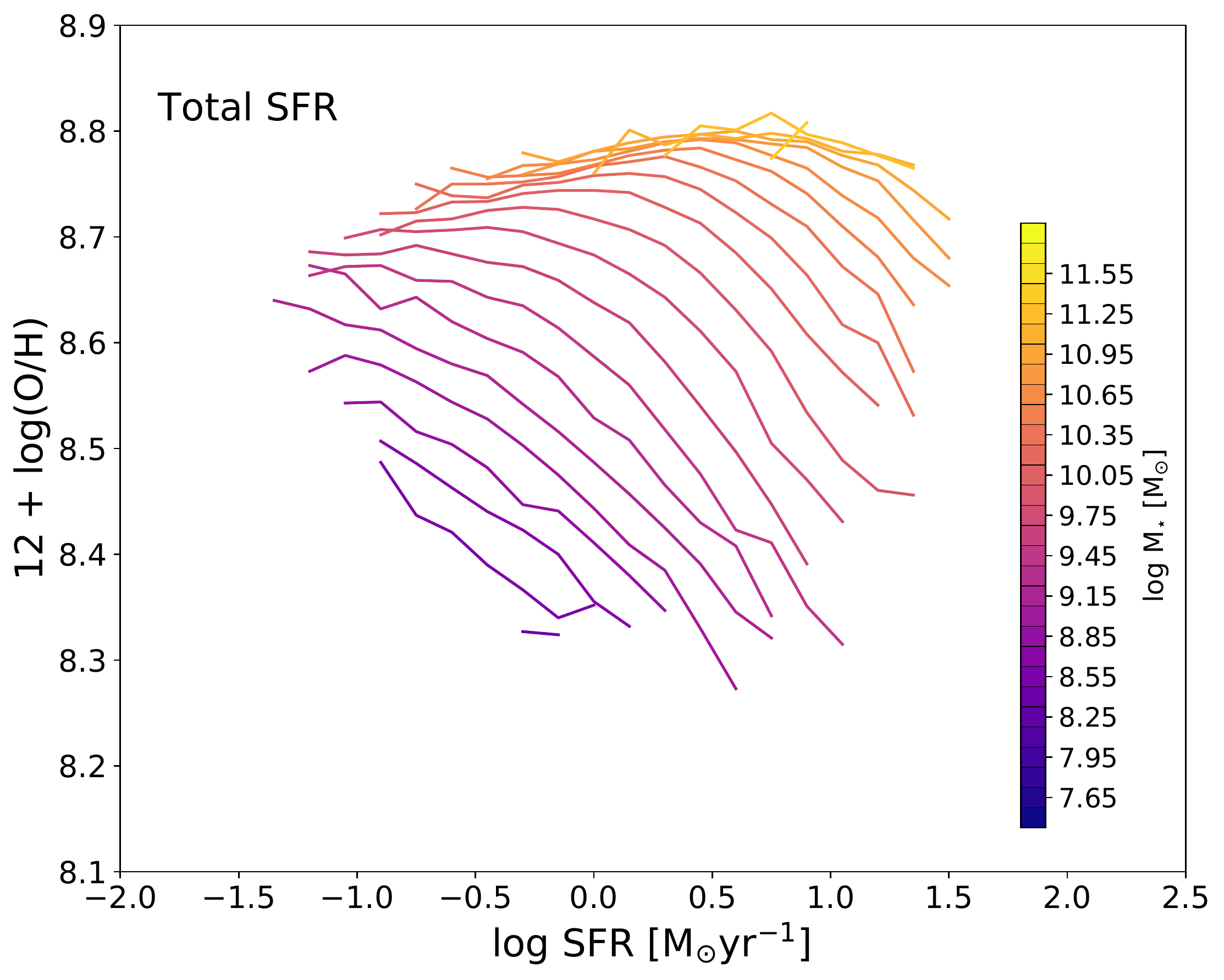}\\
	\includegraphics[width=0.45\textwidth]{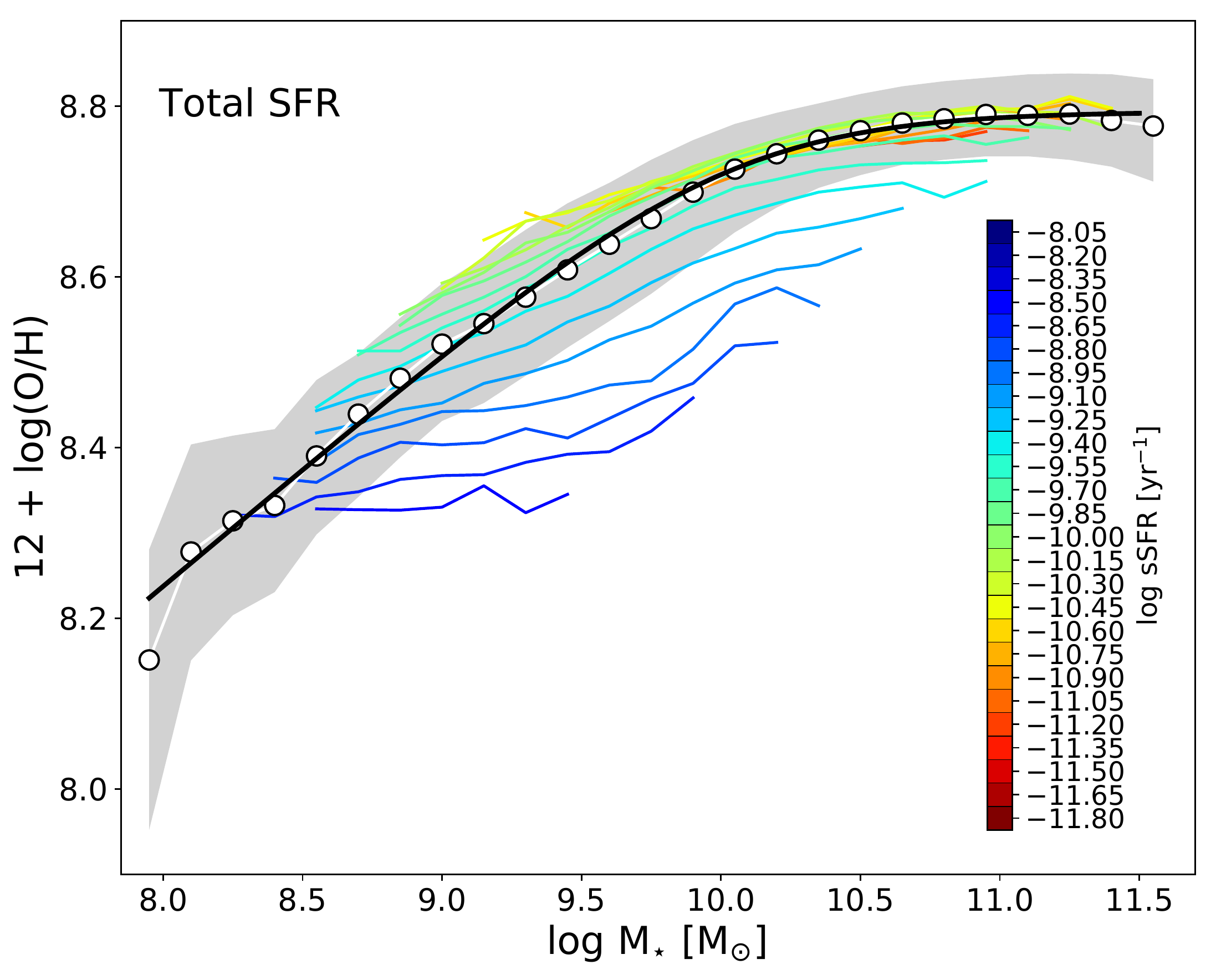}
	\includegraphics[width=0.45\textwidth]{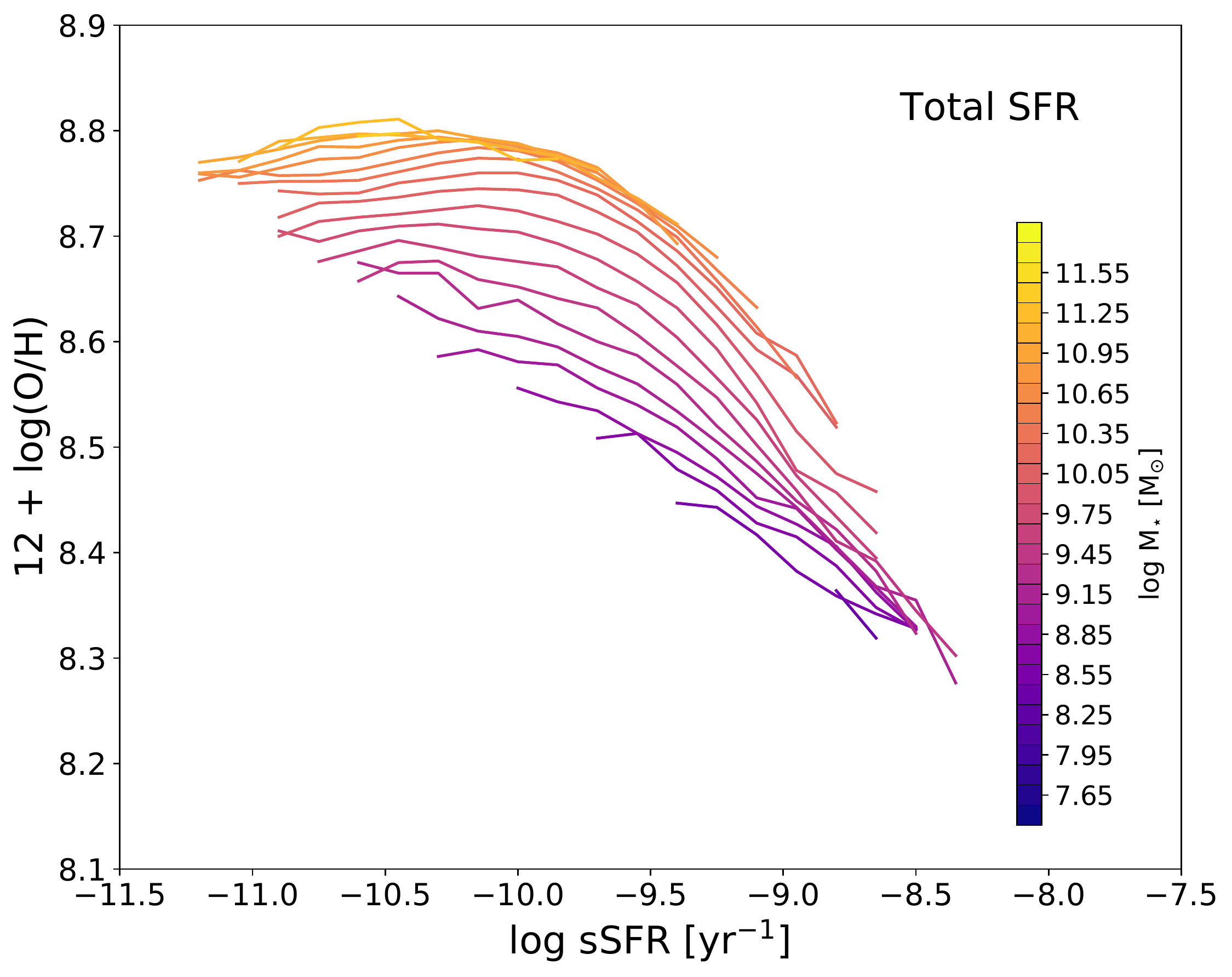}
	
	\caption{\textit{Upper Panels} : M-Z-SFR relation for our sample. Different median MZRs in (0.15 dex wide) bins of total SFR are plotted in the left panel, highlighting the secondary dependence of the mass-metallicity relation on the SFR, especially at low masses and high star formation rates.
		In the right panel instead, the relation between log(O/H) and SFR is plotted for different bins of stellar mass.
		\textit{Bottom Panels} : Same as above, assuming the sSFR=SFR/\mstar as the third variable.}
	\label{fig:fmrhacorrfinalmcmc}
\end{figure*}
\begin{figure*}
	\centering
	\includegraphics[width=0.45\textwidth]{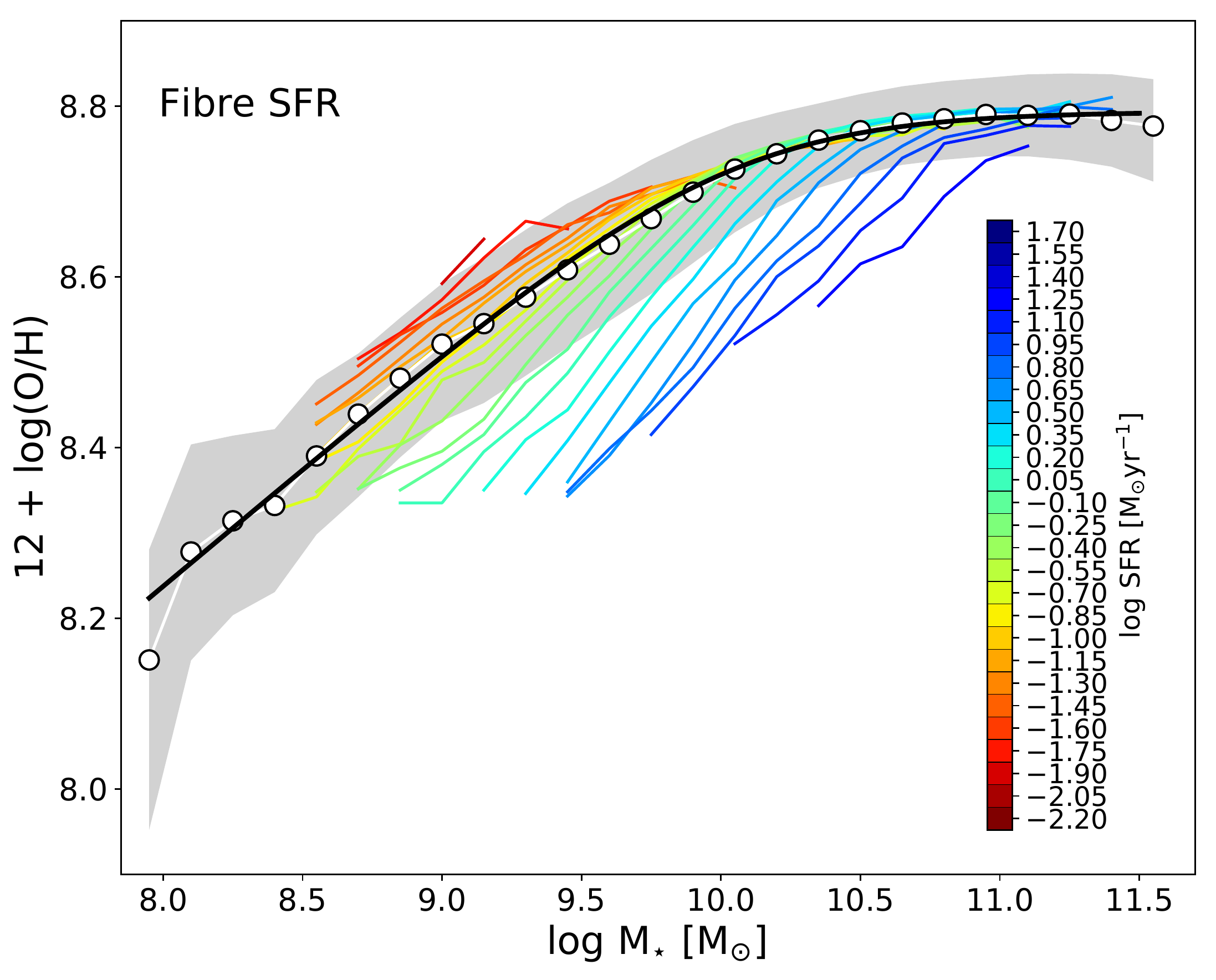}
	\includegraphics[width=0.45\textwidth]{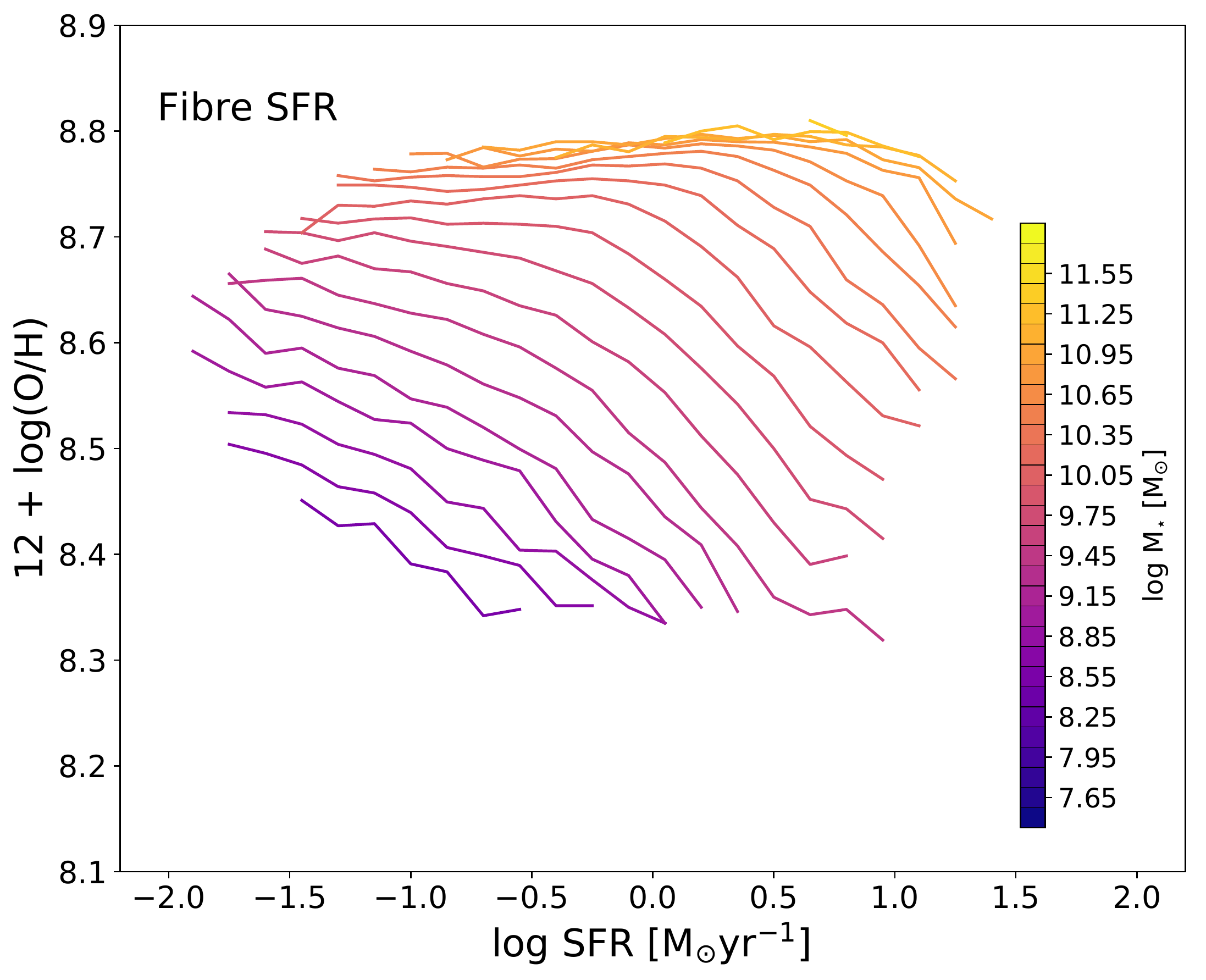}\\

	\caption{Same as upper panels of Fig.~\ref{fig:fmrhacorrfinalmcmc} for fibre-based SFRs.} 
	%In this case, we adopt an \textit{hybrid} definition of sSFR as SFR$_{\text{fibre}}$/\mstar.}
	\label{fig:m-sfr-z-no_ap}
\end{figure*}

\begin{figure*}
	\centering
	\includegraphics[width=0.97\linewidth]{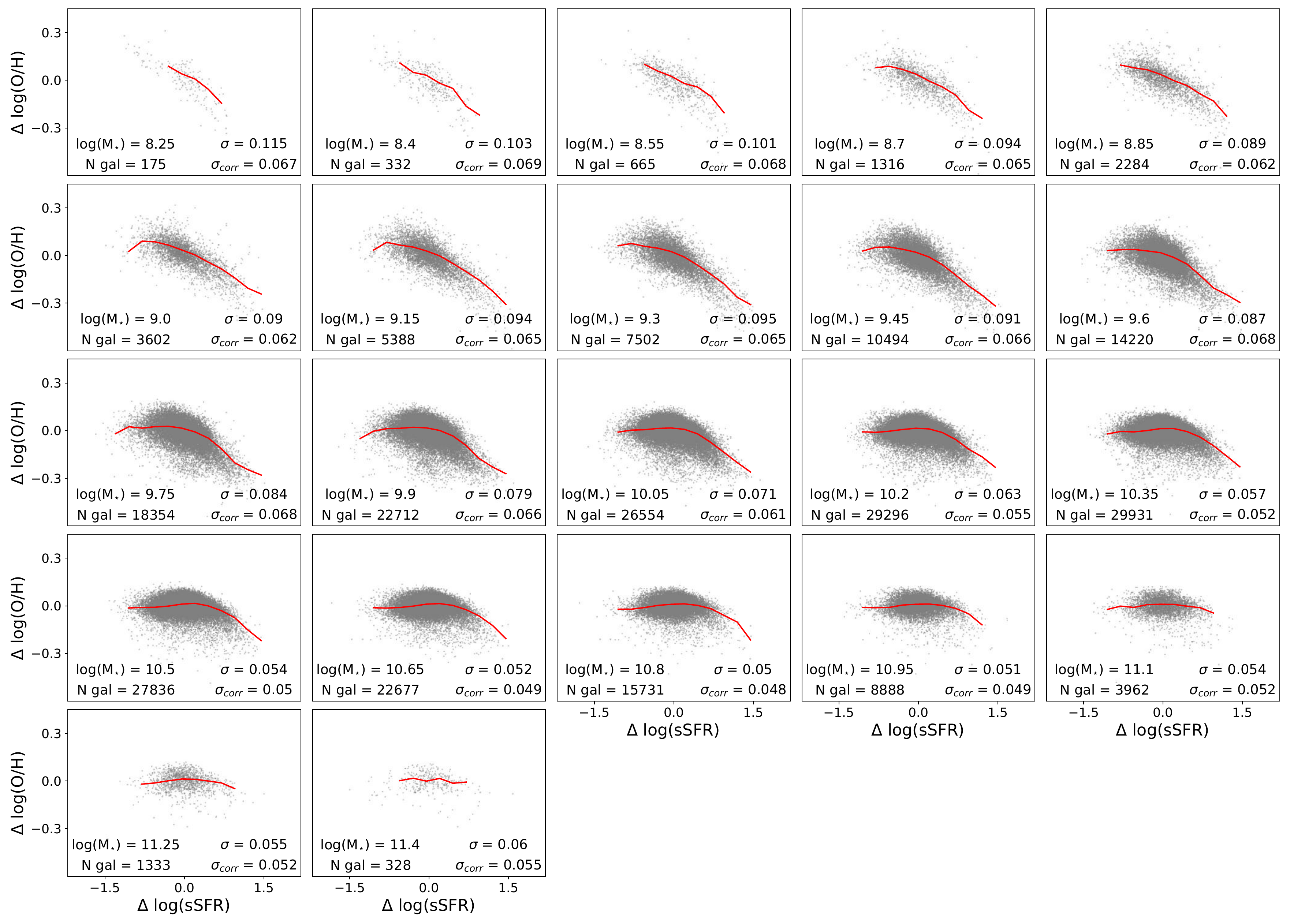}
	\caption{Metallicity residuals around the MZR as a function of $\rm\Delta$log(sSFR), i.e. the offset from the Star Forming Main Sequence, for different $0.15$ dex wide stellar mass bins. Red lines in each panel are running medians in bins of $\rm\Delta$log(sSFR).
		The number of galaxies per bin are indicated in the bottom left corner of each panel, while in the bottom right corner are reported both the metallicity dispersion of individual galaxies inside the mass bin ($\sigma$) and the dispersion after correcting for the secondary dependence on $\rm\Delta$log(sSFR) ($\sigma_{\text{corr}}$). Accounting for this secondary dependence decrease the metallicity dispersion in all \mstar bins, especially at low and intermediate masses (allowing for a reduction in the scatter by a factor up to $\sim 30\%$).
	}
	\label{fig:fmr_non_parametric}
\end{figure*}

Following the prescriptions by \cite{Salim:2014aa}, we can also visualise the dependence of metallicity on SFR at fixed stellar mass assuming a non-parametric approach.
More in detail, \cite{Salim:2014aa} proposed that the most physically motivated quantity to consider to explore the presence of secondary dependences in the MZR is the offset of the sSFR from that expected, at a given stellar mass, for a typical galaxy lying on the Star Forming Main Sequence (SFMS, \cite{Noeske:2007aa}).
In this way the implicit mass selection introduced when considering the (s)SFR as the secondary parameter driving the scatter of the MZR can be removed.  %which scales primarly with stellar mass, 
For the purposes of this analysis we have binned our data in bins of 0.15 dex in \mstar to analyse the dependence of $\rm\Delta$log(O/H) (i.e. the residuals around our best-fit MZR) on $\rm\Delta$log(sSFR), the latter quantity being defined as :
\begin{equation}
\rm\Delta \text{log}(sSFR) = \text{log}(sSFR) - \langle\text{log}(sSFR)\rangle_{M_{\star}}\ ,
\end{equation}
where the last term is the expected sSFR at a given \mstar, assumed as the median sSFR within 0.15 dex mass bins in 
our case.
The choice of such bin sizes allows us to study these trends minimizing the internal effects of the involved relationships (i.e. doing the analysis ``at fixed stellar mass") while keeping statistically meaningful number of galaxies inside each bin. The choice of narrower bin sizes does not qualitatively change the inferred results, while larger bin sizes in \mstar (or, in the worse case, not binning at all) can wash out the dependency of metallicity on SFR, due to its differential tightness in different mass regimes (e.g.  \citealt{Sanchez:2017aa,Barrera-Ballesteros:2017aa,Sanchez:2019aa}, see  \citealt{Cresci:2018aa} for an in-depth discussion of the problem).
Nothing changes in our analysis even if we consider the total metallicities (instead of the MZR residuals), thus completely removing any possible parametrisation (which means that the internal effect of the MZR in our bins is almost negligible). 

In Figure~\ref{fig:fmr_non_parametric} we plot the metallicity of individual galaxies within each bin as a function of $\rm\Delta$log(sSFR). 
We are here considering only \mstar bins including more than $100$ galaxies. The red line represents the running median of our data inside each bin.
The Z-$\rm\Delta$log(sSFR) anti-correlation is clear in the low mass bins over the entire range of sSFR values, becomes relevant only for sSFR above the SFMS for the intermediate mass bins and disappears at the highest masses.
At low masses (log\mstar $< 9.5$), this dependence can be accounted for with a linear relation across the whole $\rm\Delta$log(sSFR) range, while for increasing \mstar a sharp increase in slope starts to occur at $\rm\Delta$log(sSFR) $>0$, similarly to what shown for the different Z-sSFR curves at fixed stellar mass in the bottom right panel of  Fig.~\ref{fig:fmrhacorrfinalmcmc}.
In any case, when accounting for this secondary dependence, the dispersion in metallicity for individual galaxies within each \mstar bin is reduced by a factor of  $2\%$-$5\%$ in the highest mass bins, up to a factor of $\sim15\%$ in the intermediate mass bins and by almost $30\%$ in the low mass bins, as highlighted by the $\sigma$ and $\sigma_{\text{corr}}$ values reported within each panel of Fig.~\ref{fig:fmr_non_parametric}.
%less than the scatter reduction obtained with the \textit{median} M-Z-SFR relation for binned values, but comparable to that obtained, for the global galaxy sample, when assuming the analytical 2D projections of the FMR, as will be shown in Sec.~\ref{sec:projection}.

\subsection{The dependence of the MZR parameters on SFR}
\label{sec:m_z_sfr_params}

\begin{figure}
	\centering
	\includegraphics[width=0.95\columnwidth]{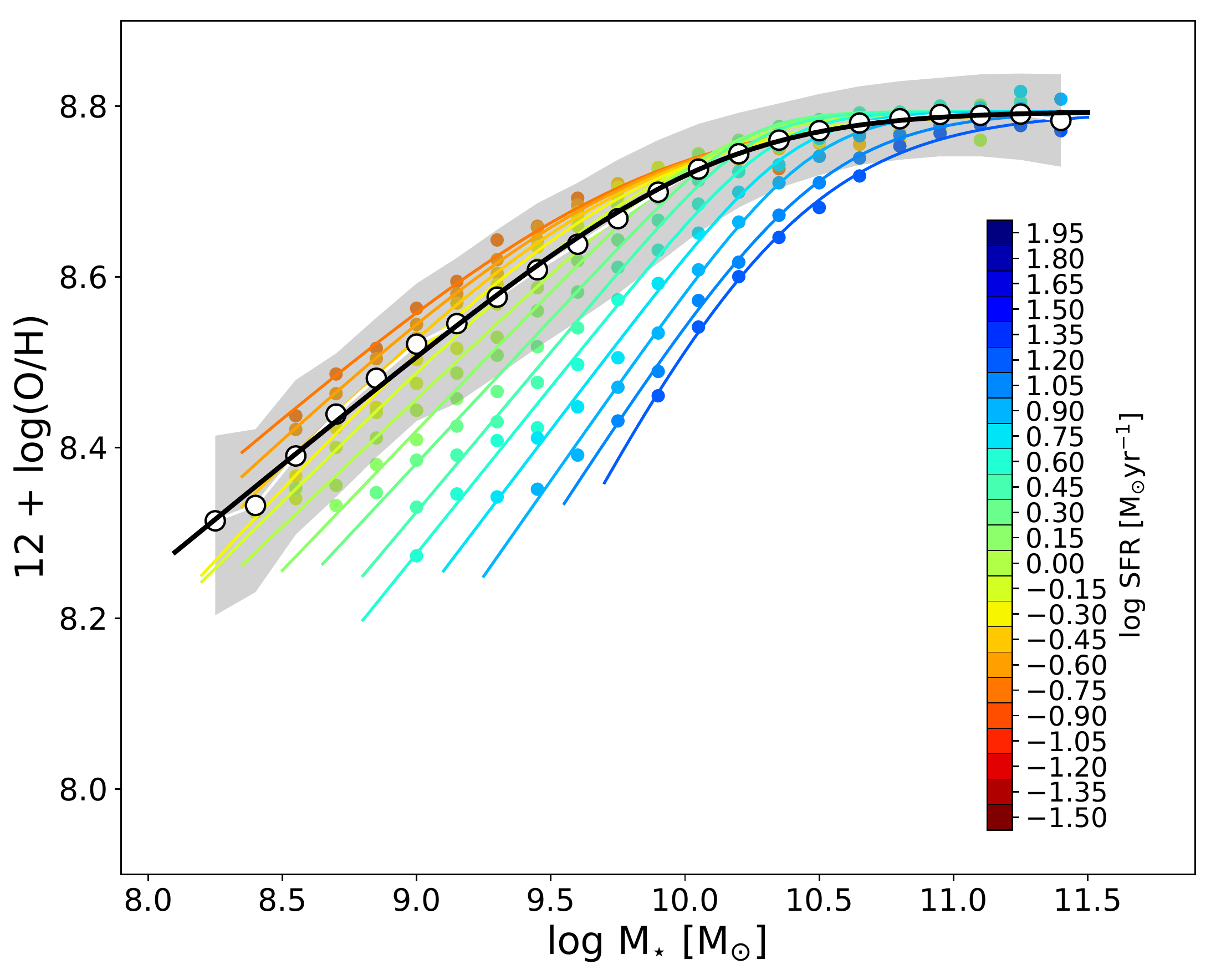}\\
	\includegraphics[width=0.48\columnwidth]{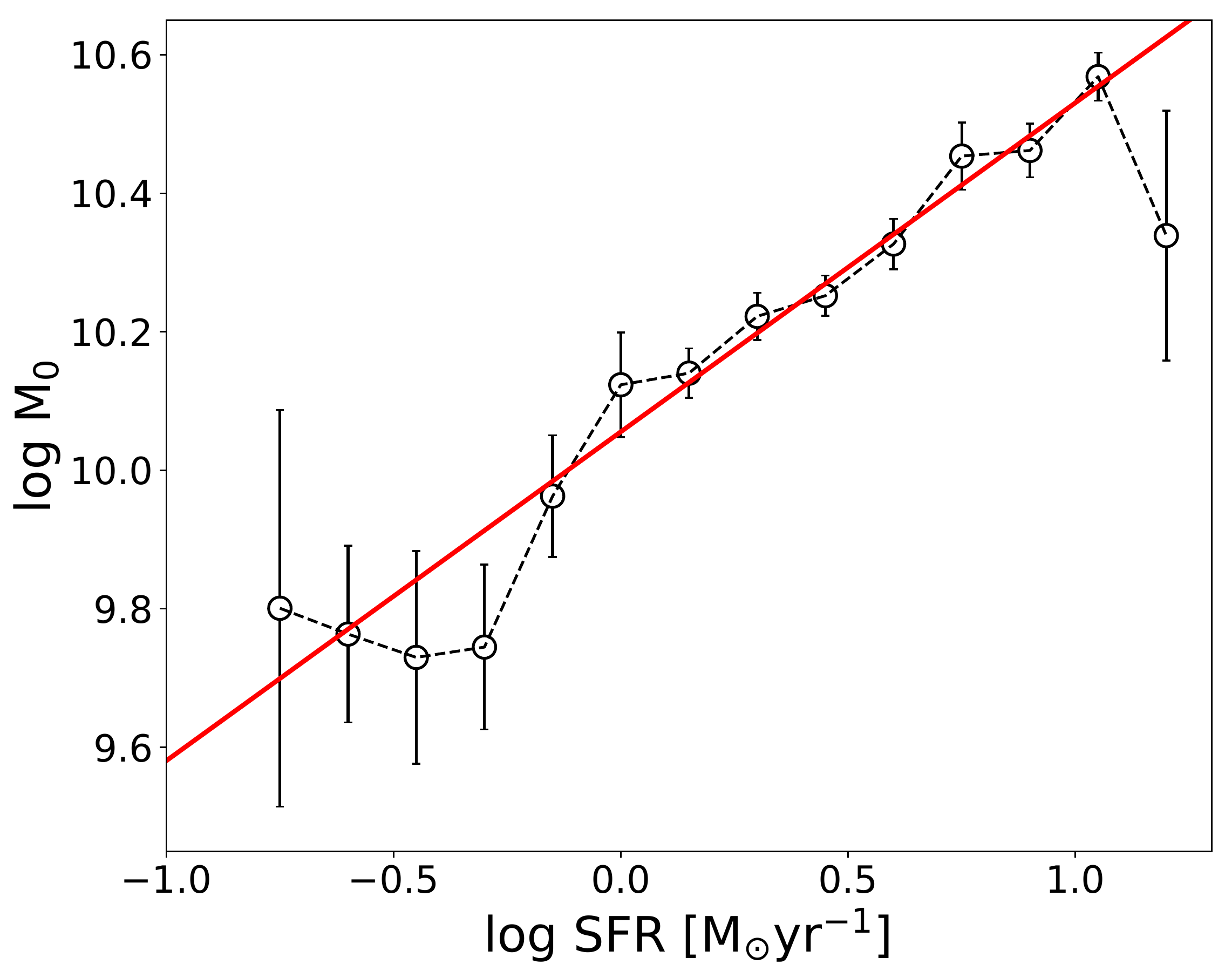}
	\includegraphics[width=0.48\columnwidth]{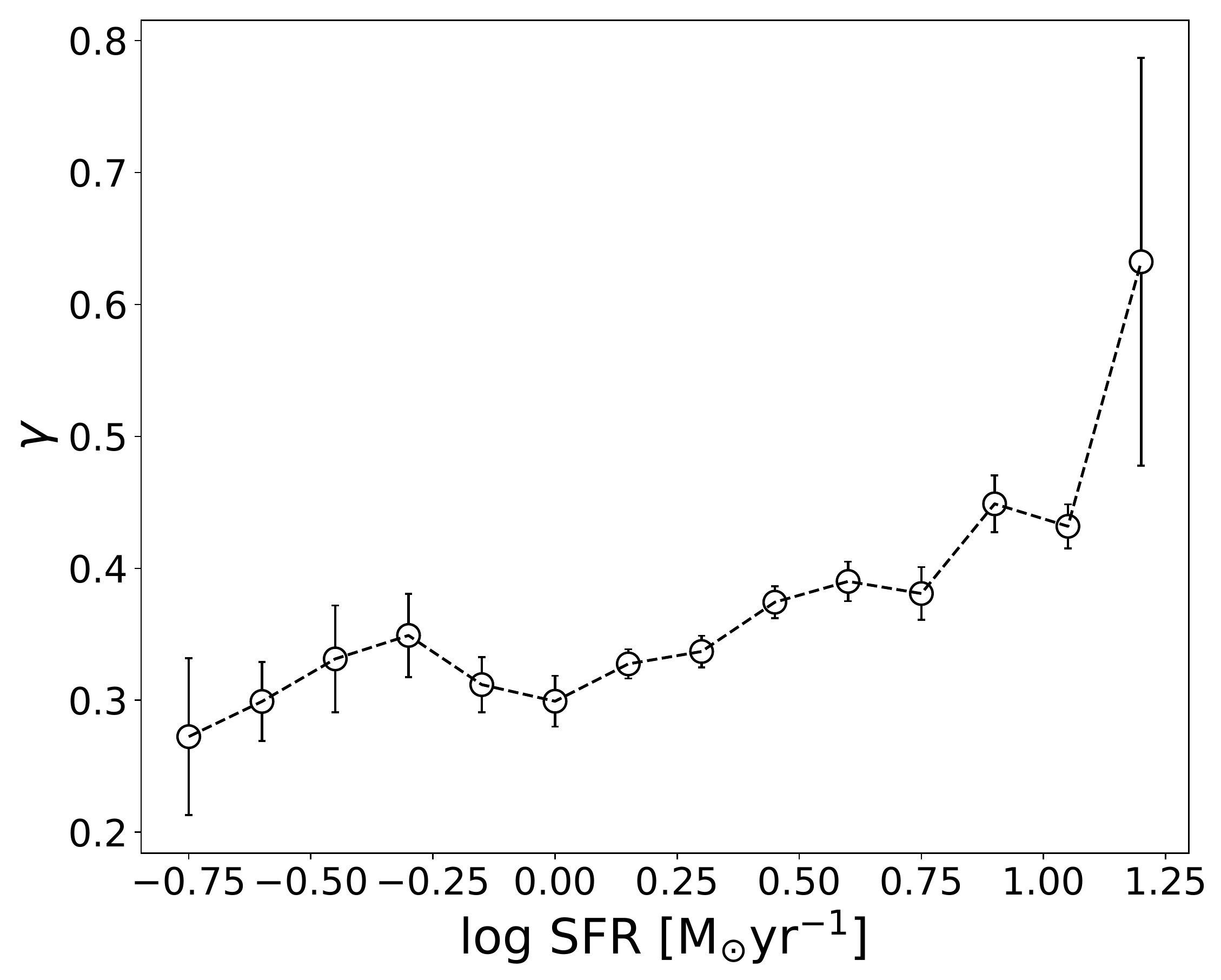}

	\caption{\textit{Upper Panel: } Mass-metallicity relations for different values of total SFR. Colored circles are median metallicities in bins of log(M$_{\star}$) and log(SFR) (color coded by SFR), while thick colored lines are the various MZR fits for different SFR values, assuming the same MZR parametrization as in Eq.~\ref{eq:mzr}.
		The white points (grey area) represent the median metallicity ($1 \sigma$ dispersion) in each stellar mass bin and the black curve is the global MZR fit as already shown in Fig.~\ref{fig:mzr_1} and ~\ref{fig:fmrhacorrfinalmcmc}.
		\textit{Bottom Panels: }Variation of the parameters of the mass-metallicity relation according to Eq.~\ref{eq:mzr} as a function of SFR, as a result of the fitted curves presented in the above panel. 
		The saturation metallicity Z$_{0}$ is fixed in the fit to its global value, while the width of the knee $\beta$ is left free to vary but does not show any clear dependence on SFR. 
		The turnover mass M$_{0}$ show a clear dependence on the star formation rate, which can be accounted for with a linear fit to the points (as shown by the red curve).
		The low-mass end slope $\gamma$, instead, only slightly increases in correspondence of the highest SFR bins.
		 %The saturation metallicity Z$_{0}$ and the width of the knee \textit{a} stays almost unchanged with varying SFR. 
	 }
	\label{fig:mzrs_sfr}
\end{figure}

To investigate more in detail the nature of the dependence of the MZR on SFR and assess the variations in its parameters, we perform a fit for all the different SFR-dependent mass-metallicity relations, assuming the same functional form presented in Eq.~\ref{eq:mzr}.
In the following analysis we primarily refer to SFR as the total-SFR (unless stated otherwise).
Moreover, we here consider only those sub-samples in SFR which allow us to robustly constrain all the MZR parameters, probing in particular both the turnover mass and the low-mass end regimes. 
For this reason we limit the analysis to those MZR curves sampled by at least $10$ points (i.e \mstar-SFR bins), each of those contains at least $25$ galaxies. 
 
A first run of the fit demonstrates that the saturation metallicity Z$_{0}$ remains constant over the entire range of SFR considered, with very small variations (of the order of $\sim 0.01$ {\rm dex}) residing well within the typical uncertainties on metallicity measurements. 
This suggests that the asymptotic limit of chemical enrichment in galaxies, which is regulated by the effective yield of metal production, modulo the impact of outflows, does not depend on the SFR.
%The trend of the \textit{a} parameter is similar, with negligible variations with varying SFR.
Therefore, the majority of the SFR-dependence of the MZR is accounted for in the variations of the slope and the turnover mass.
We make then a second run of the fit, this time fixing the Z$_{0}$ value to those derived for the global MZR, while leaving the other parameters free to vary. 
Although the trend in the $\upbeta$ parameter with SFR in the first run is almost constant, we decide to leave this parameter unconstrained to allow the various curves to adjust the shape of the knee and better catch the true values of M$_{0}$ and $\upgamma$.
Forcing $\upbeta$ to a fixed value would introduce noise, especially in the $\upgamma$-vs-SFR trend, since $\upbeta$ and $\upgamma$ are largely co-variant. 
The uncertainties on the parameters of the different SFR-dependent MZRs are estimated from the $1\sigma$ confidence intervals calculated in our fitting procedure. 
The results are shown in the upper panel of Fig.~\ref{fig:mzrs_sfr} and the best-fit parameters are reported in Table~\ref{tab:mzr_sfr_params}. 
In the lower panels of Fig.~\ref{fig:mzrs_sfr} we show how the MZR parameters behave with varying SFR: 
M$_{0}$ increases almost linearly with SFR while, in contrast, $\upgamma$ remains almost constant up to log(SFR)$=0.75$ and then shows a rise upward in the last SFR bins where, however, the uncertainties are much larger due to the poorer sampling of the low-mass end of the MZR for such SFR subsamples.
%both the low-mass end slope of the relation \textit{b} and the turnover mass M$_{0}$ show a dependence on SFR; however, this trend is much more prominent in the latter quantity than in the former.
%The slope stays almost constant or slightly increases until log(SFR)$\sim 0.75$, then steeply increases only for the highest SFR bins.

An increase in the slope $\upgamma$ with SFR would mean that the relationship between Z and \mstar is steeper, especially at low masses, for different star-forming populations.
On one hand, this could be explained as a manifestation of the chemical downsizing scenario: high mass-high SFR galaxies have already converted a large amount of their gas in stars, faster than low-mass-high-SFR galaxies, which are characterised by larger residual gas fractions. Therefore, the higher the SFR considered, the more chemically evolved high-mass galaxies are compared to low-mass ones. 
%which may be interpreted as a variation of the gas-to-stellar mass relation as a function of SFR.
On the other hand, this trend can also be ascribed to the differential impact of gas flows: dilution effects may be prominent in low-mass, high-SFR galaxies experiencing large inflows of metal poor gas, while outflows (which might eject metal enriched gas) are expected to be much more effective in small galaxies rather than 
in high-mass ones, with a relative importance that correlates with the current level of star formation. 

The observed trend in the slope is robust against the choice of different metallicity diagnostics, and is clearly present also adopting the modified version of the \cite{Zahid:2014aa} parametrisation of the MZR.
Nevertheless, the amplitude of the variation of $\upgamma$ with increasing SFR can change when considering metallicity calculated without involving the R$_{3}$ and R$_{2}$ indicators (or a combination of them, see also Appendix B).  
In fact, we have tested that, on average, the metallicity of individual galaxies in the high-mass (log(\mstar) $> 10$), high-SFR (log(SFR)$\gtrsim 0 $) regime calculated including the above-mentioned line ratios are systematically lower (by $\sim 0.025$ dex) than those calculated by using N$_{2}$ only: this causes the median metallicity of our bins in that region to be lower, causing a steepening of the slope. 
On the contrary, the metallicity in the low-SFR regime is higher on average when computed from R$_{3}$ and R$_{2}$, inducing flatter slopes and thus increasing the amplitude of the variation of $\upgamma$ with SFR.  
This effect may be introduced by the intrinsic dependence of R$_{3}$ on the ionization parameter, which is also related to the average level of star formation within a galaxy. 
Moreover, as previously stated, the behaviour of the $\upgamma$ parameter is less robustly constrained, given the not optimal sampling of the low-mass, high-SFR regime and the high covariance of $\upbeta$ and $\upgamma$. 
%The sudden decrease of \textit{b} for the highest SFR bins, instead, may involve selection effects related to the extinction; extinction level correlates with \mstar and, at high masses, is anti-correlated with SFR at fixed stellar mass.
%This means that highly star forming galaxies in the local Universe are usually highly obscured and this could more easily bring \oii, \oiii and H$\beta$ under the threshold of detectability set on H$\alpha$: if this is the case, the metallicity in these galaxies is computed from the N2 indicator (see Table~\ref{tab:diag_used}) and the cut on SNR$=3$ on that line may remove the less enriched objects, hence reducing the slope.
%In addition, since the extinction level correlates also with the metallicity, for the high mass (high-metallicity) bins we could lose the most metal objects and this, again, would contribute to a flattening of the slope. 
%Moreover, we should note also that the MZR for the highest SFR bins is not fully sampled (spanning only $\sim 1$ order of magnitude in stellar mass) and is therefore more sensitive to statistical fluctuations due to the lower number of galaxies populating that regime.

Therefore, we conclude that the $\upgamma$-vs-SFR trend shown in Fig.~\ref{fig:mzrs_sfr} could be driven by physical effects, although selection effects and the use of different metallicity diagnostics might modify the nature of the observed dependence.
The variation of M$_{0}$ with SFR is instead much more evident and robust against the various issues discussed above. 
Its possible physical interpretation is related to variations in the gas-to-stellar mass ratio \citep{Zahid:2014ab}: local high-SFR galaxies (in a way similar to the average galaxy populations at higher redshifts) are characterised by larger gas masses for a given M$_{\star}$, thus the turnover in the MZR for such populations occurs at higher stellar masses.

\begin{figure*}
	\centering
`	\includegraphics[width=0.65\textwidth]{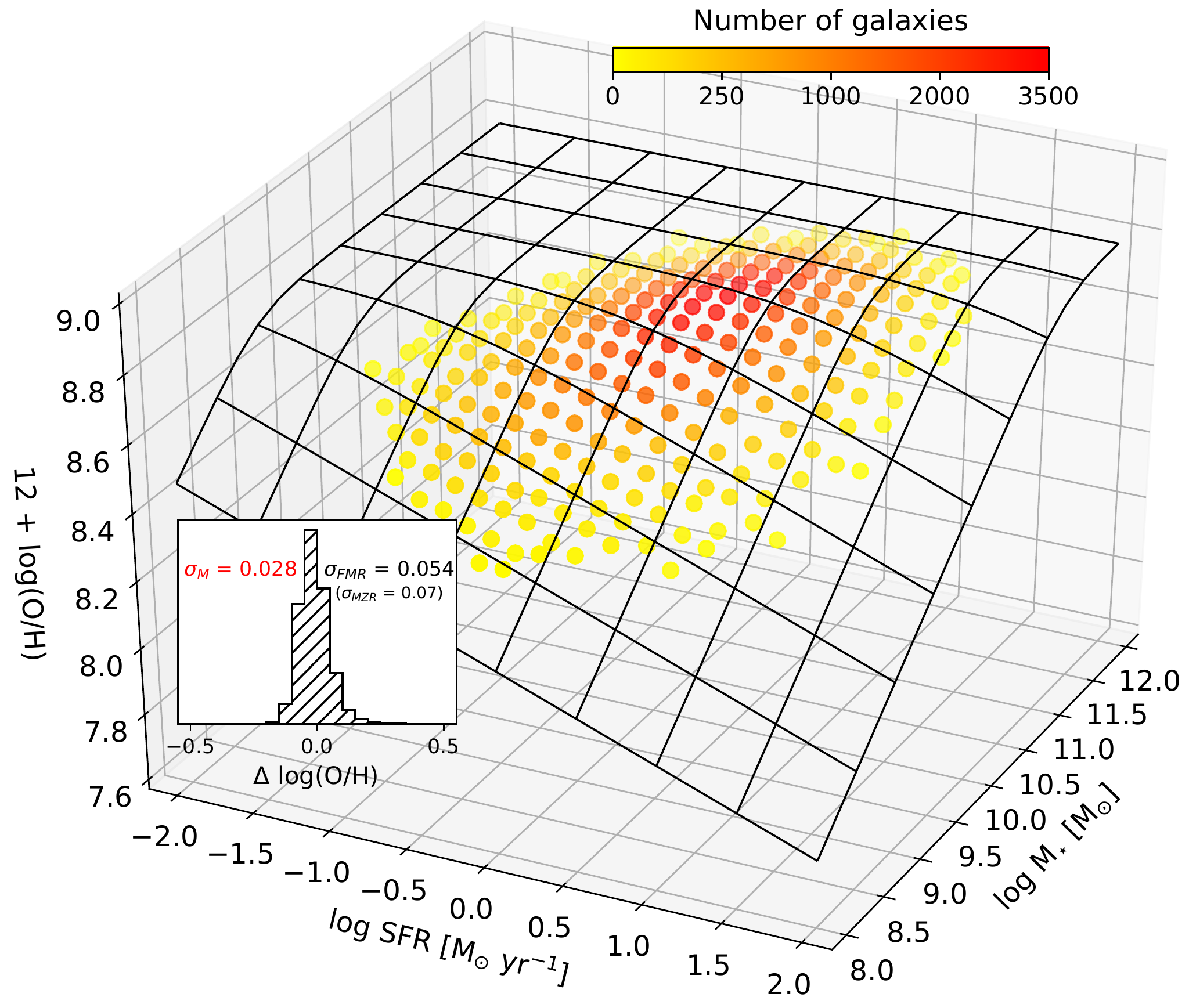}\\ 
	\vspace*{0.3cm}
	\includegraphics[width=0.65\textwidth]{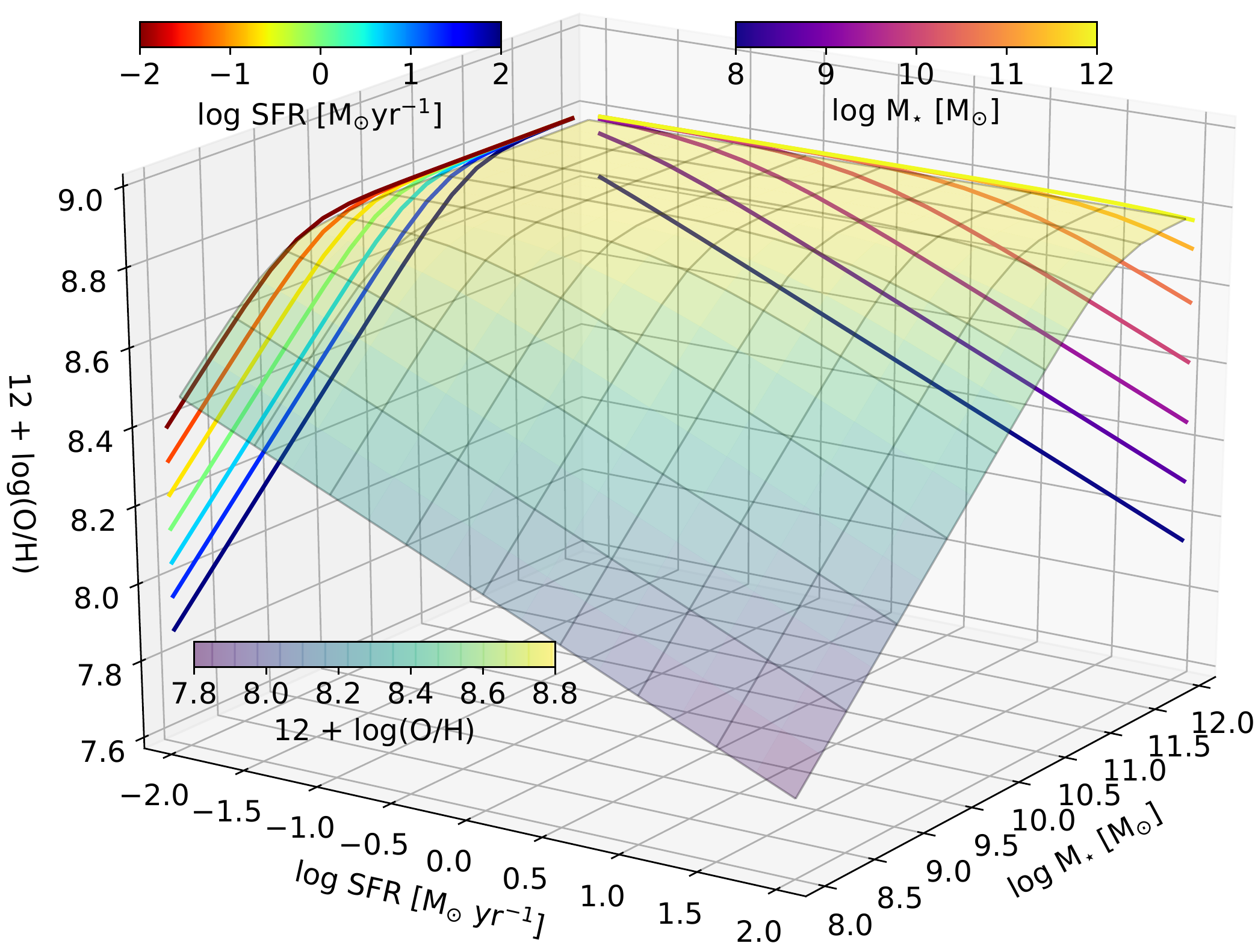}

	\caption{\textit{Upper panel}: 3D visualization of the Mass-Metallicity-SFR relation for our \mstar-SFR bins as described in Section~\ref{sec:fmr}, color coded by the number of galaxies in each bin. A sampling of the best-fit surface representing the FMR from equation~\ref{eq:FMR} is shown by the black grid, while the histogram of metallicity dispersion of individual galaxies around the surface is reported in the small box. The three quoted $\sigma$ values represent respectively the dispersion of the median-binned values around the surface ($\sigma_{\text{M}}$, in red), the scatter of individual galaxies around the surface ($\sigma_{\text{FMR}}$), and, for comparison, the scatter of individual galaxies around the best-fit MZR ($\sigma_{\text{MZR}}$).
		\textit{Bottom panel}: Graphical representation of the FMR surface, color-coded by its predicted metallicity values. The contours of the projections of the FMR onto the \mstar-Z and SFR-Z plane are also drawn, nicely reproducing the observed trends shown in Fig.~\ref{fig:fmrhacorrfinalmcmc}.
		%	\textit{Bottom left panel }: Same as above, but for the ``low-sSFR'' subsample (sSFR$<10^{-10}$ [yr$^{-1}$]). The Z-SFR dependence is weaker ($\alpha = 0.25$), but the dispersion around the relation is strongly reduced ($\sigma = 0.06$).
		%	\textit{Bottom left panel }: 2D projection for the ``high-sSFR'' subsample (sSFR$>10^{-10}$ [yr$^{-1}$]). The Z-SFR dependence is now dominant ($\alpha = 1.13$), but the dispersion is larger compared to the ``low-sSFR'' sample ($\sigma = 0.08$).
	}
	\label{fig:FMR_3d}
\end{figure*}

\begin{table}
	\centering
	\caption{Best-fit parameters for the different MZR at fixed SFR showed in Fig.~\ref{fig:mzrs_sfr}, according to the MZR parametrization of Eq.~\ref{eq:mzr}. Z$_{0}$  has been fixed in the fitting procedure to its global MZR value.
	}
	\begin{tabular}{|@{} c|c|c|c|c @{}|}\hline
		
	log(SFR) & Z$_{0}$ & log(M$_{0}$/M$_{\odot}$) & $\gamma$   & $\beta$\\ 
		\hline
		
		%\midrule
-0.75 & 8.79 & 9.81 $\pm$ 0.29 & 0.27 $\pm$ 0.06 & 1.0 $\pm$ 0.1 \\ 
-0.6 & 8.79 & 9.77 $\pm$ 0.13 & 0.30 $\pm$ 0.03 & 1.0 $\pm$ 0.1 \\
-0.45 & 8.79 & 9.74 $\pm$ 0.15 & 0.33 $\pm$ 0.04 & 1.0 $\pm$ 0.0 \\ 
-0.3 & 8.79 & 9.75 $\pm$ 0.12 & 0.35 $\pm$ 0.03 & 1.0 $\pm$ 0.1 \\ 
-0.15 & 8.79 & 9.96 $\pm$ 0.09 & 0.31 $\pm$ 0.02 & 1.3 $\pm$ 0.2 \\ 
-0.0 & 8.79 & 10.12 $\pm$ 0.08 & 0.30 $\pm$ 0.02 & 1.9 $\pm$ 0.5 \\ 
0.15 & 8.79 & 10.14 $\pm$ 0.04 & 0.33 $\pm$ 0.01 & 2.3 $\pm$ 0.4 \\ 
0.3 & 8.79  & 10.22 $\pm$ 0.03 & 0.34 $\pm$ 0.01 & 3.2 $\pm$ 0.9 \\ 
0.45 & 8.79 & 10.25 $\pm$ 0.03 & 0.38 $\pm$ 0.01 & 3.1 $\pm$ 0.6 \\ 
0.6 & 8.79  & 10.33 $\pm$ 0.04 & 0.39 $\pm$ 0.02 & 2.9 $\pm$ 0.7 \\ 
0.75 & 8.79 & 10.40 $\pm$ 0.05 & 0.41 $\pm$ 0.02 & 2.7 $\pm$ 0.6 \\ 
0.9 & 8.79  & 10.46 $\pm$ 0.04 & 0.45 $\pm$ 0.02 & 2.4 $\pm$ 0.4 \\ 
1.05 & 8.79 & 10.48 $\pm$ 0.06 & 0.49 $\pm$ 0.04 & 1.7 $\pm$ 0.2 \\ 
1.2 & 8.79  & 10.33 $\pm$ 0.19 & 0.64 $\pm$ 0.16 & 1.3 $\pm$ 0.2 \\ 
		\hline
		%\hline
	\end{tabular}
	
	\label{tab:mzr_sfr_params}
\end{table}

\subsection{A new parametrisation for the FMR}
\label{sec:new_param_fmr}

The mutual dependences between M$_{\star}$, SFR and metallicity can be easily visualised in the three-dimensional space defined by the same quantities.
\cite{Mannucci:2010aa} first observed that galaxies in the local Universe are distributed on a surface in this 3D space, the Fundamental Metallicity Relation (FMR), which is affected by a dispersion in metallicity of the order of the uncertainties associated to the measurement processes.
%More interestingly, they also observed that high-redshfit galaxies seem to follow the same relation (within the uncertainties) up to z$\sim 3$; despite hotly debated in the literature, this result clearly hold when all measurements (especially metallicity determination) are performed consistently \citep{Cresci:2018aa}.
%Therefore, they suggested that the evolution of galaxies is regulated by smooth secular processes, hence that an equilibrium condition is set between the involved physical quantities over cosmic time.
%Therefore, galaxies distribute on a thin surface in the 3D space defined by \mstar, SFR and metallicity , the Fundamental Metallicity Relation (FMR), which is affected by a metallicity dispersion of the order of the uncertainties associated to measurement processes (i.e. $\sim 0.05 dex$).
This surface is clearly revealed by the distribution of the median-binned values of \mstar, SFR and log(O/H) in the reference 3D space (upper panel of Fig.~\ref{fig:projections}); points are color-coded by the number of galaxies in each bin, to allow for a better visualization of how the surface is populated by galaxies in the local Universe.
The existence of such a surface means that, on average, the metallicity properties of galaxies can be predicted once their \mstar and SFR is known.
\begin{table*}
	
	\centering
	\caption{Best-fit parameters for the FMR parametrisation of Equation~\ref{eq:FMR}, assuming both total- and fibre-SFR. }
	\begin{tabular}{|@{}l | c|c|c|c|c @{}|}
		
%		\multicolumn{6}{c}{Fundamental Metallicity Relation  - Equation~\ref{eq:FMR}}\\ 
		\hline
		& Z$_{0}$  &  m$_{0}$ &  m$_{1}$ & $\gamma$ &  $\beta$ \\ 
		\hline
	   \text{Total SFR} & 8.779 $ \pm$ 0.005 & 10.11 $ \pm$ 0.03 & 0.56 $ \pm$ 0.01 & 0.31 $ \pm$ 0.01 & 2.1 $ \pm$ 0.4 \\
% ['8.779 $ \\pm$ 0.005 & 10.106 $ \\pm$ 0.034 & 0.559 $ \\pm$ 0.011 & 0.311 $ \\pm$ 0.011 & 2.062 $ \\pm$ 0.37']
		
	 \text{Fibre SFR} &  8.782 $ \pm$ 0.004 & 10.39 $ \pm$ 0.03 & 0.454 $ \pm$ 0.008 & 0.299 $ \pm$ 0.008 & 2.6 $ \pm$ 0.5 \\
%['8.782 $ \\pm$ 0.004 & 10.385 $ \\pm$ 0.025 & 0.454 $ \\pm$ 0.008 & 0.299 $ \\pm$ 0.008 & 2.588 $ \\pm$ 0.49']]
%		\hline		 
		\hline

		\end{tabular}
\label{tab:fmr_params}
\end{table*}

\cite{Mannucci:2010aa} originally parametrised the FMR with a second-grade polynomial surface.
However, we have shown in Sect.~\ref{sec:m_z_sfr_params} that most of the secondary dependence of the MZR can be accounted for by the variation of the turnover mass (M$_{0}$), while the small trend seen in the low-mass end slope $\upgamma$, although present, can be almost neglected up to the highest SFR considered.
Therefore, we propose a new functional form to analytically describe the FMR, by explicitly introducing the SFR dependence of the turnover mass M$_{0}$ into Equation~\ref{eq:mzr} by allowing it to vary linearly with SFR.
This new parametrization would be best suited than the simple second-order polynomial surface to account for the saturation metallicity limit at high masses and the trends in M$_{0}$ with SFR. 
We decide here not to include any explicit dependence of the slope $\upgamma$ on SFR, because even a linear trend may produce risky extrapolations in the low-mass, high-SFR regime, which is poorly sampled and thus poorly constrained by SDSS galaxies, but it is largely populated by high-redshift sources. 
This is equivalent to assuming that the dependence of the MZR on SFR can be fully accounted for by the variation of M$_{0}$; to better visualise the consequences of this assumption, we refer the reader to the analysis presented in Appendix A. %~\ref{sec:projection}. 

Our newly proposed functional form for the FMR is thus the following : 
\begin{equation}
\label{eq:FMR}
\begin{aligned}
%Z(\text{M,SFR}) =\ & Z_{0} - 1/a \ (\beta_{0} + \beta_{1}\text{log(SFR)} ) \\  
%							 & \text{log}(1 + (10^{(m_{0} + m_{1} \text{log(SFR)})}/\text{M})^{a}) \ , 
Z(\text{M,SFR}) =\ Z_{0} - \gamma/\beta \  \text{log}(1 + (\text{M}/\text{M}_{0}(\text{SFR}))^{-\beta}) \ , 
\end{aligned}
\end{equation}
where log(M$_{0}$(SFR)) $= m_{0} + m_{1} \text{log(SFR)}$ or, equivalently,\\ M$_{0}$(SFR) $= \Theta_{0} \cdot \text{SFR}^{m_{1}} $, where $\Theta_{0} = 10^{m_{0}}$.\\
The best-fit parameters obtained by fitting this equation to our median-binned data are reported in Table~\ref{tab:fmr_params} and the shape of the newly parametrised surface in the M-Z-SFR space is shown in the upper panel of Fig.~\ref{fig:FMR_3d} by the black grid.
The scatter of median values around the best-fit analytical representation of the FMR is remarkably small ($\sigma_{\text{M}} = 0.028$ \rm{dex}), with the dispersion in metallicity of individual galaxies decreasing instead to $\sigma_{\text{FMR}}=0.054$ \rm{dex} from $\sigma_{\text{MZR}}=0.07$, a reduction of $\sim 23\%$. 
The residual scatter of individual objects around the FMR is comparable with the typical uncertainties associated with the metallicity determination via the strong-line method and can not probably be reduced further at this stage.
However, it should be also stressed here that, when considering the population as a whole, a large contribution to the residual scatter comes from high mass galaxies where the effects of the FMR are less relevant; indeed, we have already shown in Section~\ref{sec:m-z-sfr} that a more significant reduction of the dispersion (up to $30\%$) is obtained when considering the role of SFR at fixed stellar mass (and in particular at low \mstar) following a non-parametric approach.

In the bottom panel of Fig.~\ref{fig:FMR_3d} instead, the FMR is graphically represented by a continuous surface, color-coded by its predicted metallicity.
On the \mstar-Z and SFR-Z planes, we show the contours of the relative 2D projections of the FMR, nicely reproducing the mutual dependencies between metallicity and SFR (at fixed \mstar) and between metallicity and stellar mass (at fixed SFR) observed in our sample and shown in Fig.~\ref{fig:fmrhacorrfinalmcmc}.

The same analysis described above on the FMR have been also performed assuming the fibre-SFR rather than the total-SFR.
Comparing the two representations of the FMR, the metallicity predicted is identical in the flat saturation region at high-mass and low-SFR while, on average, the ``fibre-based'' FMR predicts lower metallicities (of the order of $0.05$ \rm{dex}) than the ``total-SFR'' FMR as we move towards the low-mass, high-SFR region.
This is the direct, natural consequence of assigning a galaxy with fixed \mstar and metallicity to a higher SFR bin when applying the aperture corrections.
For completeness, and to allow a proper comparison of the predictions of the FMR presented in this paper with the largest possible variety of data (i.e. from fiber and/or IFU spectroscopy, on local and high-z galaxies), we therefore provide also the best-fit parameters of the FMR based on SFR measurements inside the fibre: these are reported in Table~\ref{tab:fmr_params}.

The metallicity predictions provided by the FMR of equation~\ref{eq:FMR} span a wider region of the M$_{\star}$, SFR parameter space than that covered by the calibration sample of local SDSS galaxies. Indeed, the FMR is often used to predict the metallicity outside the ranges of mass and SFR where it is defined, especially in high redshift studies, where galaxies are characterised, on average, by higher SFRs compared to local ones. 
This means that, when comparing  metallicities observed in high-z sources, even when rescaled to the proper \Te\ abundance scale, one is often forced to rely on extrapolations of the locally calibrated FMR, mainly in the high-SFR regime, and this effect should be carefully taken into account when trying to assess and interpret the evolution of the FMR with redshift.

In order to give an estimate of the uncertainties associated with the extrapolations of the FMR represented by equation~\ref{eq:FMR} 
(implicitly assuming the validity of the underlying physical background), we run an MCMC by letting the different FMR parameters to randomly vary within their uncertainties following a normal distribution.
We thus generate $1000$ different realizations of the FMR and compute the dispersion in the metallicity predictions at each fixed value on a \mstar and SFR grid. 
The results of this test are shown in Fig.~\ref{fig:fmr_extrapol}. 
For comparison, the M$_{\star}$-(total)SFR binning scheme adopted in this work, which defines the region of the parameter space covered by local galaxies, is superimposed in black.
The higher uncertainties (of the order of $\gtrsim0.3$ \rm{dex}) are found to occur in the low-mass, high-SFR regime (the region of the FMR with the lowest predicted metallicity); this is not surprising, as it is mainly driven by the uncertainties on the low-mass end slope.
We also note that this represents the region of most extreme extrapolation, where we do not have any constraint from the observed data.
We stress here that  Fig.~\ref{fig:fmr_extrapol} gives only a rough estimate of the typical uncertainty associated with the hereby presented analytical form of the FMR. 
All the potential uncertainties related to measurement errors of the involved quantities, as well as those arising from the choice of a different parametrisation, are not considered at this stage.
%In this sense, they can be considered as lower-limit of the true un
However, every observational study aiming at comparing measured metallicities with extrapolation of the  FMR provided in this work should consider this minimum level of uncertainty in the interpretation of the results.

\begin{figure}
	\centering
	\includegraphics[width=0.98\columnwidth]{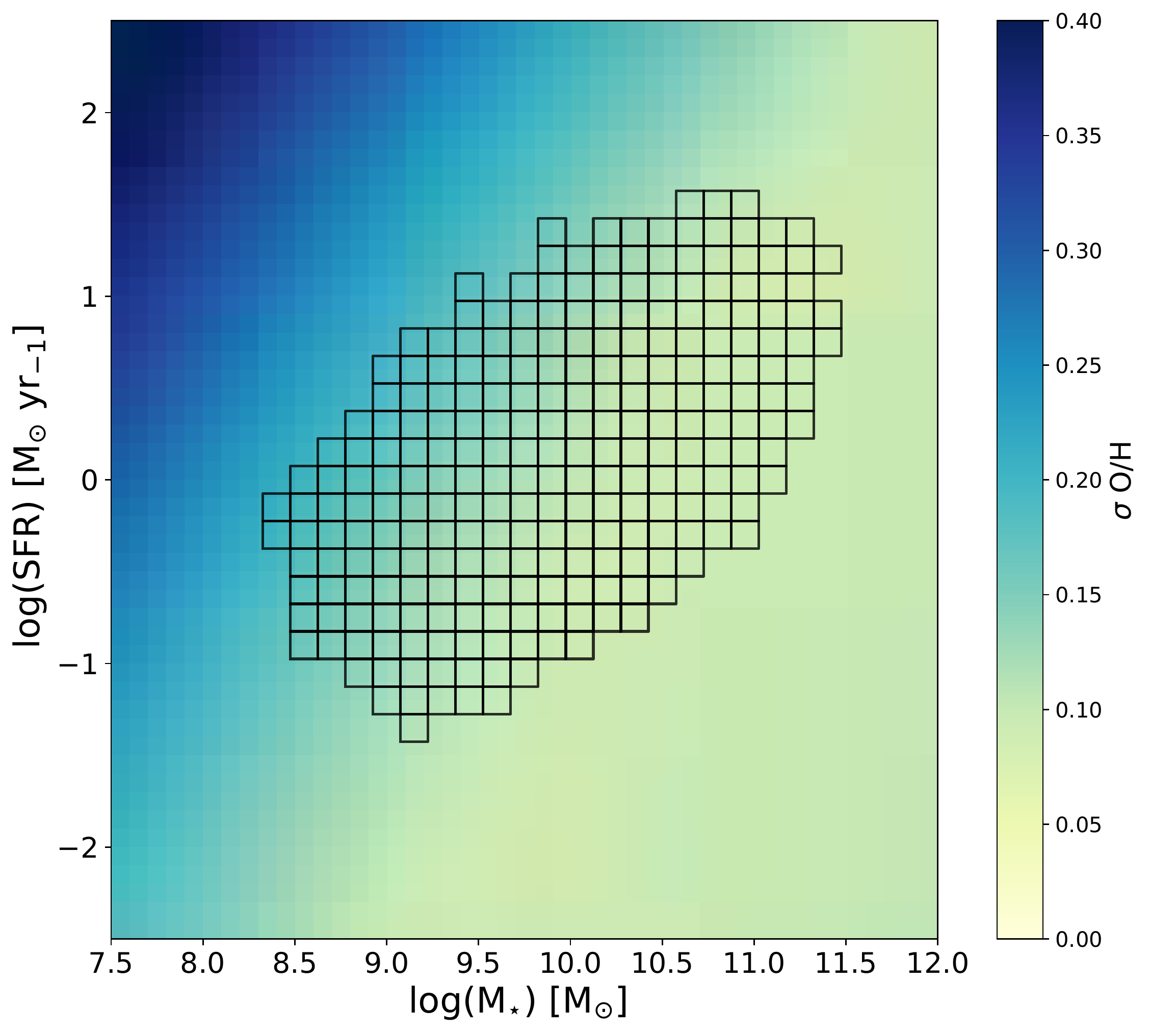}
	\caption{Minimum uncertainties associated to the metallicity predictions provided by the FMR parametrisation of equation~\ref{eq:FMR}, as computed from $1000$ different realisations obtained varying the FMR parameters in an MCMC simulation. The \mstar-SFR binning grid defined in this work by local SDSS galaxies is superimposed in black. 
	The higher level of uncertainty ($\gtrsim 0.3$ \rm{dex}) is obtained in the low-mass, high-SFR regime, outside the region covered by the local sample. 
	These values are propagated from the errors on the best-fit parameters of the FMR and do not take into account other sources of uncertainty, like those associated to the measurements of involved quantities and those simply related to extrapolating a given functional form outside the regime covered by the data.
	}
	\label{fig:fmr_extrapol}
\end{figure}

\section{Summary and Conclusions}
\label{sec:disc}

We have analysed the scaling relations between stellar mass, metallicity and star formation rate in the local Universe in light of the metallicity calibrations introduced by \cite{Curti:2017aa}, and complemented in this work, which are fully based on the \Te\ abundance scale defined for SDSS galaxies (Fig.~\ref{fig:met_cal_recap}).
The main conclusions reached in this paper can be summarised as follows .
\begin{itemize}
	
	\item We have parametrised the mass-metallicity relation with a new functional form (Equation~\ref{eq:mzr} and upper panel of Fig.~\ref{fig:mzr_1}) which allows us to control the width of the knee ($\upbeta$) and better capture the value of the turnover mass (M$_{0}$), which we find to occur at log(\mstar/\msun)$=10.02 \pm 0.09$. The low-mass end slope of our MZR is $\gamma = 0.28 \pm 0.02$ and approaches the saturation metallicity at Z$_{0}=8.793 \pm 0.005$ (i.e. about $1.27$ times Z$_{\odot}$). The dispersion in metallicity of individual galaxies around the median relation is $0.07$ \rm{dex}.

	\item A comparison between our new MZR and previous assessments in the literature is shown in the bottom panel of Fig.~\ref{fig:mzr_1}. Our MZR deviates from those based on abundances predicted by grids of photoionisation models, in the sense of a lower normalisation of the high-mass regime of $\sim 0.3$ \rm{dex}, whereas showing good consistency with different determinations of the relation based on \Te-metallicities.
	The agreement with \cite{Andrews:2013aa} in particular is remarkable, despite a small divergence in the low mass regime. 
	The MZR by \cite{Yates:2019aa} presents a systematic offset towards lower abundances at fixed stellar mass, possibly due to selection effects driven by the requirement of a [\ion{O}{iii}]$\lambda4363$ detection, or by the different average SFR (and size) of the studied sample. 
	Our strong-line MZR is also consistent with the independent measurements of chemical abundances obtained by means of spectroscopy of blue and red supergiants in very nearby galaxies \citep{Davies:2017ab} at high masses, while slightly deviating from the median relation below log(\mstar) $< 9.5$, nevertheless confirming the good agreement between the \Te\  scale for abundances of the ISM and that defined by stellar abundances as measured in young, massive stars.
	
	\item The MZR shows a clear dependence on star formation rate, both considering total SFR or SFR within the fibre, which is
	more evident in low \mstar - high SFR regimes (Fig.~\ref{fig:fmrhacorrfinalmcmc}, left panels). 
	The dependence of metallicity on SFR, at fixed stellar mass, can be also visualised by plotting log(O/H) against (s)SFR for median-binned values (Fig.~\ref{fig:fmrhacorrfinalmcmc}, right panels) or the MZR residuals vs $\rm \Delta$sSFR (i.e. the distance from the SFMS) for individual galaxies in narrow mass-bins (Fig.~\ref{fig:fmr_non_parametric}, as originally suggested by \citealt{Salim:2014aa}).
	The anti-correlation between metallicity and SFR appears strong at low masses, decreasing for increasing \mstar until disappearing at high masses.
	When accounting for the Z-SFR dependency, the scatter in each individual mass bin is reduced, for individual galaxies, by a factor $\sim 15\%$ at intermediate masses and up to a factor of $\sim30 \%$ in some of the lowest mass bins.
	However, intense star-forming galaxies (log(sSFR) $\lesssim -9.5$) maintain the dependence between log(O/H) and (s)SFR at almost all masses.
	
%	\item We studied discuss fibre- and total- SFR....
	
	\item We have parametrised the M-Z-SFR relation with the same functional form adopted for the MZR, investigating the dependence of its main parameters on SFR (Fig.~\ref{fig:mzrs_sfr}).
	The turnover mass M$_{0}$ shows a clear trend with varying SFR, while the saturation metallicity does not change.
	The turnover mass increases with SFR, as a possible consequence of the different gas-to-star mass ratio in highly star forming galaxies (as also suggested by \citealt{Zahid:2014aa}).
	The variations of the slope $\gamma$ with SFR is much more shallow and may be affected by how different populations of galaxies react to the effects of outflows; however, this latter quantity is also sensitive to the choice of the metallicity diagnostics and to selection effects.
	% (see also Appendix~\ref{sec:app_systematics}).
	
	\item The scatter of galaxy population is reduced when considering a relation in the three-dimensional space defined by \mstar, metallicity and SFR, the so-called Fundamental Metallicity Relation (FMR). 
	We explicitly introduced the dependence of M$_{0}$ on SFR to derive a new functional form of the FMR (equation~\ref{eq:FMR} and Fig.~\ref{fig:FMR_3d}). The scatter around this new relation is only $0.028$ dex for median values in bins of \mstar and SFR and $0.054$ for the global population of individual galaxies, a reduction of $\sim22\%$ compared to the scatter around the MZR only.
	However, we note that a large contribution to the residual global scatter comes from highly populated high \mstar bins, where the effects of the FMR are less relevant.

   \item The new parametrisation of the FMR provided here represents a local benchmark to be compared with chemical evolution models and observations (of both local and high-redshift galaxies) which are tied to the \Te\ abundance scale.
	An estimate of the (minimum) uncertainties associated to the metallicity predictions of the new FMR is presented in Fig.~\ref{fig:fmr_extrapol}: the uncertainties increase up to $\sim0.3$ \rm{dex} in the low-mass, high-SFR regime, outside the region sampled by local galaxies.
	
%	A potential simplification comes from considering only the variations in the turnover mass (equation~\ref{eq:FMR}): despite being less representative of the median values, the simplified FMR does not introduce sensible scatter in the overall population, while providing easy and reasonable extrapolations in the low-mass, high-SFR regime (populated by the majority of high redshift sources).	

%	\item In order to quantify the strength of the secondary dependence of the MZR on SFR, we can search for the 2D projection which minimizes the scatter around the Z vs $\mu_{\alpha}$ (i.e \mstar - $\alpha$log(SFR)) relation. The best-fit $\alpha$ value for the global sample ($\alpha=0.44$) reveals a tighter correlation between Z and SFR than determined in \cite{Mannucci:2010aa}.
%	However, the reduction of the dispersion is more prominent in the low-sSFR regime ($\sigma=0.057$ for log(sSFR)$ < -10$ galaxies), where the secondary dependence on SFR is weaker ($\alpha=0.23$), compared to the high-sSFR regime ($\sigma=0.18$ for log(sSFR)$ > -10$ galaxies), where the dependence between Z and SFR becomes tighter ($\alpha=0.57$).
%	This confirms that the relative importance of the SFR in the relation changes not only as a function of stellar mass, but also as a function of the (s)SFR itself.
	
%	\item ...
	
\end{itemize}

\section*{Acknowledgements}
MC and RM acknowledges support from the ERC Advanced Grant 695671 ``QUENCH'' and support from the Science and Technology Facilities Council (STFC).
FM and GC acknowledge funding from the INAF PRIN-SKA 2017 program 1.05.01.88.04.
We greatly appreciate the MPA/JHU group for making their catalog public.
Funding for the SDSS and SDSS-II has been provided by the Al- fred P. Sloan Foundation, the Participating Institutions, the National Science Foundation, the U.S. Department of Energy, the National Aeronautics and Space Administration, the Japanese Monbukagakusho, the Max Planck Society and the Higher Education Funding Council for England.

This research made use of Astropy, a community-developed core Python package for Astronomy \citep{The-Astropy-Collaboration:2018aa}.

\bibliographystyle{mnras}
\setlength{\labelwidth}{0pt}
\bibliography{/Users/mirkocurti/Google_Drive_pro/astro_biblio.bib}

\section*{Supporting Information}

\section*{Appendix A: The projection of minimum scatter}
\label{sec:projection}

Following \cite{Mannucci:2010aa}, we search for the 2D-projection of the FMR that reduce the metallicity scatter around the relation Z- $\mu_{\alpha}$, where $\mu_{\alpha} = \text{log}(M_{\star}) - \alpha\ \text{log}(\text{SFR})$.
%this also represents a simple way to give a quantitative estimate of the strength of the secondary dependence of the MZR on SFR.
%This was initially proposed by \cite{Mannucci:2010aa} in the form of a linear relation between O/H and a new quantity defined as : 
%\begin{equation}
%\label{eq:projection}
%\mu_{\alpha} = \text{log}(M_{\star}) - \alpha\ \text{log}(\text{SFR}) \,
%\end{equation} 
In this framework, $\upalpha$ is the parameter which quantifies the strength of the Z-SFR correlation at fixed stellar mass : for $\upalpha = 0$ the relation reduces to the MZR, meaning no correlation between metallicity and SFR, while larger values of $\upalpha$ would imply stronger correlation between metallicity and SFR.
We fit our M-SFR bins (we here consider total-SFR only) against log(O/H) according to the same functional form used for the MZR (i.e. Eq.~\ref{eq:mzr}), where the x-variable is now $\mu_{\alpha}$: the results are show in Fig.~\ref{fig:projections} and reported in the first row of Table~\ref{tab:proj_min_scatter}.
We obtain a best-fit $\upalpha$ value of $0.55$, larger than what found by \cite{Mannucci:2010aa} ($\upalpha=0.32$) but lower than other previous estimates (e.g. $\upalpha=0.66$ as found  by \citealt{Andrews:2013aa}); however, it is well known that the relative strength of the secondary dependence of the MZR on the SFR can be strongly affected by many factors, primarily related to selection biases and the choice of the metallicity calibrations.
Moreover, we have seen in the previous sections how the tightness of the Z-SFR relation changes as a function of the SFR itself; 
this induce a change in the slope of the different MZRs at fixed SFR, whose effect can be clearly seen in the upper panel of Fig.~\ref{fig:projections}, where the residuals around the best-fit still correlate with SFR as highest star-forming galaxies (blue points) would require steeper slopes to be reproduced compared to low star-formers (red points).
Indeed, to obtain a perfect and linear 2D projection on the Z-$\mu_{\alpha}$ plane would require only a variation in M$_{0}$, such that the slopes of the different MZRs at fixed SFR do not intercept when rotating the FMR around the ``log(O/H) axis''.
This can be also visualised by plotting the metallicity against \mstar normalised to the turnover mass M$_{0}$ for each SFR subsample (similarly to what done by \citealt{Zahid:2014ab} for samples at different redshifts), fixing at the same time the slope $\upgamma$: if the evolution of the MZR with SFR resides entirely in the variation of M$_{0}$, such a change of variable should remove the scatter around the new relation (similarly to what achieved by the 2D projection on  Z-$\mu_{\alpha}$). This is shown in the upper right panel of Fig.~\ref{fig:projections}, where we find qualitatively a similar behaviour as in the left panel: the majority of the SFR-dependence is removed, but a residual effect, related to the variation in the slope $\upgamma$ with the SFR, is still present.
The same result is obtained also assuming the modified \citealt{Zahid:2014ab} functional form of the MZR presented in Eq.~\ref{eq:zahid14}.

Therefore, the two upper panels of Fig.~\ref{fig:projections} both show that accounting for the variation in M$_{0}$ only is not enough to completely remove the SFR dependency of the MZR (although accounting for its primary source) or, in other words, that the same result can not be achieved by a simple 2D-projection of the FMR, due to the fact that the variation of the strength of the Z-SFR correlation with SFR itself is responsible for the change in the slope for the different MZRs at fixed star formation rate.
%\textbf{spiegare l'effetto della variazione della slope sulla proiezione e di come assumere solo M0 variabile equivalga alla proiezione stessa se plotto Z vs b*log(M/M0) }
%Therefore, it is perhaps more interesting to investigate how the tightness of the Z-SFR anti-correlation varies as a function of the SFR itself.

However, from what we have seen in Fig.~\ref{fig:fmrhacorrfinalmcmc}, two different regimes can be approximately identified where the shape of the Z-SFR relation, at fixed stellar mass, changes substantially, steepening at all masses for sSFR $\gtrsim 10^{-9.5}$ yr$^{-1}$.
Thus, we can divide our total sample into two subset of galaxies, the ``high sSFR'' sample (log(sSFR)$>-9.5$) and the ``low sSFR'' sample (log(sSFR)$<-9.5$) respectively, bin them in \mstar and SFR and perform the fit again, whose results are now shown in the bottom panels of Fig.~\ref{fig:projections} and reported in Table~\ref{tab:proj_min_scatter}.
For the low-sSFR sample, the strength of the Z-SFR dependence is weaker, as expected ($\upalpha = 0.22$); however, the 2D projection now considerably reduces the scatter of median metallicities for this galaxy population around the new relation for all SFRs. 
%with the dispersion of the median-binned values being only $0.009$, and the dispersion of individual galaxies equal to $0.057$ dex.% when considering the dispersion around the MZR of the same population of galaxies (i.e., those with ...) is ....
For the high-sSFR sample instead,  the Z-SFR dependence is now stronger ($\upalpha = 0.65$), but the small number of objects populating this regime in the local Universe (only the $\sim 15\%$ of the sample analysed in this work) and the largest scatter of these galaxies in each M-SFR bin do not allow to obtain the same level of reduction of the dispersion as for the low-sSFR sample. %Moreover, in this latter case, the lack of high mass galaxies in the subsample (being high-sSFR galaxies mainly low mass objects which have recently experienced accretion of large amount of gas) do not allow to set proper constraints on the turnover and asymptotic value of the relation; the 2D projection of the high-sSFR subsample can thus be nicely parametrized also with a linear fit, whose slope and intercept are also given in Table ...

%now dominates over the M-Z dependence ($\upalpha = 1.13$),
% 
%
%\begin{figure*}
%	\centering
%	\includegraphics[width=0.45\textwidth]{figures/fmr_Ha_corr_final_MCMC}
%	\includegraphics[width=0.45\textwidth]{figures/sfr_oh_Ha_corr_final_MCMC}\\
%	\includegraphics[width=0.45\textwidth]{figures/fmr_sSFR_Ha_final_MCMC}
%	\includegraphics[width=0.45\textwidth]{figures/ssfr_oh_sSFR_Ha_final_MCMC}
%	
%	\caption{\textit{Upper Panels} : M-Z-SFR relation for our sample. Different median MZRs in (0.25 dex wide) bins of SFR are plotted in the left panel, highlighting the secondary dependence of the mass-metallicity relation on the SFR, especially at low masses and high star formation rates.
%	In the right panel instead, the relation between log(O/H) and SFR is plotted for different bins of stellar mass.
%	\textit{Lower Panels} : Same as above, assuming the sSFR as the third variable.}
%	\label{fig:fmrhacorrfinalmcmc}
%\end{figure*}

\begin{figure*}
	\centering
	\includegraphics[width=0.45\textwidth]{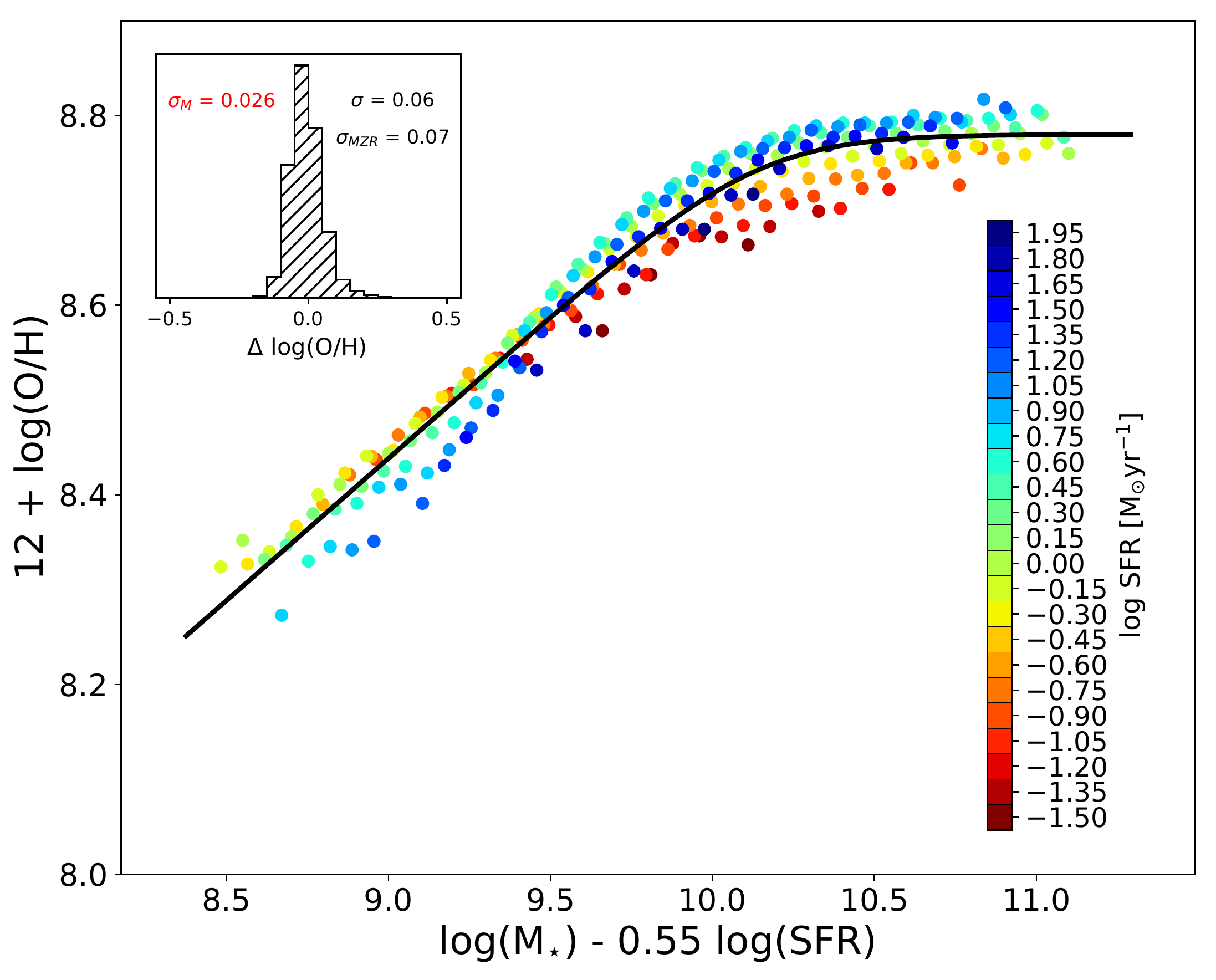}
	\includegraphics[width=0.45\textwidth]{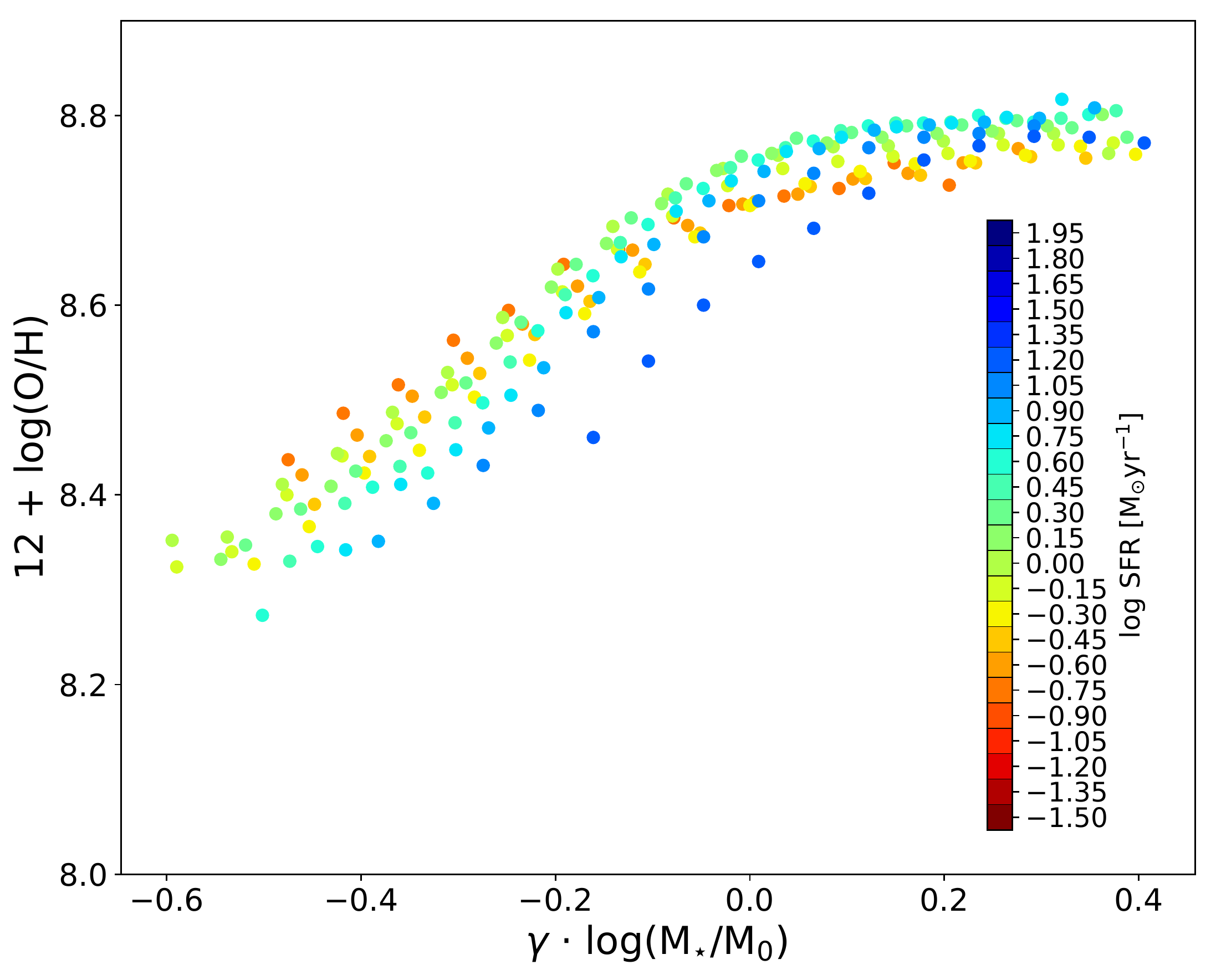}\\
	
	\includegraphics[width=0.45\textwidth]{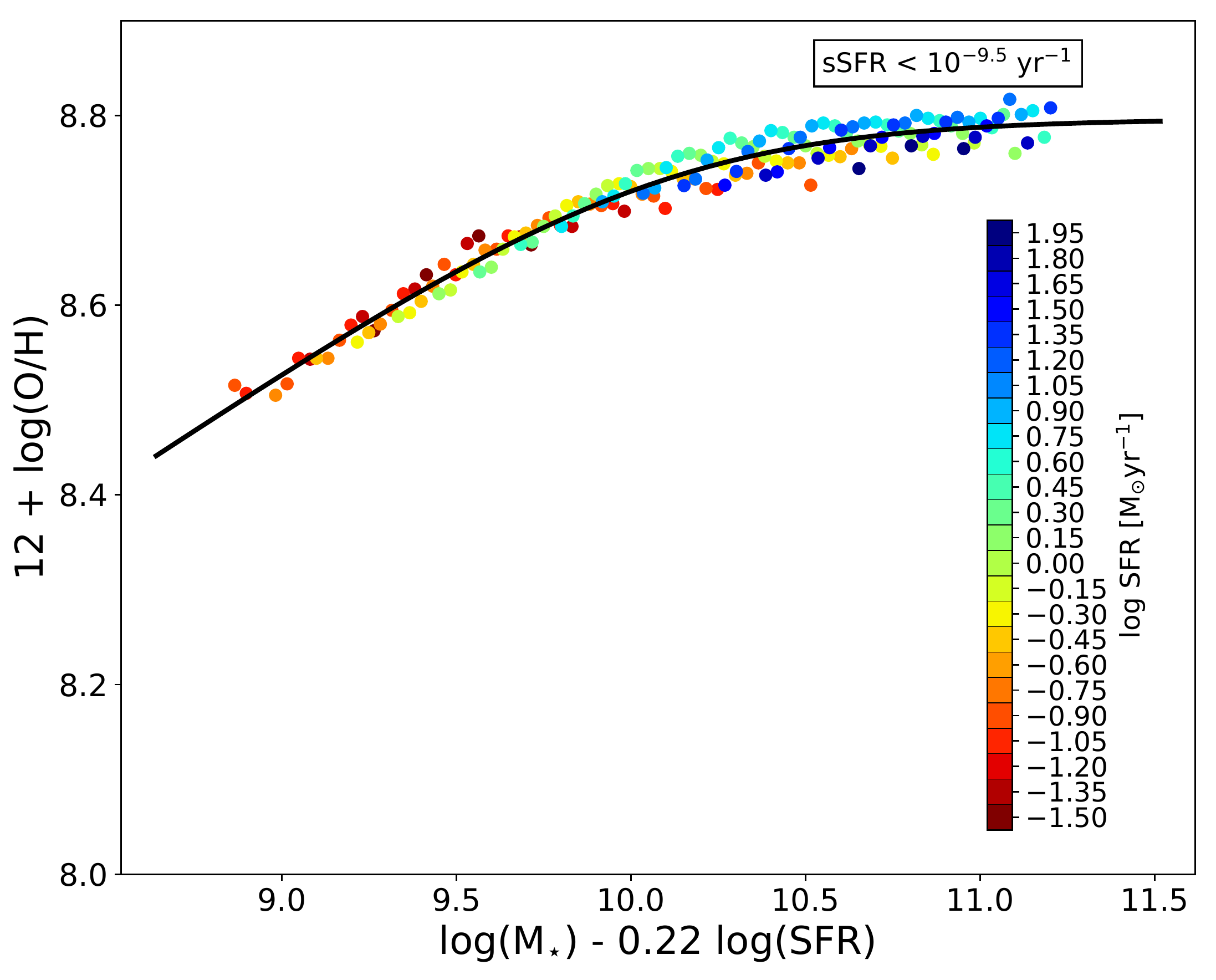}
	\includegraphics[width=0.45\textwidth]{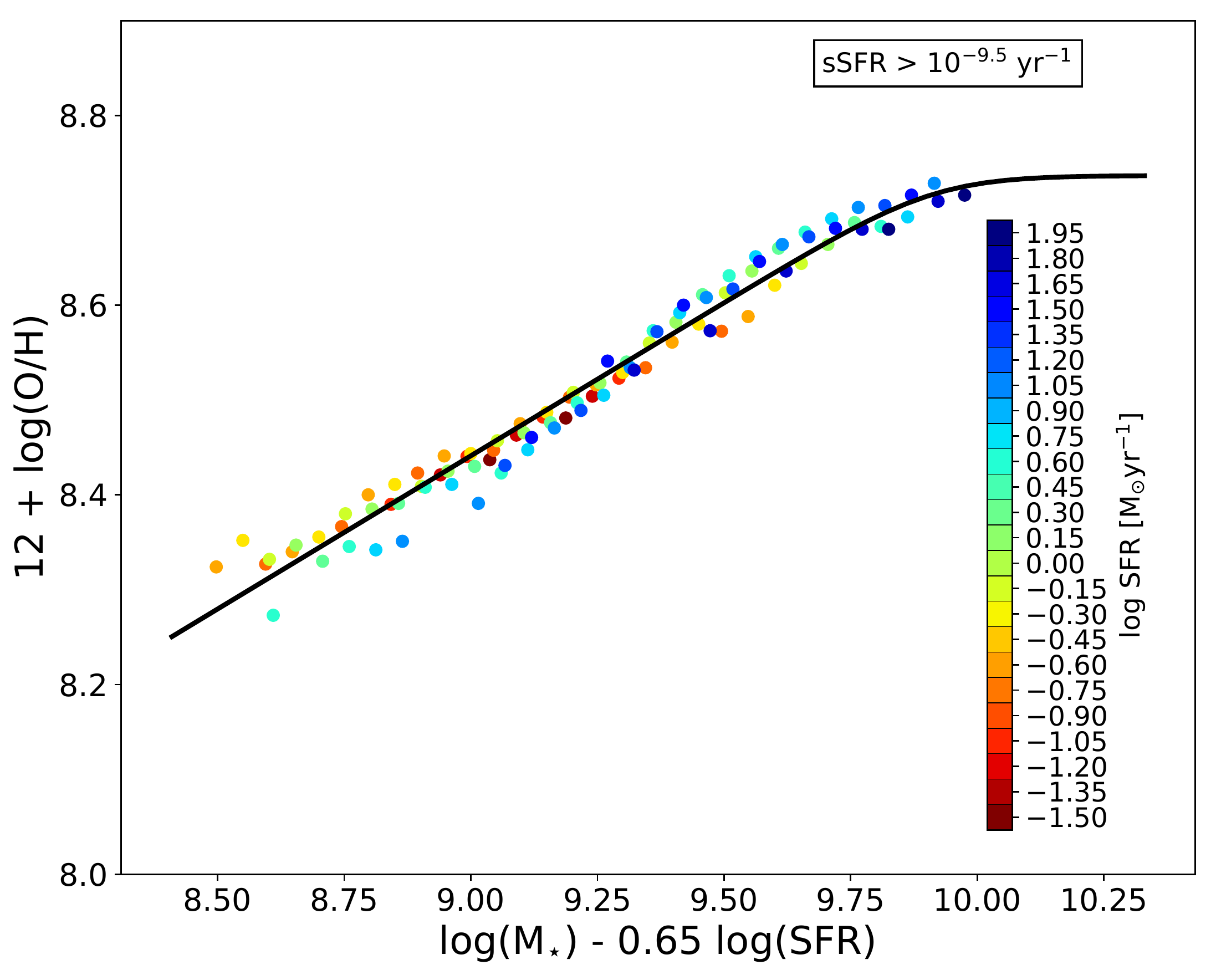}
	
	\caption{
		\textit{Upper left panel}: 2D projection of the M-Z-SFR relation on the Z-$\mu_{\alpha}$ plane, with $\alpha = 0.55$. This projection minimizes the scatter of the median-binned metallicities around the relation defined as 12+log(O/H) = log(M$_{\star}$)$-\alpha$log(SFR). All the points are color coded according to their total SFR. In the small box, the histogram of metallicity dispersion of individual galaxies around the surface is reported, together with the dispersion of the median-binned values around the best-fit relation ($\sigma_{M}$, in red), the scatter of individual galaxies around the best-fit relation, and the scatter of individual galaxies around the MZR.
		The variation in the slope for the different SFR regimes is visible, which can not simply be accounted for by a projection on the Z-$\mu_{\alpha}$ plane.
		\textit{Upper right panel}: Median metallicity plotted against stellar mass normalised to the turnover mass M$_{0}$, for each SFR subsample. The residual trend with SFR  seen at low masses is indicative of the variation of the slope $\gamma$ with SFR. 
		\textit{Bottom left panel}: 2D projection for the ``low-sSFR'' subsample (sSFR$<10^{-9.5}$ [yr$^{-1}$]). The Z-SFR dependence is weaker ($\alpha = 0.22$), but the dispersion around the relation is strongly reduced.
		\textit{Bottom right panel}: 2D projection for the ``high-sSFR'' subsample (sSFR$>10^{-9.5}$ [yr$^{-1}$]). The Z-SFR dependence is tighter ($\alpha = 0.65$), but the dispersion is slightly larger compared to the ``low-sSFR'' sample.
	}
	\label{fig:projections}
\end{figure*}

\begin{table*}
	\centering
	\caption{Best fit parameters for the 2D projections of the FMR on the Z-$\mu_{\alpha}$ plane, for different SFR regimes. }
\begin{tabular}{|@{}l | c|c|c|c|c @{}|}

%			\multicolumn{6}{c}{Fundamental Metallicity Relation   - Projections onto the \mstar-$\mu_{\alpha}$ plane}\\ 		
			\hline
			
			&  $\alpha$ & Z$_{0}$ &  $\mu_{0}$ &  $\gamma$  &  $\beta$ \\ 
			\hline
			\text{Global Sample} &  0.55 $\pm$ 0.01 & 8.780     $\pm$ 0.004 & 10.14    $\pm$ 0.03 & 0.30      $\pm$ 0.01 & 2.4  $\pm$ 0.4 \\
			\text{sSFR}$< 10^{-9.5}$ Gyr$^{-1}$  & 0.22 $\pm$ 0.02 & 8.796     $\pm$ 0.005 & 10.1    $\pm$ 0.1 & 0.25      $\pm$ 0.03 & 1.1      $\pm$ 0.2 \\			
			\text{sSFR}$> 10^{-9.5}$ Gyr$^{-1}$  & 0.65 $\pm$ 0.01 & 8.74     $\pm$ 0.04 & 9.9    $\pm$ 0.1 & 0.32   $\pm$ 0.01 & 5.0   $\pm$ 3.0 \\
			
			\hline
\end{tabular}	

\label{tab:proj_min_scatter}
\end{table*}

\section*{Appendix B: Systematics of the metallicity calibrations}
\label{sec:app_systematics}

It is well known that indirect, strong-line abundance diagnostics do not always agree with one another when used individually.
Although the calibration set presented in Fig.~\ref{fig:met_cal_recap} is characterised by a good level of self-consistency (see Figs.10 and 11 of \citealt{Curti:2017aa}), it is impossible to completely remove the underlying discrepancies, given the intrinsically different physical mechanisms responsible for the dependences between the various line-ratios considered and the oxygen abundance.
Therefore, in order to assess the amplitude of possible systematics in our analysis, here we study the MZR and the M-Z-SFR relation as obtained adopting only nitrogen-based and only oxygen-based diagnostics.
We note here that in this case, on top of the original sample selection described in Sect.~\ref{sec:selection}, we had to apply a SNR cut also on the \nii, \oii\ and \oiii\ lines in order to perform a meaningful comparison.

In the upper panels of Fig.~\ref{fig:diag_comparison} we show the MZR obtained adopting only a combination of the R$_{2}$ and R$_{3}$ diagnostics on the left, whereas the MZR based only on the N$_{2}$ diagnostic is shown on the right.
In both panels, the ``original'' MZR derived by combining all the diagnostics together is also shown in blue for reference.
In the central panels of the same Figure, the M-Z-SFR relation (as in Fig.~\ref{fig:fmrhacorrfinalmcmc}) in shown in the two different cases as well.
It can ben seen that the N$_{2}$ calibration provides lower abundances (on average) than the combination of R$_{2}$ and R$_{3}$.  
This translates into a slightly lower normalisation of the MZR, which produces in particular a lower asymptotic metallicity and a steepening of the slope in the low-mass end.
Moreover, the strength of the Z-SFR relation at fixed stellar mass in increased when considering the R$_{2}$+R$_{3}$ metallicity, especially in the high-mass, high-SFR regime, as a possible consequence of the different impact that the ionisation parameter has on the different abundance diagnostics, in particular R$_{3}$ (see also the discussion in Sect.~\ref{sec:m_z_sfr_params}).

Overall, these effects are of the order of $\sim 0.025-0.03$ {\rm dex} and appear more prominent when considering the N$_{2}$ diagnostic alone. 
Nonetheless, they do not prevent the detection of the secondary Z-SFR dependence, which is clearly visible in both representations.
This can be also clearly seen in the bottom panels of Fig.~\ref{fig:diag_comparison} where we plot, as a function of stellar mass, the difference between the median metallicity in bins of \mstar (white points) and bins of \mstar-SFR (colored points) computed with the two different combination of diagnostics (i.e. [R$_{2}$,R$_{3}$] and N$_{2}$) and the ``original'' metallicity adopted throughout the paper (inferred involving all the diagnostics simultaneously).
However, we note that for \mstar lower than $10^{9}$\msun and for \mstar$>10^{10}$\msun and log(SFR)$\gtrsim 1$ the two predictions can diverge up to $\sim 0.05$\rm{dex}, modifying, as we have seen, the shape of the low-mass end of the MZR and/or the amplitude of the secondary Z-SFR dependence.
Therefore, we stress the importance that the simultaneous combination of multiple emission line diagnostics has in minimising these potential systematics effects.

\begin{figure*}
	\centering

	\includegraphics[width=0.45\textwidth]{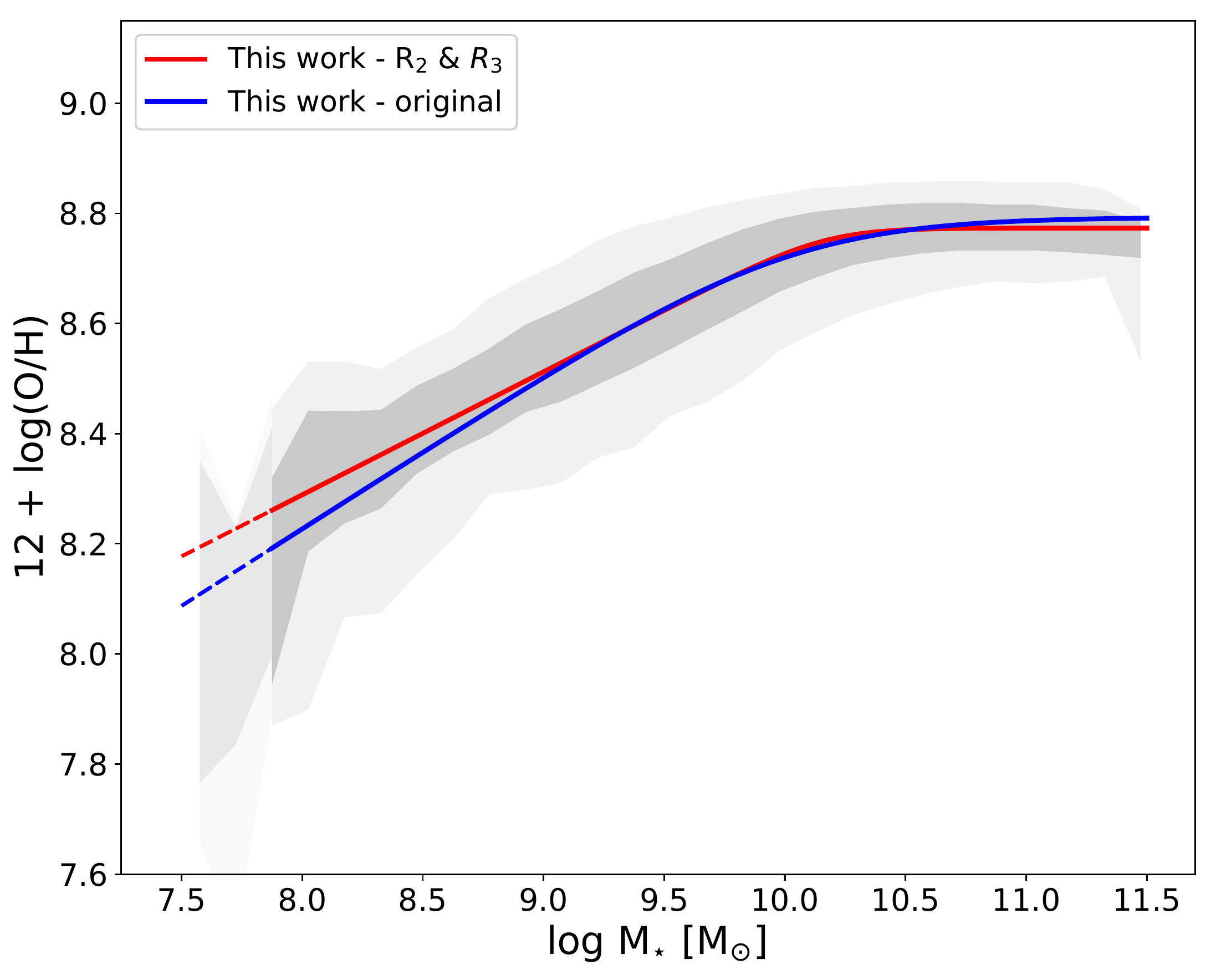}
	\includegraphics[width=0.45\textwidth]{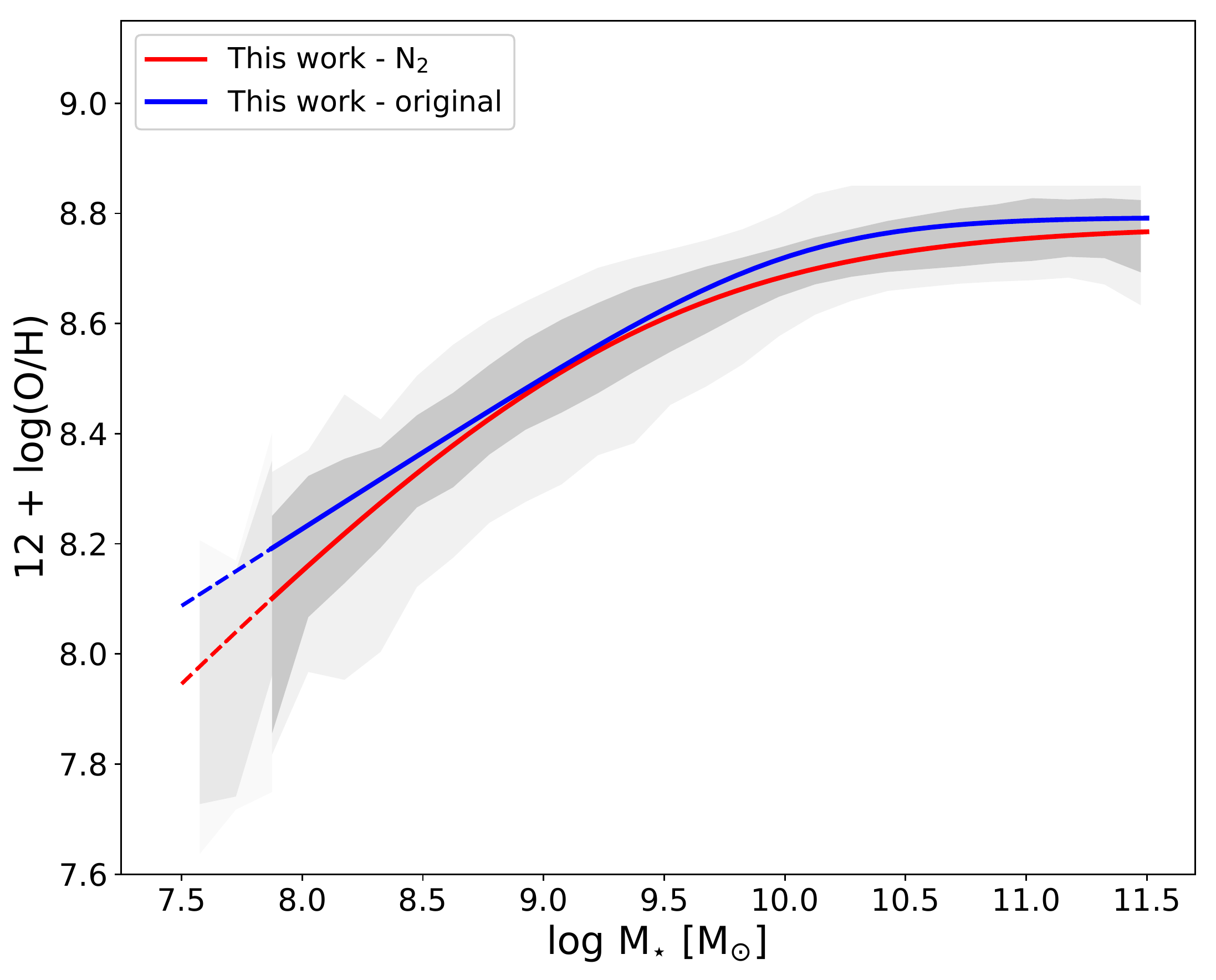}\\
	\includegraphics[width=0.45\textwidth]{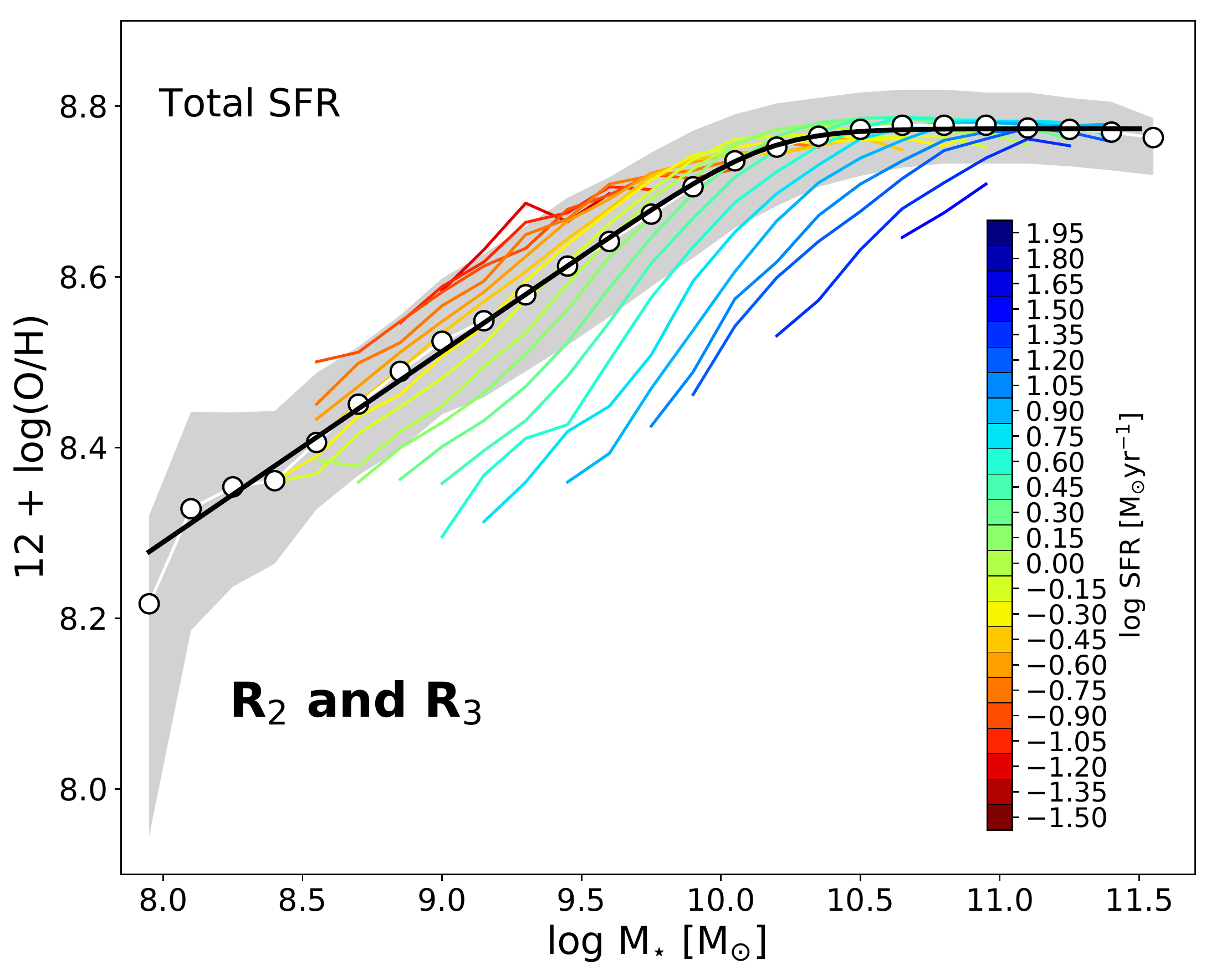}
	\includegraphics[width=0.45\textwidth]{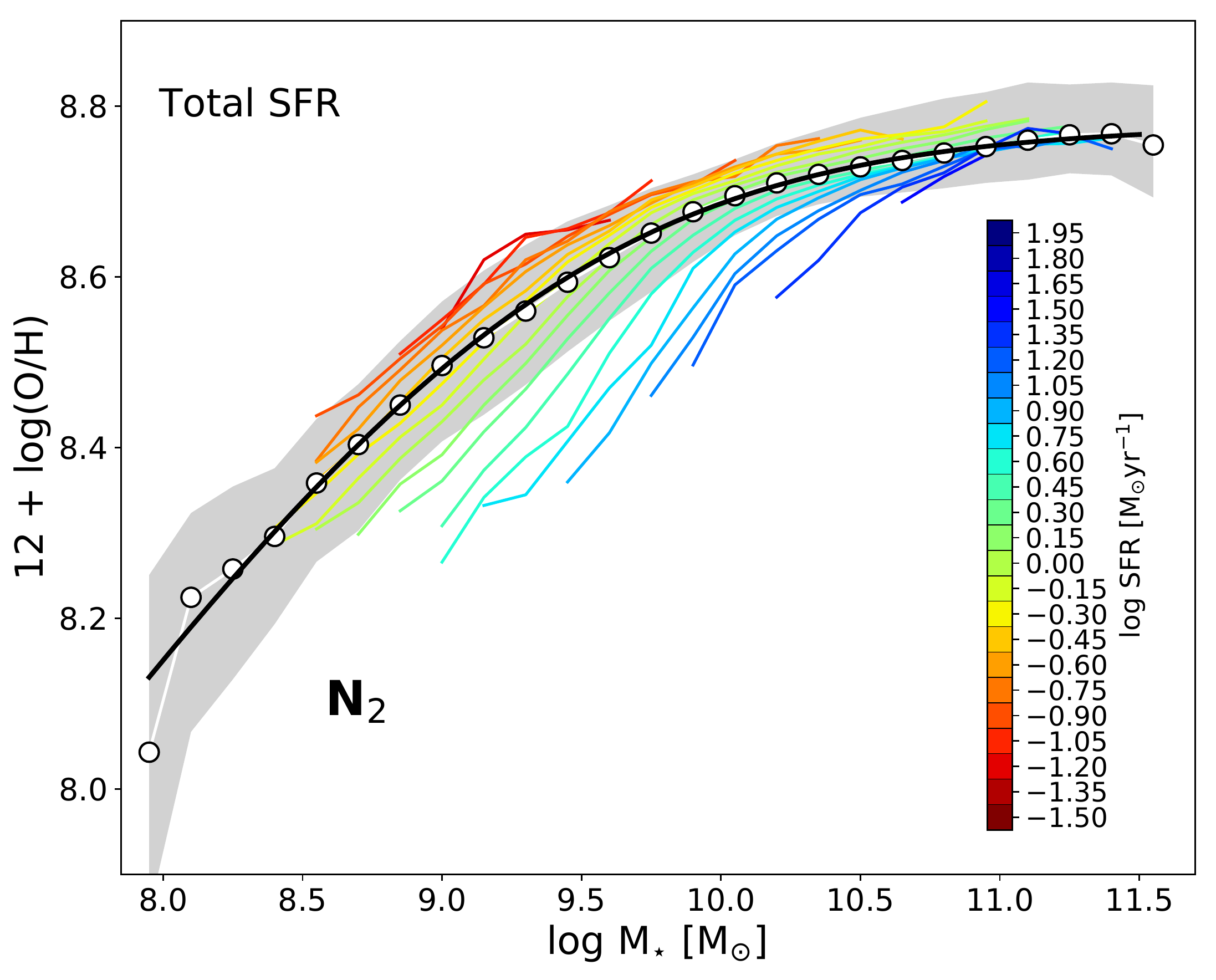} \\
	\includegraphics[width=0.45\textwidth]{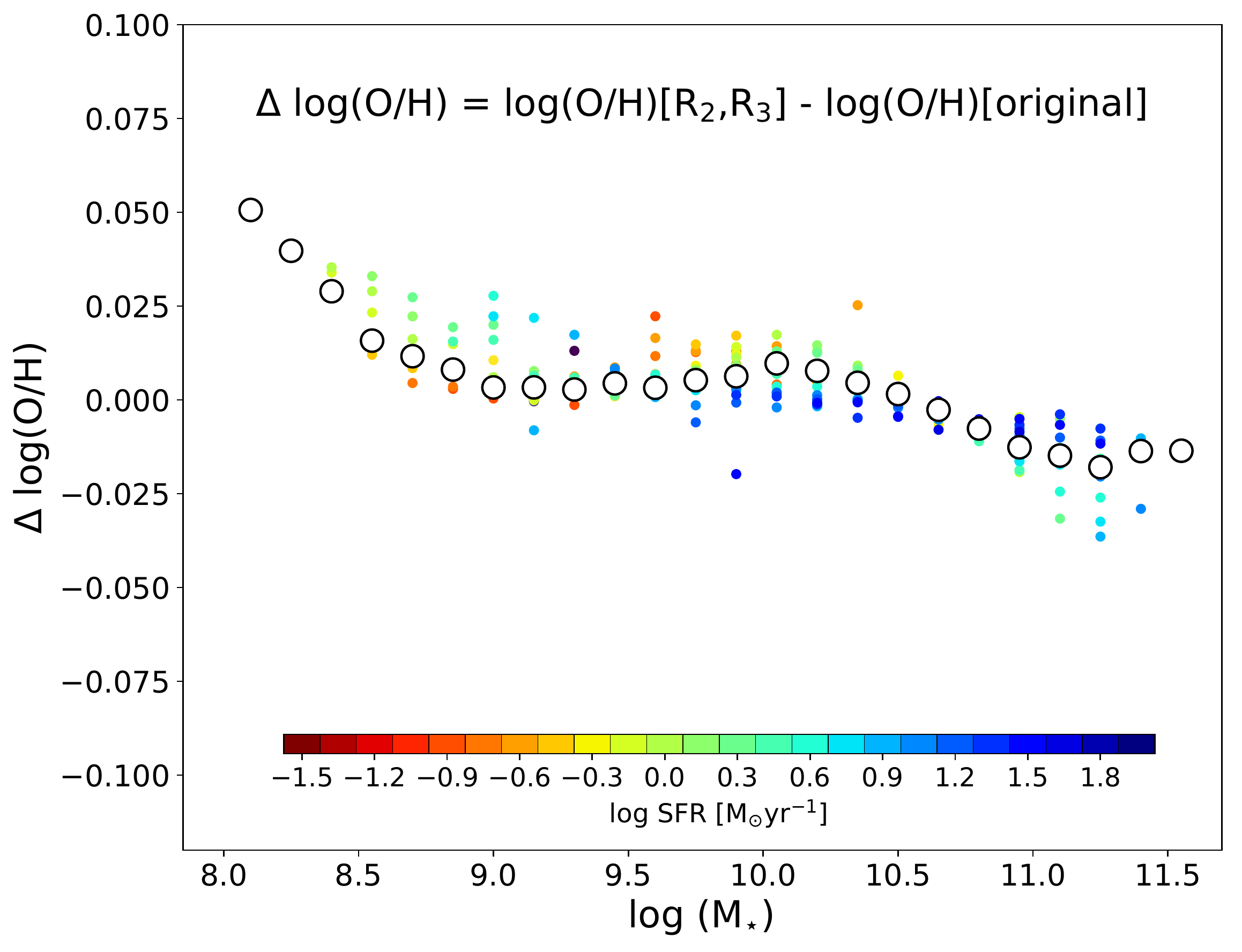}
	\includegraphics[width=0.45\textwidth]{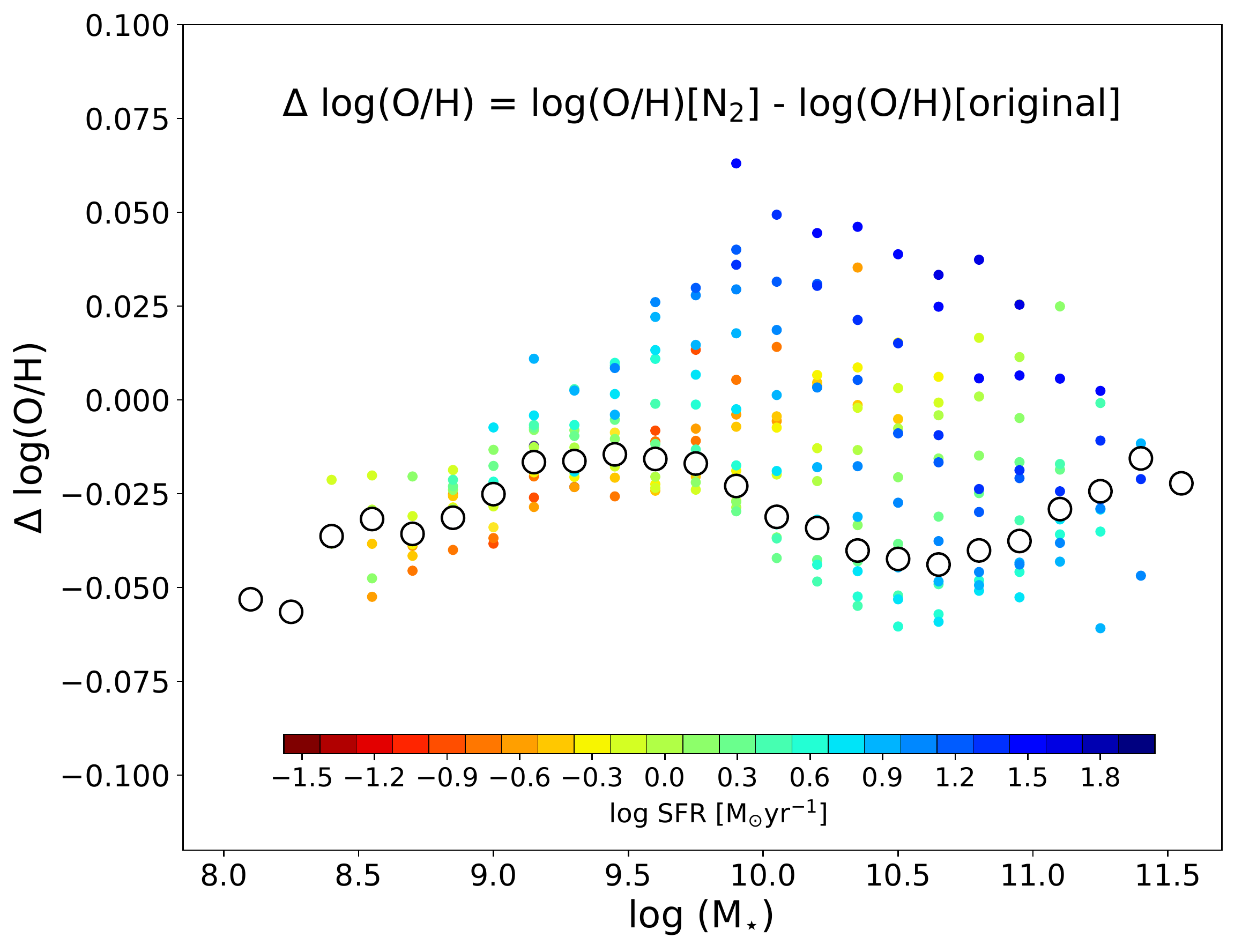} 

	\caption{Mass-Metallicity and M-Z-SFR relations derived adopting only oxygen-based (i.e. R$_{2}$ and R$_{3}$, left panels) or nitrogen-based (i.e. N$_{2}$, right panels) diagnostics. In the top panels, the ``original'' MZR considered throughout the paper is shown for reference in blue.
	In the bottom panels we plot, as a function of stellar mass, the difference between the metallicity in bins of stellar mass (white points) and bins of \mstar-SFR (colored points) computed with the relative combination of diagnostics and the ``original'' metallicity adopted throughout the paper (i.e. inferred involving all the diagnostics simultaneously).
	Symbols and colours are the same as in Fig.~\ref{fig:mzr_1} and Fig.~\ref{fig:fmrhacorrfinalmcmc}.
	}
	\label{fig:diag_comparison}

\end{figure*}

\end{document}